\documentclass{aastex61}
\usepackage{amsmath}
\usepackage{amssymb}
\usepackage{lineno}
\usepackage[caption=false]{subfig}

\usepackage[outdir=./]{epstopdf}

\makeatletter
\def\input@path{{./picts/}}
\makeatother

\graphicspath{{./picts/}}

\newcommand\mb{\mathbf}
\newcommand{\bea}{\begin{eqnarray}}
\newcommand{\eea}{\begin{eqnarray}}
\newcommand{\beg}[1]{\begin{equation}\label{#1}}
\newcommand{\done}{\end{equation}}

\newcommand{\vecB}{\textbf{B}}
\newcommand{\vecJ}{\textbf{J}}

\newcommand{\vecv}{\textbf{v}}

\newcommand{\curl}[1]{\nabla\times{#1}}
\newcommand{\divv}[1]{\nabla\cdot{#1}}

\numberwithin{equation}{section}
\newcommand{\lare}{Lare3D }

\DeclareFontFamily{OT1}{pzc}{}
\DeclareFontShape{OT1}{pzc}{m}{it}{<-> s * [1.10] pzcmi7t}{}
\DeclareMathAlphabet{\mathpzc}{OT1}{pzc}{m}{it}

\newcommand{\bi}{\begin{itemize}}
\newcommand{\ei}{\end{itemize}}

\newcommand{\JEL}[1]{\textcolor{black}{#1}}

\newcommand{\RR}[1]{\textcolor{black}{#1}}
\shorttitle{Solar Eruptions} \shortauthors{Leake et al.}

\begin{document}

\title{The Role of Reconnection in the Onset of Solar Eruptions}

\correspondingauthor{James Leake}
\email{james.e.leake@nasa.gov}

\author[0000-0003-0072-4634]{James E. ~Leake}
\affiliation{Heliophysics Science Division \\ NASA Goddard Space Flight Center \\ 8800 Greenbelt Rd.\\ Greenbelt, MD 20771, USA}
\author[0000-0002-4459-7510]{Mark G. ~Linton}
\affiliation{U.S. Naval Research Laboratory, Washington, DC, USA.}
\author[0000-0003-0176-4312]{Spiro K. ~Antiochos}
\affiliation{University of Michigan, \\ Department of Climate and Space Sciences and Engineering, \\ Ann Arbor, MI 48109, USA}


\begin{abstract}

Solar eruptive events such as coronal mass ejections and eruptive flares are frequently associated with the emergence of magnetic flux from the convection zone into the corona. We use three dimensional magnetohydrodynamic numerical simulations to study the interaction of coronal magnetic fields with emerging flux and determine the conditions that lead to eruptive activity.  A simple parameter study is performed, varying the relative angle between emerging magnetic flux and a pre-existing coronal dipole field. We find that in all cases, the emergence results in a sheared magnetic arcade that transitions to a twisted coronal flux rope via low-lying magnetic reconnection. This structure, however, is constrained by its own outer field, and so is non-eruptive in the absence of reconnection with the overlying coronal field. The amount of this overlying reconnection is determined by the relative angle between the emerged and pre-existing fields. 
The reconnection between emerging and pre-existing fields is necessary 
\JEL{to generate sufficient} expansion of the emerging structure so that flare-like reconnection below the coronal flux rope becomes strong enough to trigger its release.
Our results imply that the relative angle is the key parameter in determining whether the resultant active regions exhibit eruptive behavior, and is thus a potentially useful candidate for predicting eruptions in newly emerging active regions. More generally, our results demonstrate that the detailed interaction between the convection zone/photosphere and the corona must be calculated self-consistently in order to model solar eruptions accurately.
\end{abstract}




\section{Introduction}
\label{sec:intro}
The widely-accepted explanation for the frequently-observed eruptions from the Sun's atmosphere, such as coronal mass ejections (CMEs) and coronal jets,  is that free energy stored in magnetic fields residing above the solar surface is somehow released explosively to accelerate particles, plasma, and magnetic flux into the heliosphere. The magnetic structures responsible for the free energy are referred to as ``filament channels," and consist of low-lying flux that straddles a polarity inversion line (PIL) and has a strong ``shear" component along the PIL. The bulk of the free energy is due to this filament channel shear \citep[e.g.,][]{Gibson18}. The filament flux may be twisted as well, however, the amount and importance of this twist remains an issue of debate by both observers and modelers \citep{Patsourakos_2020}. Also an issue of debate is the mechanism responsible for the explosive energy release. In one class of theories, magnetic reconnection, a resistive process, destabilizes the force balance between the filament flux and overlying ``strapping" field \citep[e.g.,][]{1999ApJ...510..485A}; whereas, in another class an ideal instability or loss-of-equilibrium causes the eruption \citep[e.g.,][]{Forbes1990,Forbes1991,Demoulin2010,Kliem2014,Fan2017}. A major complication in distinguishing between the two classes is that reconnection plays an important role in all models, because it is the process responsible for the extreme plasma heating and particle acceleration observed as a solar flare. Most eruptions are accompanied by a flare, especially the large energetic eruptions that originate from the strong magnetic concentrations of \textit{Active Regions} (ARs). The difference between the two classes of models, therefore, is that in the resistive models flare reconnection accelerates the eruption \citep[e.g.,][]{Karpen_2012, Wyper_2017}, whereas in the ideal models the outward acceleration drives the flare reconnection \citep[e.g.,][]{Fan2017}. Given that reconnection and plasma heating is observed to be so prevalent during eruptions, it has not yet been possible to distinguish the eruption mechanism directly from observations \citep{Chen_2011}.

In principle, the eruption could be determined by sufficiently accurate observations of the free energy buildup and the resulting filament channel structure. Three general processes have been proposed for filament channel formation \citep[see][]{Mackay2015}: shear flows    and/or the closely related helicity condensation model \citep{Antiochos_2013,Dahlin_2021}, flux cancellation \citep{Gaizauskas1998,Wang2007}, and flux emergence \citep{2004ApJ...610..588M, Fan_2009}. The key point here, as reviewed in \citet{Patsourakos_2020}, is that the ideal eruption mechanisms require that the filament field have sufficient twist to trigger a kink-like or torus instability \citep{Shafranov56,1989ApJ...338..453C,Kliem_2006}.
Since the formation of twist generally requires flux cancellation \citep[e.g.,][]{Ballegooijen1989}, it would seem possible to identify the eruption mechanism by closely observing filament channel evolution. The problem, however, is that the amount of twist required is fairly small: one turn or even less for a strongly sheared channel \citep{Aulanier2002}, and the actual photosphere ubiquitously exhibits a complex mixture of emergence, submergence/cancellation, and shear flows, so that, again, the observations have not been able to pin down definitively the eruption mechanism.

As a result, there has been intense interest and work during the past few decades on determining the eruption mechanism by numerical simulations \citep[see][and references therein]{Chen_2011}. In these simulations the governing equations of magnetohydrodynamics (MHD) are integrated in time to model the build up and release of magnetic energy in the corona. Due to the great disparity between characteristic wave speeds and spatial scales of the photosphere and corona, these models typically span only the upper atmosphere of the Sun, the solar corona, and include the effects of the denser, cooler, lower atmosphere, i.e., the photosphere and chromosphere, only as a boundary condition at the base of the model. 

Classic examples of this approach to investigating solar eruptions have been the many simulations of the Magnetic Breakout Model, which uses reconnection as the eruption mechanism \citep{Antiochos_1998, Antiochos99}. In typical Breakout simulations  magnetic energy is built up in the corona by prescribed surface motions at the lower boundary, either large-scale shearing flows \citep[e.g.,][]{Karpen_2012} or small-scale randomized motions leading to helicity condensation \citep[e.g.,][]{Dahlin_2021}. The normal component of the coronal field is held fixed at this lower boundary, which is presumed to represent the high-beta photosphere, and the coronal flux is assumed to be multipolar so that a separatrix surface and null point are present. Note that even a bipolar AR implies the presence of a coronal separatrix surface and null when the field of the bipole is embedded in the solar global dipole \citep[e.g.,][]{2008ApJ...683.1192L}. The initial conditions in most Breakout simulations are taken to be the minimum-energy potential field ($\mb{J}\equiv \nabla\times\mb{B}/\mu_{0}=0$ where $\mathbf{B}$ is the magnetic field), but the imposed boundary motions inject magnetic free energy into corona, leading to the formation of a sheared-arcade filament channel with current sheets at the overlying separatrix surface and null between the filament flux system and neighboring flux systems. In the Breakout Model, magnetic reconnection (often called Breakout reconnection) at the separatrix current sheet is necessary so that the sheared filament flux can expand upward sufficiently for a vertical current sheet to form below/inside this flux system. The onset of magnetic reconnection in this vertical current sheet (often loosely described as flare reconnection) greatly accelerates the eruption, leading to the ejection of the shear and the relaxation of the corona back down to a near-potential state \citep[e.g.,][]{Karpen_2012,Wyper_2017}. A related, but physically distinct reconnection-driven eruption model is the so-called tether-cutting, which posits that with sufficient boundary shear a vertical current sheet can form spontaneously in the filament channel, without the need for Breakout reconnection, leading to an eruption \citep[e.g.,][]{Moore2001}. Recent simulations using the same approach as the Breakout ones just described appear to support this scenario \citep{Jiang2021}. 

This approach of simulating only the coronal evolution with some photospheric boundary condition has also been applied to modeling ideally driven eruptions \citep[e.g.,][]{Torok2004, Torok2005,Kliem_2006}. Perhaps the most widely simulated eruptions of this class are those that invoke the torus instability \citep[e.g.,][]{Shafranov56}. In this model the filament channel is presumed to be twisted and have a net electric current  so that there is an upward hoop force that must be balanced by the tension/pressure of the overlying strapping fields \citep{Kliem_2006}. In typical torus simulations a twisted flux rope is either directly inserted into the corona as an initial condition \citep[e.g.][]{Torok2018}, or else created by first applying shear flows and then flux cancellation by enhanced diffusion at the PIL \citep[e.g.,][]{Amari2003}. The system can then be driven to instability by multiple methods, usually by modifying the lower boundary condition in the lower corona to induce cancellation and reduce the overlying tension.

Note that for both the reconnection and ideal models described above, the energy buildup and ultimate trigger of the eruption is the driving of the corona by the photosphere. This driving, however, is not calculated rigorously, but is simply input as some initial conditions and/or boundary conditions. Consequently, it is not possible to determine which model is viable and which is not; all the models ``work" given the appropriate assumptions. Our simulations below, on the other hand, attempt to calculate the self-consistent photosphere-corona interaction so that any eruption that occurs (or does not) is due to the physics of this interaction rather than to ad-hoc assumptions. 

The most important physical process in the photosphere-corona interaction, and the origin of ARs and many other forms of activity, is the emergence of magnetic flux from the solar convection zone, where it is built up by dynamo processes, operating either at the base of the convection zone, or in a shallow layer near the surface \citep{Charbonneau_2014}. Some dynamo models exhibit the production of twisted magnetic wreathes that buoyantly rise towards the surface \citep{Nelson_2014}. In addition, analysis of some emerging ARs suggest a twisted magnetic flux rope is responsible for their observed magnetic field patterns \citep{Luoni_11,Poisson_15,Poisson_16}. \RR{More recently, \citet{2021NatCo..12.6621M} applied an analysis technique called fieldline-winding, which is a normalization of magnetic helicity that removes the magnetic flux dependence, to both simulations of emerging twisted flux ropes and newly emerging solar active regions. The similarity of the fieldline-winding in the simulations and the observations indicated that the active regions analyzed exhibited magnetic topology consistent with the emergence of twisted flux ropes, and that this emergence is what determines the topology of the region. 
}
Replacing ad-hoc boundary conditions or initial conditions in the corona-only models described above with MHD modeling of dynamic flux emergence from the convection zone is essential for determining the mechanism driving solar eruptions. 

There have been numerous simulations of the emergence of magnetic flux ropes from the convection zone into the corona. These previous studies \citep{Matsumoto1998,Fan_2001,Manchester_2004, Fan_2009, Leake_2013, Leake_2014} showed that due to the stratification at the top of the convection zone and lower atmosphere, buoyant flux ropes do not emerge bodily into the solar atmosphere, but undergo significant deformation and re-organization to produce, most often, a filament channel consisting of a sheared magnetic arcade, but sometimes a coronal flux rope, and exhibit various phenomena that are observed in real ARs, such as sunspot rotation, shearing motions, and converging and diverging flows \citep{Manchester_2004,Luoni_11,Poisson_15,Poisson_16,Toriumi2017,Knizhnik_2018}. For those simulations in which the subsurface flux rope emerges into a field-free corona, the vast majority do not result in eruption. On the real Sun, however, flux always emerges into pre-existing coronal field.
Early work examining the interaction of emerging fields and a horizontal coronal field was presented in \citet{Galsgaard_2005}, focusing on jets as a result of the magnetic reconnection. A later important result related to the onset of eruptive activity was presented in \citet{Archontis_2008}. They postulated that the emerging flux rope's twist field (which in many cases is necessary to survive the evolution in the convection zone), which emerges first, plays the role of a `strapping' or `envelope' field that counteracts any upward magnetic forces created by the partial emergence described above. They added a horizontal coronal field which was designed to reconnect with this envelope field and found that the partially emerged field formed a coronal flux rope that eventually underwent a fast rise to the top of the simulation domain. 

Following on this study, \citet{Leake_2014} showed that magnetic reconnection between emerging magnetic fields and simple arcade coronal magnetic fields can lead to eruptions consistent with the magnetic Breakout picture. The conclusion is that the Magnetic Breakout Model can indeed be extended from a corona-only model and can operate within a paradigm where magnetic flux emerges from the convection zone and through the lower layers of the solar atmosphere. Even more important than demonstrating eruption, \citet{Leake_2013} showed that if there were no reconnection between the emerging and pre-existing field, then eruption did not occur and the system settled into a quasi-static equilibrium. This is a key result. It proves that reconnection is the driver of the eruption and not just a consequence. The determining parameter for whether reconnection, and eruption, occurred or not was the relative orientation between the subsurface flux rope and the overlying coronal field. The \citet{Leake_2013,Leake_2014} study considered only two orientations, the ones most favorable and least favorable for reconnection. In this paper a parameter study varying this orientation is performed in order to determine, in detail, the effects of magnetic reconnection between emerging magnetic fields and pre-existing magnetic fields on the likelihood of eruption. For this parameter study the simple, idealized system of  \citet{Leake_2014} is used, consisting of an initial subsurface twisted flux rope and a bipolar coronal arcade. One of the major advantages of idealized simulations is that one can span parameter spaces to a) test the robustness of the model, b) attempt to relate the fundamental mechanisms directly involved in the phenomena under investigation to the free parameters of the system, and c) relate these parameters to observable quantities which may aid in predictive capabilities. 

The numerical integration of the governing equations and their boundary and initial conditions are detailed in \S \ref{sec:numerics}, along with the parameter study. \S \ref{sec:results_new} presents the major results of the simulations and \S \ref{sec:discussion_new} contextualizes the results and highlights outstanding issues.

\section{Numerical Setup}
\label{sec:numerics}

\subsection{Governing Equations}
To model the emergence of magnetic flux though the solar surface, from the convection zone into the corona, and its interaction with pre-existing magnetic flux above the surface, the visco-resistive MHD equations with gravitational forces are evolved in time using the Lagrangian Remap (\lare) code \citep{Arber_2001}. For reference, the version of the code used is v2.5. The equations are presented in Lagrangian form here:\beg{mass}
\frac{D\rho}{Dt} = -\rho \divv{\vecv}, 
\done
\beg{momentum}
\frac{D\vecv}{Dt} = -\frac{1}{\rho}\Big((\curl{\vecB})\times\vecB+\nabla P +\rho\textbf{g}+\divv{\textbf{S}}\Big), 
\done 
\beg{energy}
\frac{De}{Dt} = \frac{1}{\rho}\Big(-P\divv{\vecv} + \zeta_{ij}S_{ij}+\frac{\eta}{\mu_0^2}(\curl{\vecB})^2\Big), ~ \textrm{and}
\done 
\beg{induction}
\frac{D\vecB}{Dt} = (\vecB\cdot\nabla)\vecv - \vecB(\divv{\vecv}) - \curl{\eta\vecJ}.
\done 

In these equations, $\rho$ is the mass density, $\vecv$ is the velocity, $\vecB$ is the magnetic field, and $e$ is the specific energy density. $\mu_0$ is the permeability of free space, which is used to relate the magnetic field to the current density $\mu_0\vecJ=\curl{\vecB}$. Resisitivity is denoted by $\eta$, the gravitational acceleration at the surface, $\textbf{g}_{sun} = -274\;\mathrm{ms^{-2}}\hat{\mb{z}}$, and \textbf{S} is the stress tensor with components given by
\beg{stresstensor}
S_{ij} = \nu \Big(\zeta_{ij}-\frac{1}{3}\delta_{ij}\divv{\vecv}\Big).
\done 
The viscosity is $\nu$, $\delta_{ij}$ is the Kronecker delta function, and $\zeta$ can be written in terms of the components of the Jacobian matrix $\nabla\vecv$ as 
\beg{upzeta}
\zeta_{ij} = \frac{1}{2}\Big(v_{ij} + v_{ji}\Big)
\done 
where $v_{ij}$ are components of the tensor $\nabla\mathbf{v}$.
The gas pressure is defined through the mass density, Boltzmann constant $k_B$, temperature $T$ and reduced particle mass $\mu_m$ as
\beg{idealgas}
P = \frac{k_B\rho T}{\mu_m},
\done 
where $\mu_m=1.25m_p$, with $m_p$ the proton mass. The temperature is related to the specific energy density via
\beg{eandP}
e = \frac{T}{\gamma-1},
\done 
where $\gamma=5/3$ is the ratio of specific heats. \lare  defines the plasma variables $e$ and $\rho$ on cell centers, magnetic field components at cell faces, and the velocity components at the cell vertices, and the numerical scheme preserves the divergence of $\vecB$ to machine precision. The resistivity and viscosity are set to $1.458\times10^1\;\mathrm{\Omega}\;\mathrm{m}$ and $3.350\times10^3\;\mathrm{kg}\;\mathrm{m^{-1}}\;\mathrm{s^{-1}}$, respectively, which in normalized units (see below) are both equal to 0.01.

In addition to the bulk velocity described above, LaRe3D uses a shock capturing scheme with an artificial scalar  pressure which is comprised of a linear and quadratic function of the grid cell size $L$ \citep{Wilkins_1980}:
\begin{equation}
q = c_{1}\rho c_{f}L|s| + c_{2}L^2\rho s^2
\end{equation}
where $c_{f}$ is the local fast magnetosonic speed and 
$L$ and $s$
are the grid length and strain rate projected, cell by cell, onto the local
direction of acceleration. In particular, for acceleration $\partial_t\vecv$ 
and a grid cell of size $(d_x,d_y,d_z)$, $L=(d_x,d_y,d_z)\cdot\partial_t\vecv/|\partial_t\vecv|$,
and $s=\partial_t\vecv\cdot\nabla\vecv\cdot\partial_t\vecv/|\partial_t\vecv|^2$. 
Note, $q$ is set to
zero for $s\ge0$, so that this only applied in cells where a shock has already formed, or could be in the process of forming.
This viscous pressure $q$ is added to the pressure P above in the momentum equation. This approach preserves shock viscosity and allows the determination of the shock heating.  As discussed in \citet{Arber_2001} values of $c_{1},c_{2}=0.5,0.8$ are used in this study.

\subsection{Normalization}
The MHD equations  are solved in dimensionless units, and are scaled by choosing quantities to normalize the physical parameters. While ideal MHD $(\eta,\nu=0)$ in the absence of gravity and other source terms is scale-free, visco-resistive MHD with gravitational forces has inherent scales. For these equations, one is free to choose three normalizing constants. Here those constants are $B_0 = 0.13 \mathrm{T}$ ,  $L_0 = 1.7\times10^5 \mathrm{m}$, and the value of gravitational acceleration at the solar surface $\mb{g}_{sun}$ above. 
These choices constrain the normalizing values for the gas pressure
\beg{prescl}
P_0=\frac{B_0^2}{\mu_0} = 1.345\times10^4 \;\mathrm{Pa},
\done 
mass density 
\beg{rhoscl}
\rho_0 = \frac{B_0^2}{\mu_0L_0|\mb{g}_{sun}|} = 2.887\times10^{-4}\; \mathrm{kg\;m^{-3}},
\done
velocity 
\beg{velscl}
v_0 = \sqrt{L_0|\mb{g}_{sun}|} = 6.825\times10^3\;\mathrm{m\; s^{-1}},
\done
time
\beg{timescl}
t_0 = \sqrt{\frac{L_0}{|\mb{g}_{sun}|}} = 24.909\; \mathrm{s},
\done 
temperature
\beg{temscl}
T_0 = \frac{m_pL_0|\mb{g}_{sun}|}{k_B} = 5.637\times10^3\;\mathrm{K},
\done 
current density
\beg{curscl}
J_0 = \frac{B_0}{L_0\mu_0} = 0.609 \;\mathrm{A\;m^{-2}},
\done 
viscosity
\beg{viscscl}
\nu_0 = P_0t_0 = 3.350\times10^5\;\mathrm{kg}\;\mathrm{m^{-1}}\;\mathrm{s^{-1}},
\done 
resistivity, 
\beg{resistscl}
\eta_0 = \mu_0v_0L_0 = 1.458\times10^3\;\mathrm{\Omega}\;\mathrm{m}.
\done 
and energy 
\beg{energyscl}
E_0 = \rho_0 {v_0}^2 L_0^3 =  6.607\times10^{19}\mathrm{J}.
\done
However, while the equations are evolved in normalized units, unless specified otherwise, all results will be presented in dimensional units.

\begin{figure}
\begin{center}
\includegraphics[width=0.45\textwidth]{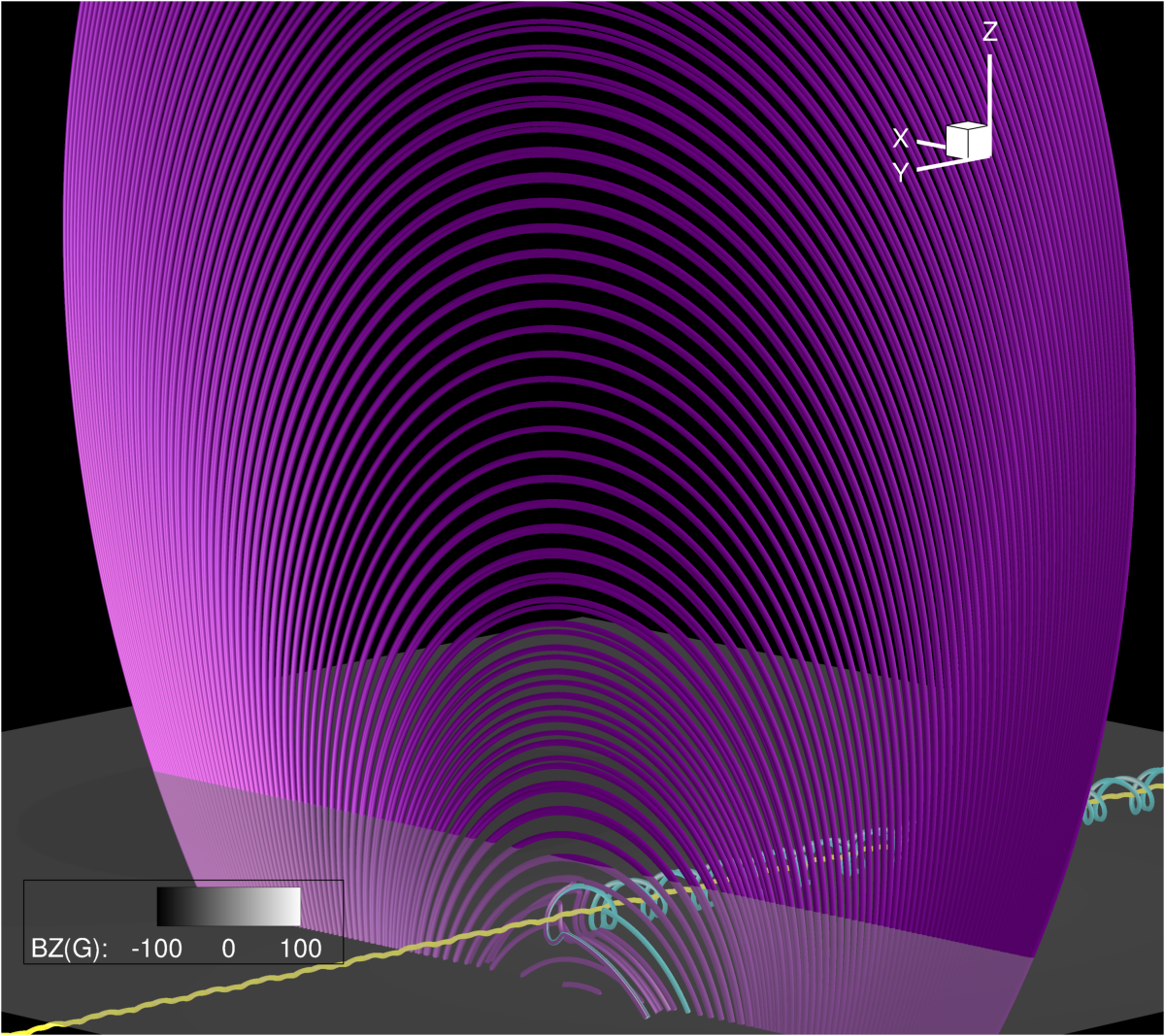}
\includegraphics[width=0.5\textwidth]{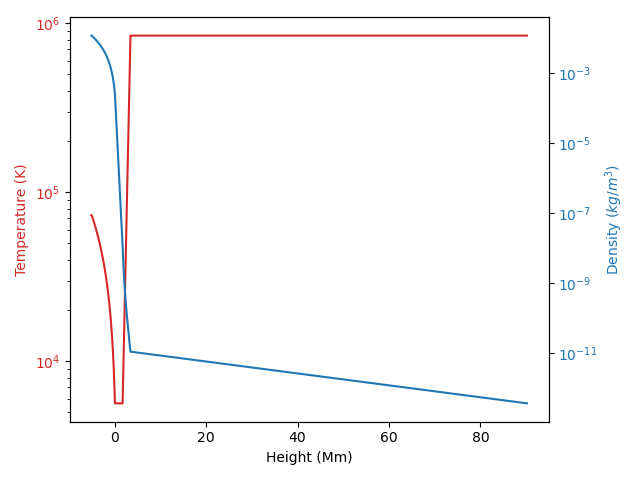}
\caption{Left Panel (a):  Sub-domain of the simulation at $t=0$ showing select fieldlines at regular intervals along the vertical line $x=y=0$. 
The yellow fieldline belongs to the convection zone flux rope, and purple fieldlines belong to the background dipole field. The cyan fieldlines connect at one end to the dipole field, and at the other end to the flux rope. The fieldlines are seeded at the same locations in later figures which highlight coronal structures and are not optimized to show the rope's magnetic flux density. Also shown is a semi-opaque contour of $B_z$ at the model photosphere, which will be used to observe the emergence in later plots.  Right
Panel (b): Initial hydrostatic 1D atmosphere.\label{fig:IC}}
\end{center}
\end{figure}

\subsection{Domain and Boundary Conditions}

\JEL{The simulation's numerical grid has a non-constant resolution with higher resolution located in the center of the domain in the horizontal direction, and higher resolution in the convection zone and lower atmosphere in the vertical direction. First, a constant grid $(x_0,y_0,z_0)$ \JEL{at} high resolution, i.e., grid size $(d_{x,0}, d_{y,0}, d_{z,0})$, is defined. Then, in each direction,  a new spatially-varying \JEL{grid size} $(d_{x}, d_{y}, d_{z})$ is defined as a function of this original grid, with the resolution \JEL{decreasing} horizontally towards the side boundaries and vertically towards the top boundary. This new resolution is then used to defined the new grid $(x,y,z)$.}  

\JEL{The original high resolution grid in these simulations is ${(x,y)}_{0} ~\epsilon ~ [-{L_{(x,y)}}_0/2,{L_{(x,y)}}_0/2] = [-128,128]L_{0}$, and $z_{0} ~\epsilon ~  [-30,162]L_{0}$, with $[n_x,n_y,n_z] = [512,512,768]$, so that $(d_{x,0},d_{y,0},d_{z,0}) = (0.5,0.5,0.25)L_{0}$.}

The new \JEL{grid size} in $x$ (and $y$, substituting $y$ for $x$ below) is defined by 
\begin{equation}
d_x =  d_{x,0} + f_{x}\tanh{\frac{(|x_0|-l_x)}{w_x}}
\end{equation}
where $f_x=6$, $l_x=L_{x,0}/2.1$, and $w_x=L_{x,0}/10$. This creates a new spatially varying resolution equal to $d_{x,0}$ at the center and increasing to its maximal value at the edges of the new grid.

\JEL{For $z$, the original grid is only stretched upwards (as opposed to in both directions for $x,y$):
\begin{equation}
d_z =  d_{z,0} + f_{z}\tanh{\frac{(z_0-l_z)}{w_z}}
\end{equation}}
\JEL{ where $f_x=2$, $l_z=70L_0$, and $w_x=L_{x,0}/20$.}

\JEL{This process results in a final domain with horizontal resolution ($x$ and $y$) varying from 0.085 Mm at 0 to 0.714 Mm at the side boundaries, and vertical resolution varying from 0.0425 Mm below and just above the surface to 0.2125 Mm at the top boundary. This resultant domain has extents $X\times Y\times Z = [-47.1,47.1]\times[-47.1,47.1]\times[-5.1,90.2]$ Mm, see Figure \ref{fig:IC}. This is approximately twice the domain previously run in \citet{Leake_2014} and allows for a better expansion of the emerging fields. }

At the boundaries, all velocity components and normal gradients of the other MHD variables are set to zero.  \JEL{In addition, a damping region is applied to the velocity in the vicinity of all four horizontal boundaries, $|x|,|y| > l_{damp,x,y}$, and the vicinity of the top boundary, $z>l_{damp,z}$, where $l_{damp,x}=l_{damp,y}=\frac{3L_x}{8}L_{0} = 35.3 ~ \textrm{Mm}$ and $l_{damp,z}= 7\max{(z)}/8 =78.9 ~ \textrm{Mm}$.}

\JEL{Within each of these 5 boundary damping regions, the velocity is divided by a factor $F$ every timestep. For example, for the two damping regions near the two $x$ boundaries:
\begin{equation}
F = 1 + dt\frac{|x|-l_{damp,x}}{\frac{L_x}{2}-l_{damp,x}}, ~ |x|>l_{damp,x}
\end{equation}
with the same formula applied to the two regions near the $y$ boundaries (substituting $y$ for $x$ above). The $dt $ dependence reduces the possibility of instability. Hence, the factor is 1 at $|x,y|=l_{damp,x,y}$, a distance of $Lx/8$ from each side boundary, and increases to $(1+dt)$ at each side boundary. } \JEL{For the damping region located at the top boundary
\begin{eqnarray}
F & = & 1 + dt \frac{z - l_{damp,z}}{\max{(x)} - l_{damp,z}}, ~ z > l_{damp,z}.
\end{eqnarray}
} This damping helps reduce reflected outwardly propagating perturbations from interacting with the solution in the interior.

\subsection{Initial Conditions and parameter study}

The simulations are initialized with a hydrostatic background atmosphere with a convection zone ($z<0$), a photosphere/chromosphere ($0<z<10L_0$), transition region ($10L_0<z<20L_0$) and corona ($20L_0<z$). In the convection zone, the atmosphere is polytropic, with the temperature following
\JEL{
\beg{temperature}
T_{cz}(z) = T_0 \Big(1-\frac{z}{L_{0}}\frac{\gamma-1}{\gamma}\Big).
\done 
}
In the photosphere, $T(z) = T_0$, $\rho(z) = \rho_0$ and $P(z) = P_0$. In the corona, the temperature is 
\beg{Tcorona}
T_{cor}(z) = 150 T_0.
\done
The photosphere and corona are connected by the transition region, where the temperature increases from $T_{0}$ to $T_{cor}$ via
\beg{Ttr}
T_{tr}(z) = T_0\Big(\frac{T_{cor}}{T_0}\Big)^{(z-10L_0)/10L_0}.
\done
\newline
The equation $\nabla P = -\rho g$ is then integrated  in space, using the ideal gas law, Equation \ref{idealgas}, with the conditions $(\rho(z=0),T(z=0),P(z=0))=(\rho_0,T_0,P_0)$, to obtain the density and pressure profiles throughout the atmosphere. The initial hydrostatic atmosphere is shown in Figure \ref{fig:IC}(b). 

To this 1D atmosphere, the magnetic field due to a magnetic dipole moment outside the simulation domain is added: 
\begin{eqnarray}
\mb{B}_{dipole} & = & \nabla\times\mb{A}_{dipole} \\
\mb{A}_{dipole} & = & \frac{\mu_{0}}{4\pi}\frac{\mb{m}\times\mb{r}}{r^3}
\end{eqnarray}
where $r = \sqrt{x^2+y^2+(z-z_{d})^2}$ and the magnetic moment is aligned along the $x$ direction:  $\mb{m}=d\hat{\mathbf{x}}$. \JEL{The strength of the dipole is $d=15000 B_{0}L_{0}^3/\mu_{0} = 7.6 \times 10^{24} ~ \textrm{A}\textrm{m}^2$. This gives a representative dipole field strength at the model photosphere of 20 G ($2\times10^{-3}$ T).}

\begin{figure}
\begin{center}
\includegraphics[width=0.6\textwidth]{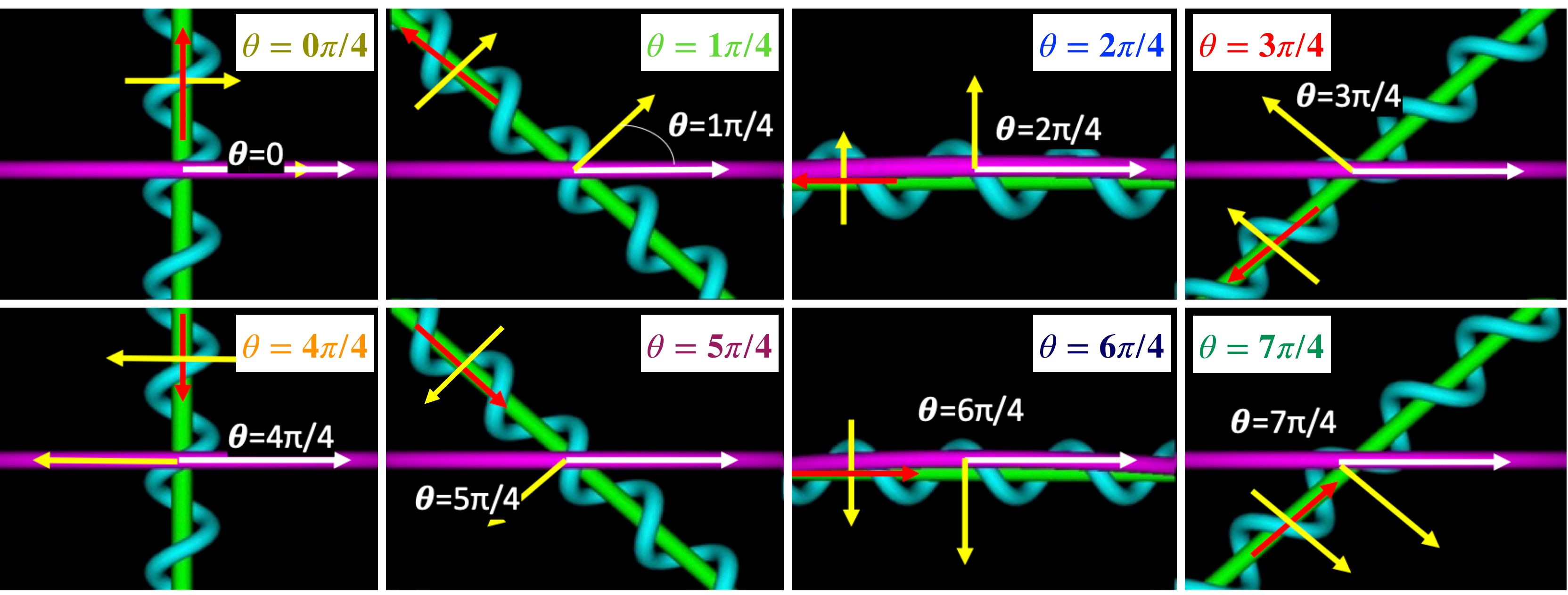}
\includegraphics[width=0.3\textwidth]{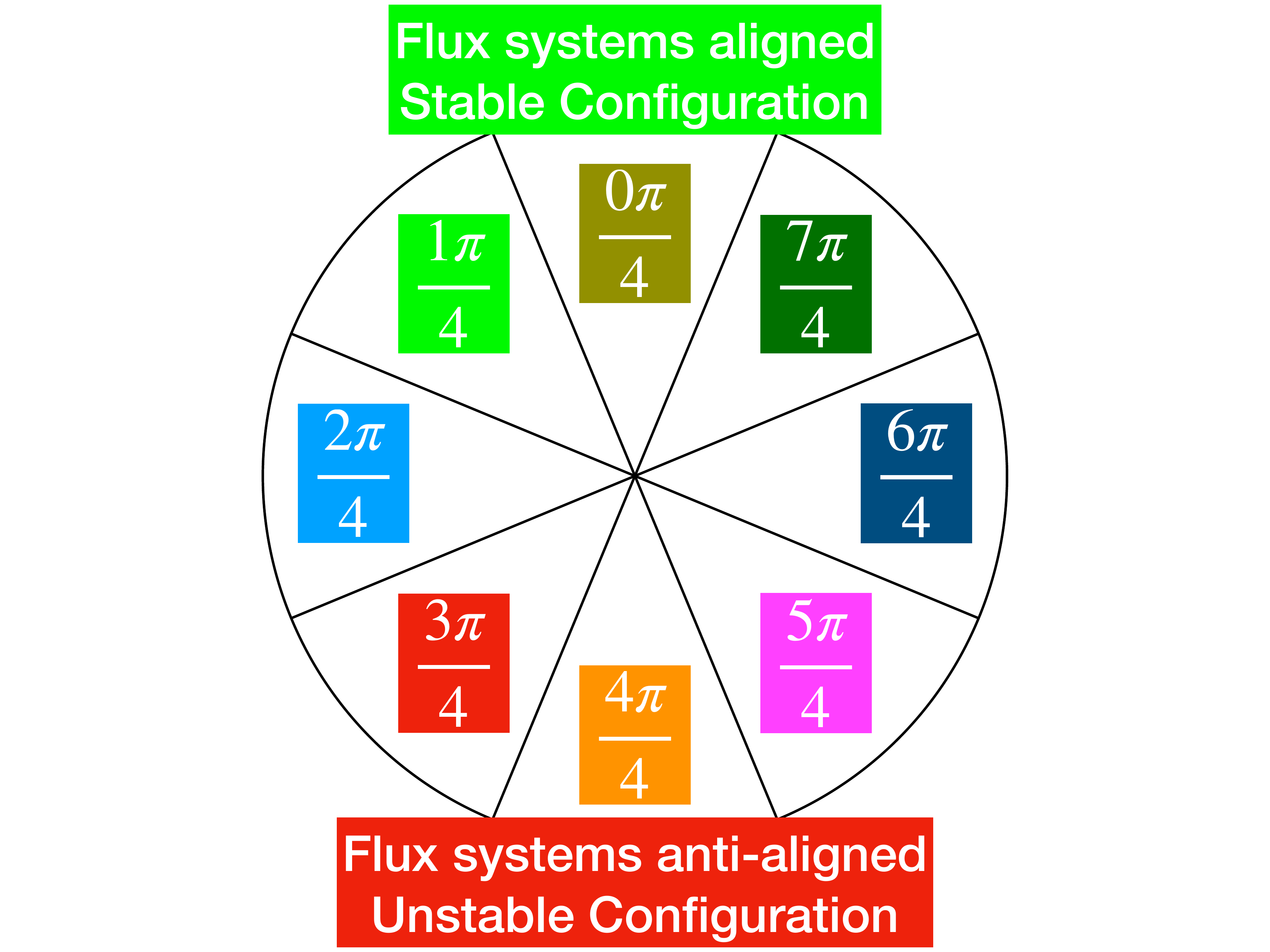}
\caption{Schematics showing alignment of initial convection zone flux rope and overlying field, and definition of $\theta$. In each panel, the green field-line is the axis of the convection zone rope and the red arrow is the axial field's direction. The cyan field-line is a sample twisted field-line away from the rope's axis. The yellow arrow is the direction of the twist component of the convection zone rope above the axis, and is replicated to highlight the angle between it and the white arrow. The magenta field-line belongs to the dipole field above the rope and the white arrow is the direction of this dipole field. 
\label{fig:orientation}}
\end{center}
\end{figure}

In addition to this magnetic field, a twisted cylindrical magnetic flux rope is placed in the convection zone. This flux rope will have an axial component and a twist component. The relative orientation of the flux rope and the dipole will be varied in this study, so for each different simulation the flux rope is \JEL{ defined using a new cartesian co-ordinate system $(x_r,y_r,z_r)$ created by rotating the original system about the $z$-axis by an angle $\theta$:
\begin{eqnarray}
\begin{bmatrix} 
x_{r} \\
y_{r} \\
z_{r} \\
\end{bmatrix}
=
\begin{bmatrix}
     \cos{\theta} & \sin{\theta} & 0  \\
     -\sin{\theta} & \cos{\theta} & 0 \\   
     0 & 0 & 1\\ 
  \end{bmatrix}
 \begin{bmatrix} 
x \\
y\\
z \\
\end{bmatrix}.
\end{eqnarray}
}


The cylindrical flux rope is then defined in localized cylindrical coordinates
$(r_{1},\phi_{1},y_{1})$ where
\begin{eqnarray}
x_{r} & = & r_{1}\cos{\phi_1} \\
y_{r} & = & y_{r} \\
z_{r} & = & r_{1}\sin{\phi_1}
\end{eqnarray}
and $r_{1} = \sqrt{x_r^2+(z_r-z_{rope})^2}$ where \JEL{$z_{rope}$} is defined below.

\JEL{The form for the magnetic flux rope is then 
\begin{eqnarray}
\mb{B}_{rope} & = & B_{t}\exp{(-\frac{r_{1}^2}{w^2})}\hat{\mb{y}}_{1} + qr_{1}B_{t}\exp{(-\frac{r_{1}^2}{w^2})}\mb{\hat{\phi}}_{1}
\end{eqnarray}
where \JEL{$q=-1/w$} and so the rope is initially marginally stable to the kink instability \citep{Linton_1996}.}

The Lorentz force due to the magnetic field is matched by a gas pressure profile such that 
\begin{equation}
\nabla p_{1} = \mathbf{J}_{rope}\times\mb{B}_{rope}.
\end{equation}
\JEL{In general, one is free to perturb the density and the temperature in combination to achieve a required pressure perturbation.} As in previous work, here the density is modified as  follows
\begin{equation}
\rho_{1}(r_1) = -\rho_{0}(z)\frac{p_1(r)}{p_0(z)}\exp{\left(-{(\frac{y_1}{\lambda})}^2\right)}
\end{equation}
so that $\rho_{1}$ is maximally negative at the middle of the rope, and zero at its ends. This generates an $\Omega$-shaped rising flux rope in broad agreement with the rising wreaths seen in the dynamo simulations of \citet{Nelson_2014}. The parameter realization for the flux rope parameters are: strength $B_{t} = 5B_0 = 0.65 ~ \textrm{T} = 6500 ~ \textrm{G}$ ; width $w=2.5L_0 = 0.425 ~ \textrm{Mm}$; height $z_{rope}=-12 L_0 = -2.04 ~ \textrm{Mm}$, buoyant length $\lambda=10L_0 = 1.7 ~ \textrm{Mm}$.

Figure \ref{fig:orientation} is a schematic diagram showing how the $\theta$ parameter defines the relative orientation of the two magnetic flux systems (flux rope and dipole). The magnetic flux rope has an azimuthal (or twist) component perpendicular to its axis (green fieldlines), and this has opposite signs above and below the rope axis. The parameter angle $\theta$ determines the relative angle between the twist component above the axis (cyan fieldlines, yellow arrows) and the dipole field (magenta fieldlines, white arrows), as shown in the left 8 panels in Figure \ref{fig:orientation}. In \citet{Leake_2013, Leake_2014} the values of $\theta=0$ and $\theta=4\pi/4$ were investigated, showing that when $\theta=0$, the rope's twist field above the axis is aligned parallel to the dipole field, and so there is no significant reconnection between emerging field and coronal field. As a result, the emerging structure created a stable coronal configuration. Conversely for $\theta=4\pi/4$, the rope twist field above the axis was aligned antiparallel to the dipole field, leading to significant reconnection  between the two flux systems, allowing a rise of the emerging field and eventually an eruption. This type of reconnection is here-after called \textit{Breakout reconnection}, as it is analogous with the reconnection necessary for an eruption in the magnetic Breakout Model.  These simulations improve upon the previous study, as they include a larger domain, extending vertically to 90.2 Mm, compared to  approximately 34 Mm in \citet{Leake_2013, Leake_2014}. They also  investigate other parameter realizations of $\theta$ than in \cite{Leake_2013, Leake_2014}.

Based on those results of \citet{Leake_2013,Leake_2014} one can make a prediction of how the simulations will evolve for all 8 values of $\theta$. This is shown on the right panel of Figure \ref{fig:orientation}, the green values of $\theta=[7,0,1] \pi/4$ have mostly twist field parallel to the dipole field, and should be stable. The red values of $\theta=[3,4,5] \pi/4$ have mostly twist field anti-parallel to the dipole field, and should be unstable. The blue intermediate values of $\theta=[2,6] \pi/4$ have twist field normal to the dipole field, and clearly require further analysis. It will be shown that the evolution during emergence complicates this simple prediction.

\section{Results}
\label{sec:results_new}
\subsection{Recap of general evolution for eruptive scenario}

\begin{figure}
\begin{center}
\includegraphics[width=0.31\textwidth]{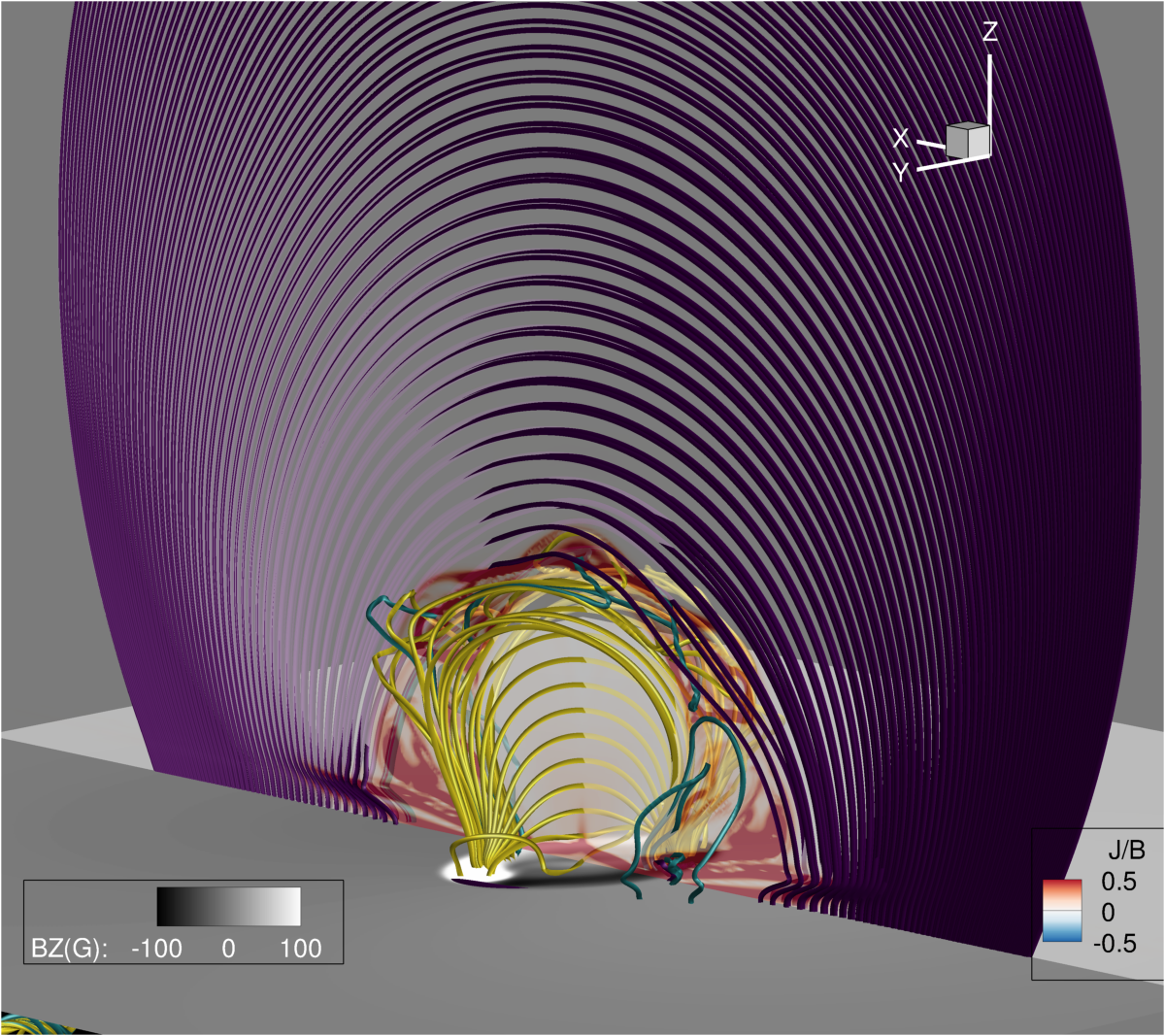}
\includegraphics[width=0.31\textwidth]{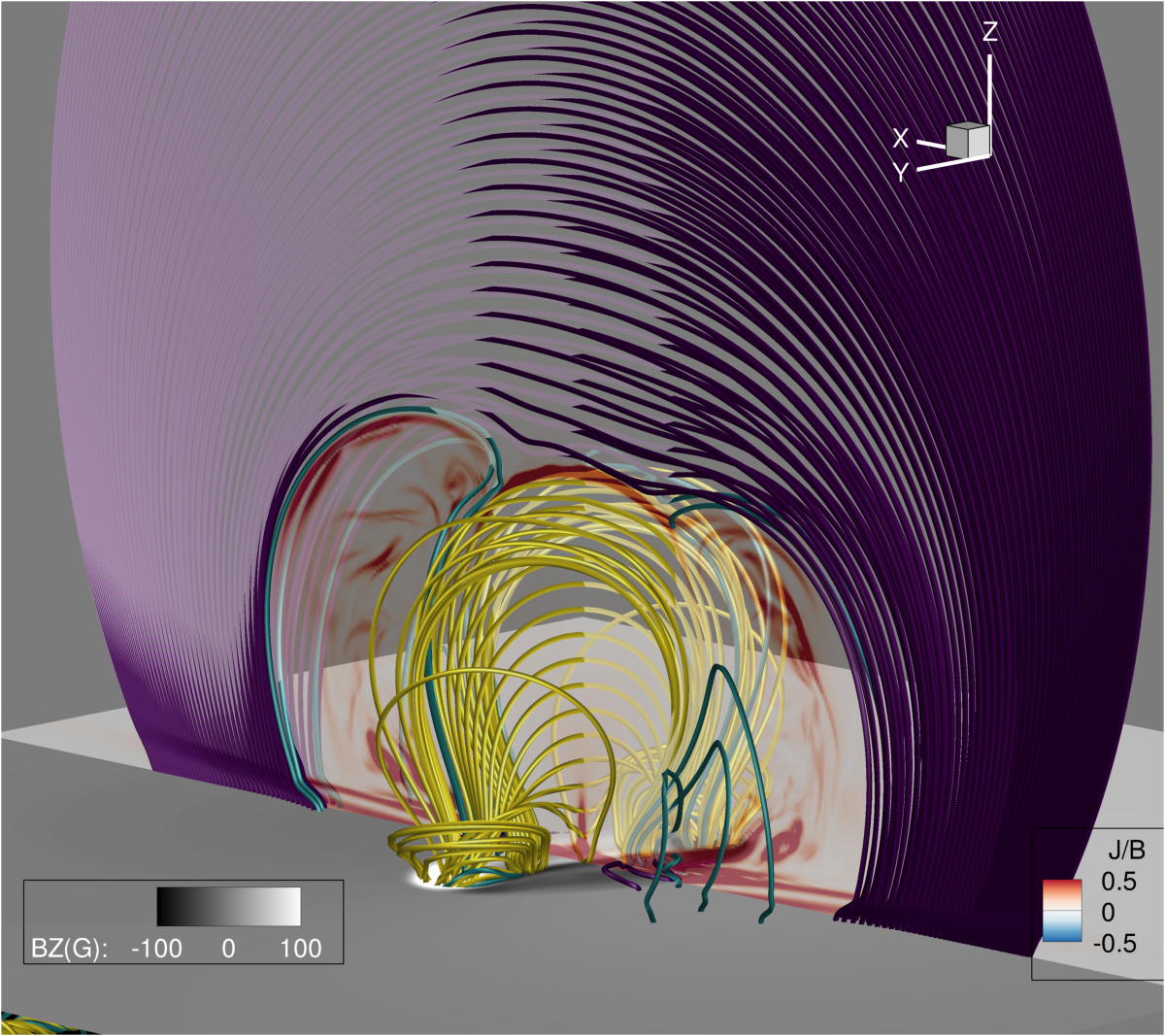}
\includegraphics[width=0.31\textwidth]{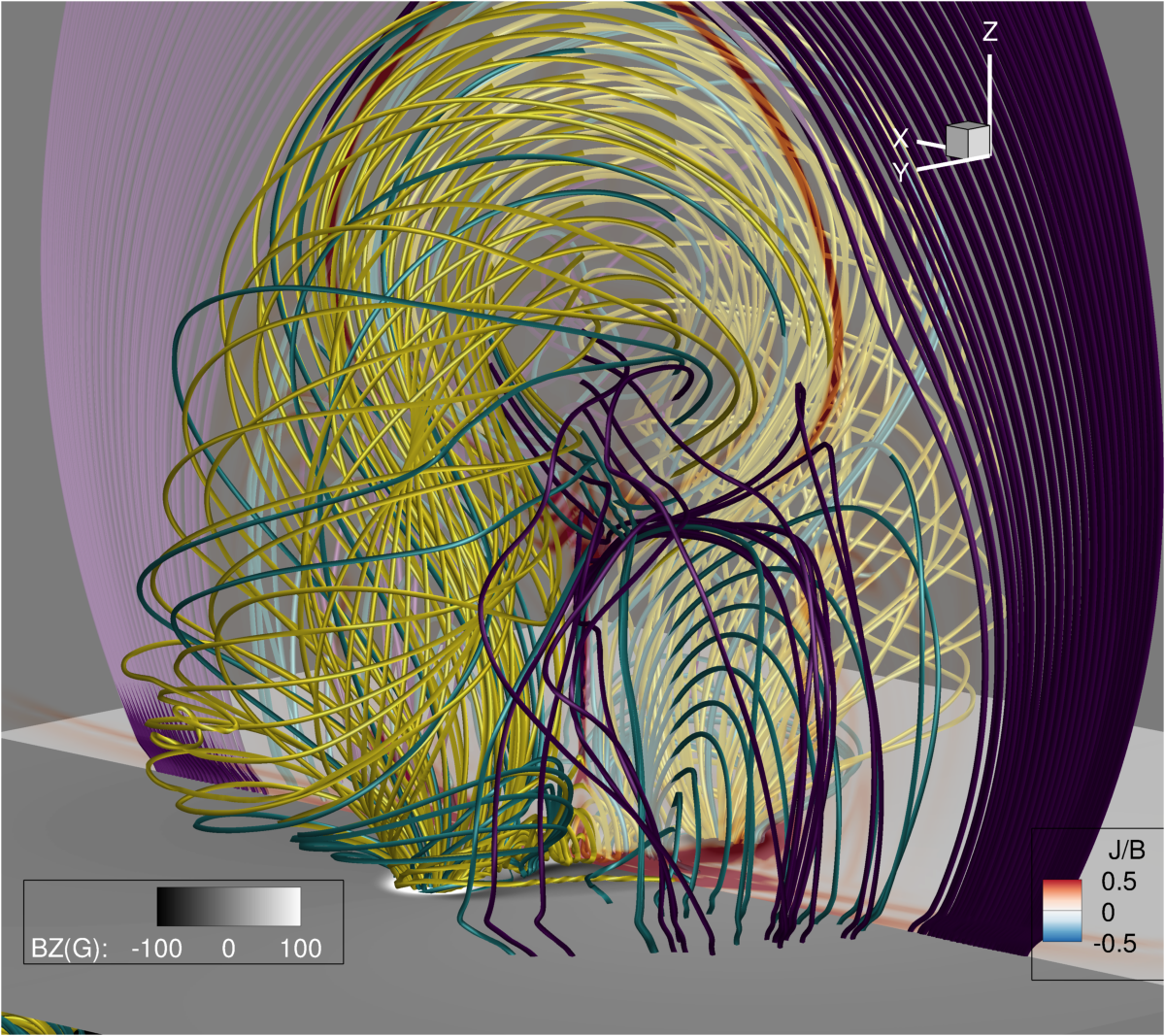} \\
\includegraphics[width=0.31\textwidth]{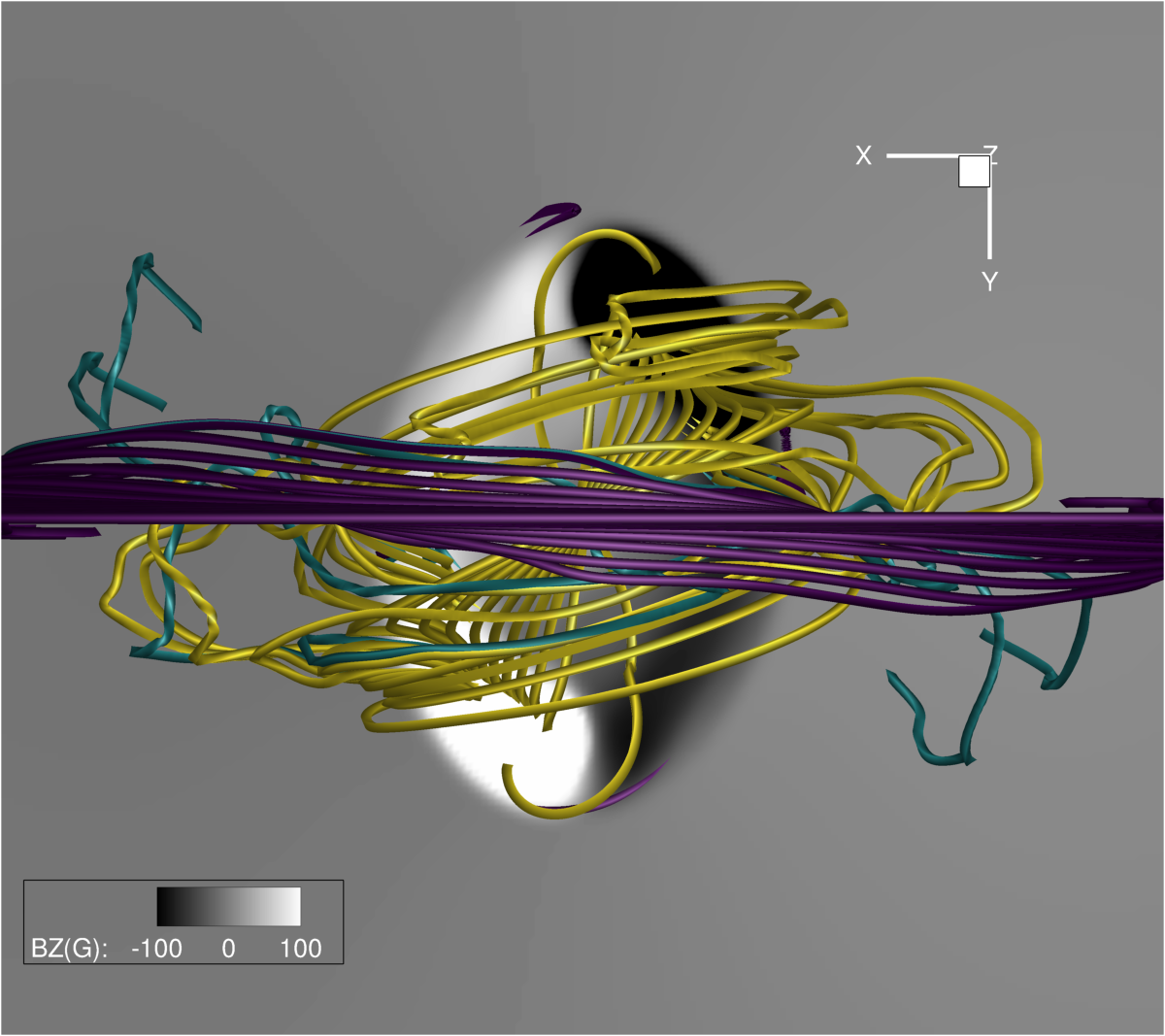}
\includegraphics[width=0.31\textwidth]{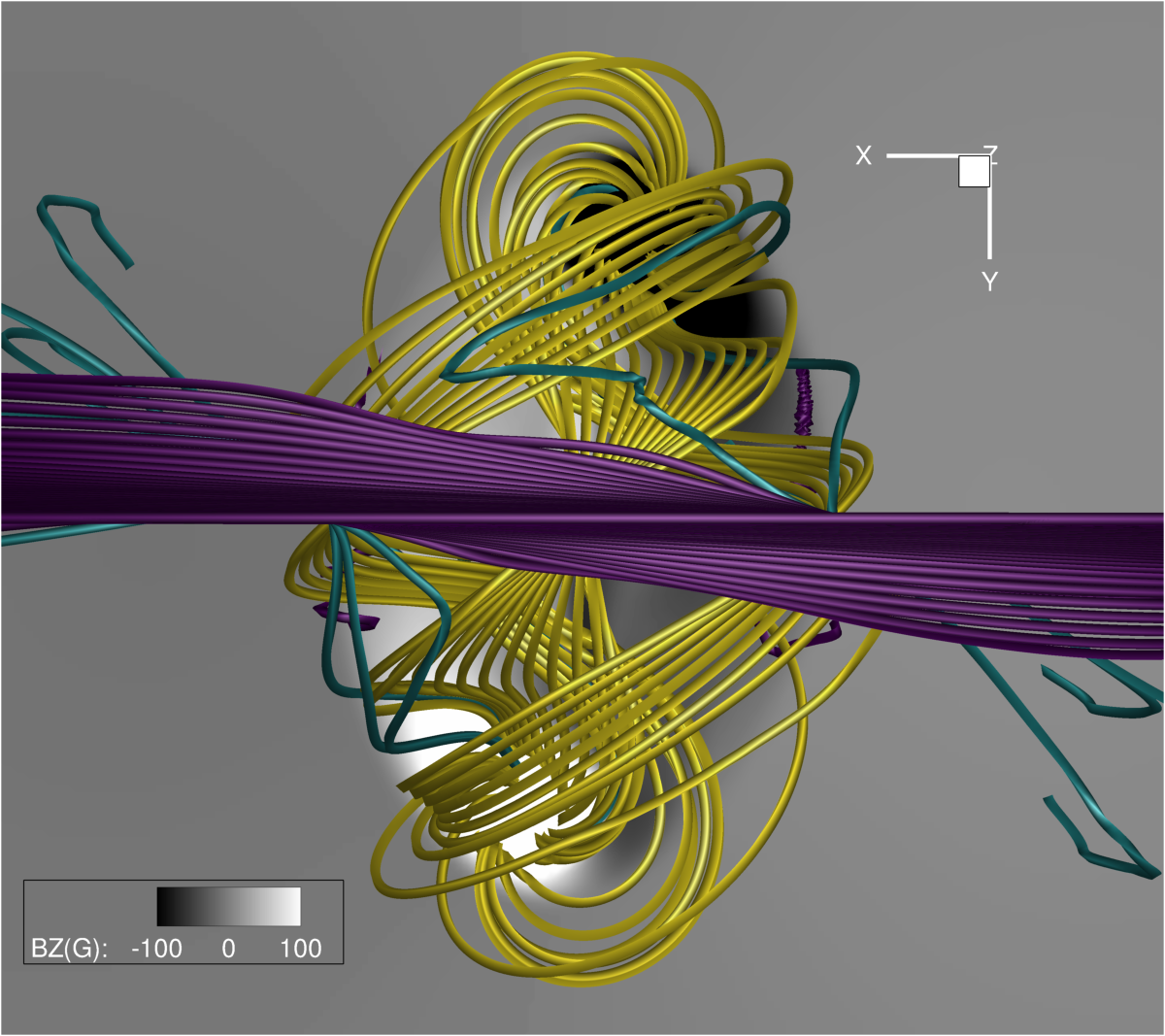}
\includegraphics[width=0.31\textwidth]{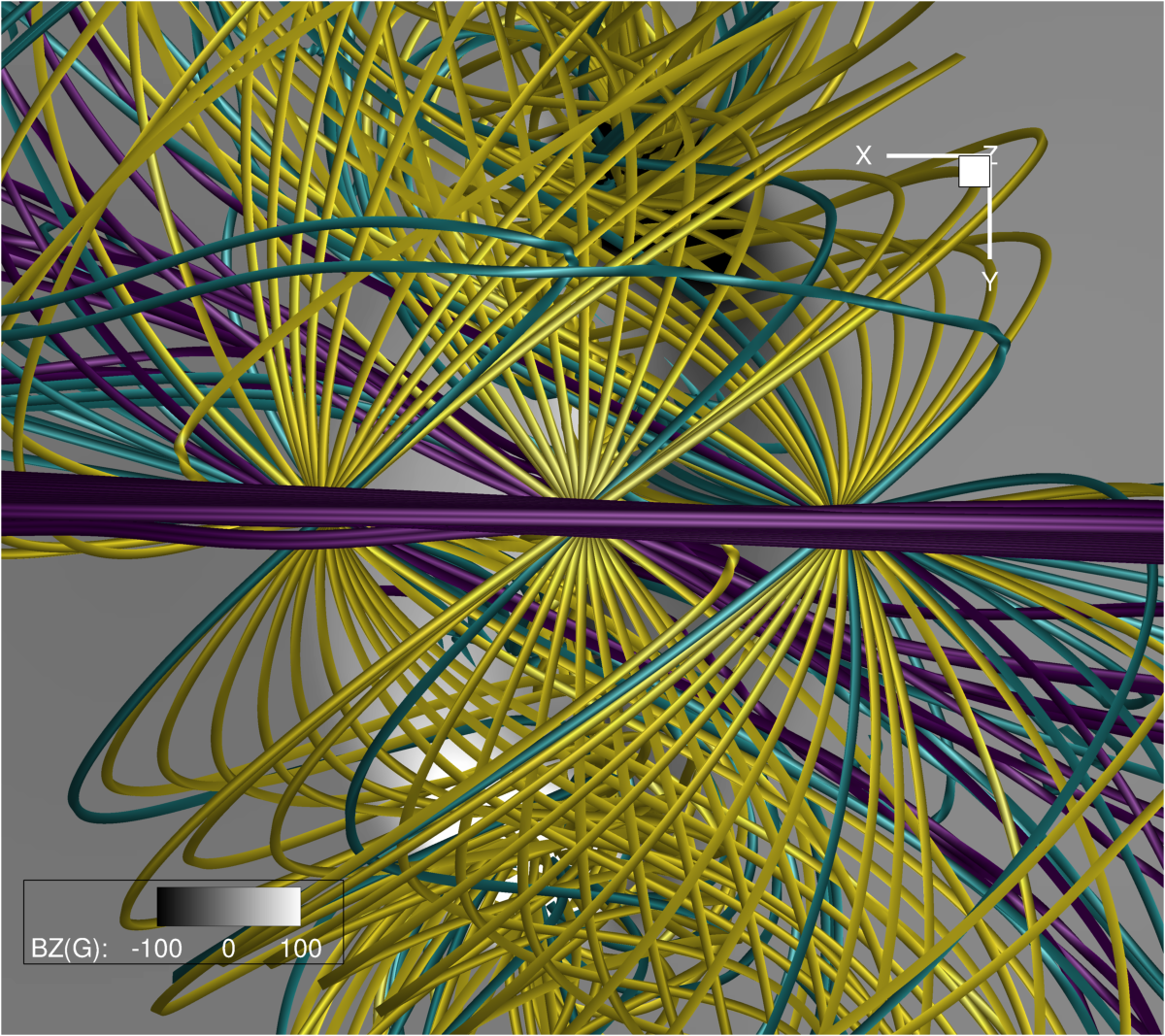} \\
\caption{Evolution of fieldlines in Simulation 4, at times $100,142,184 t_{0}$. Fieldlines are colored by connectivity. Purple fieldlines connect to the dipole field at both ends. Yellow fieldlines connect to the flux rope at both ends. Cyan fieldlines connect at one end to the dipole and one end to the flux rope. Also shown are  mid-plane slices of $|\mb{J}/J_0|/|\mb{B}/B_0|$ (top panels) and horizontal slices of $B_{z}$.   \label{fig:Sim24_3Dview1}}
\end{center}
\end{figure}

\begin{figure}
\begin{center}
\includegraphics[width=0.31\textwidth]{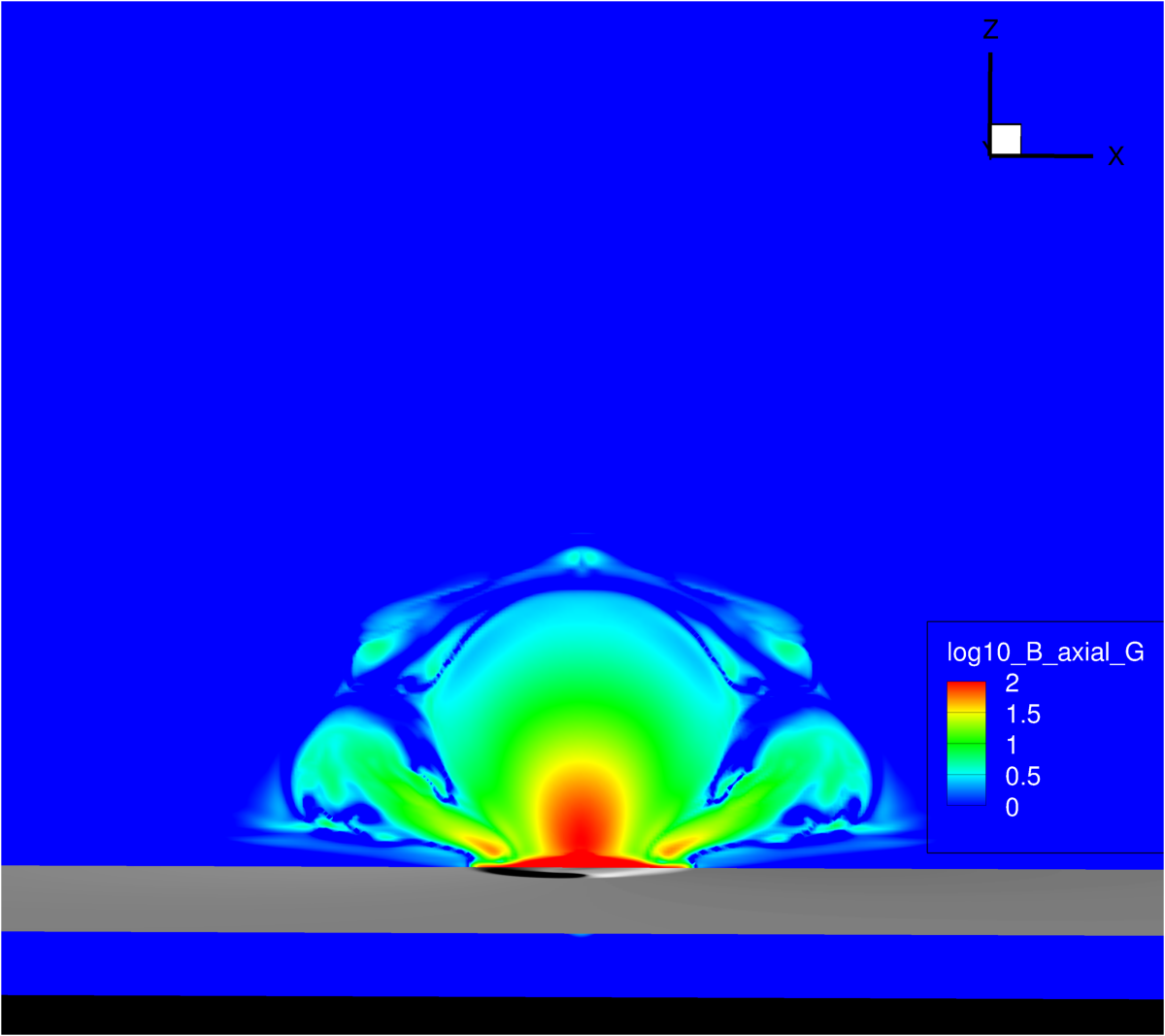}
\includegraphics[width=0.31\textwidth]{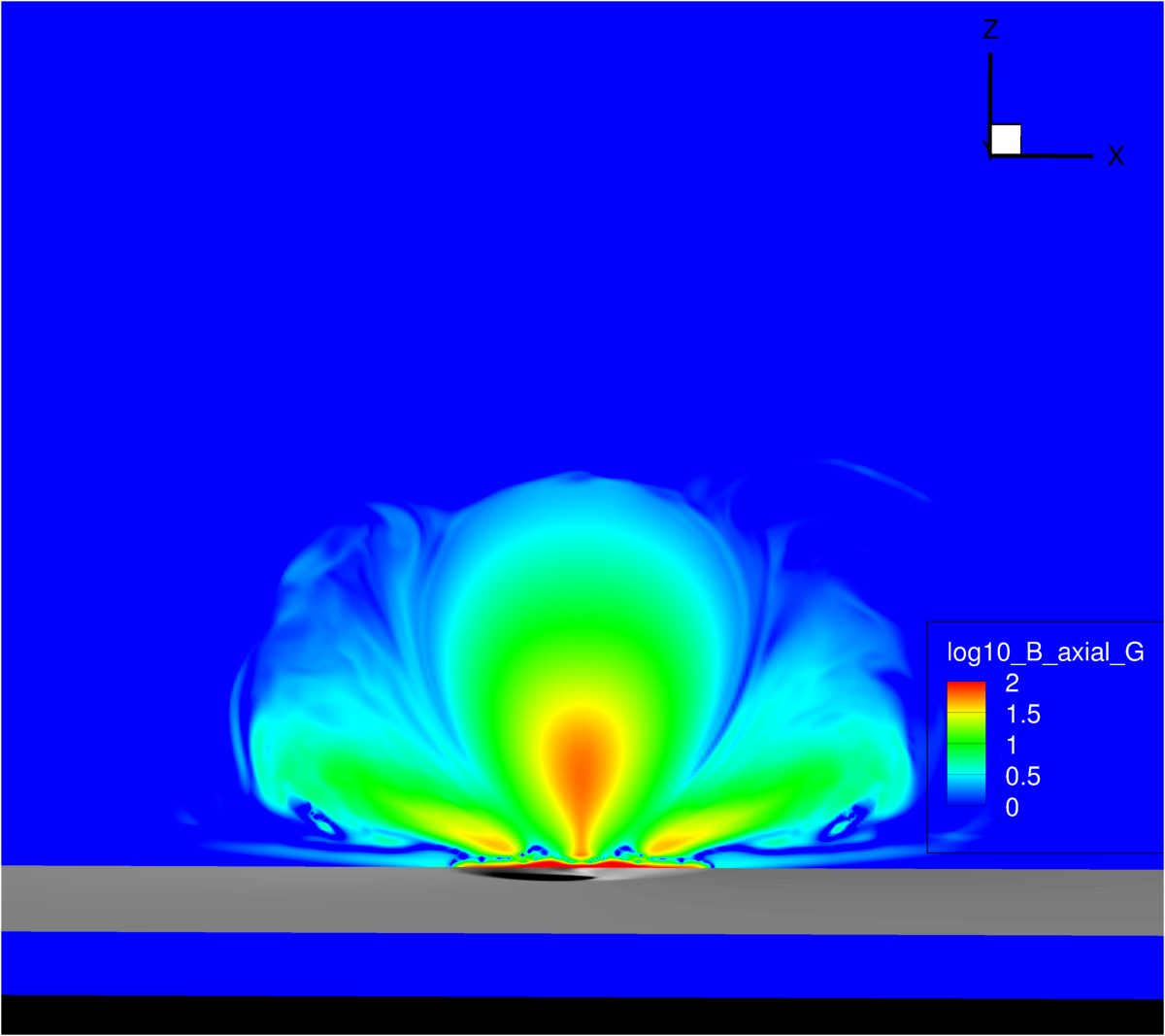}
\includegraphics[width=0.31\textwidth]{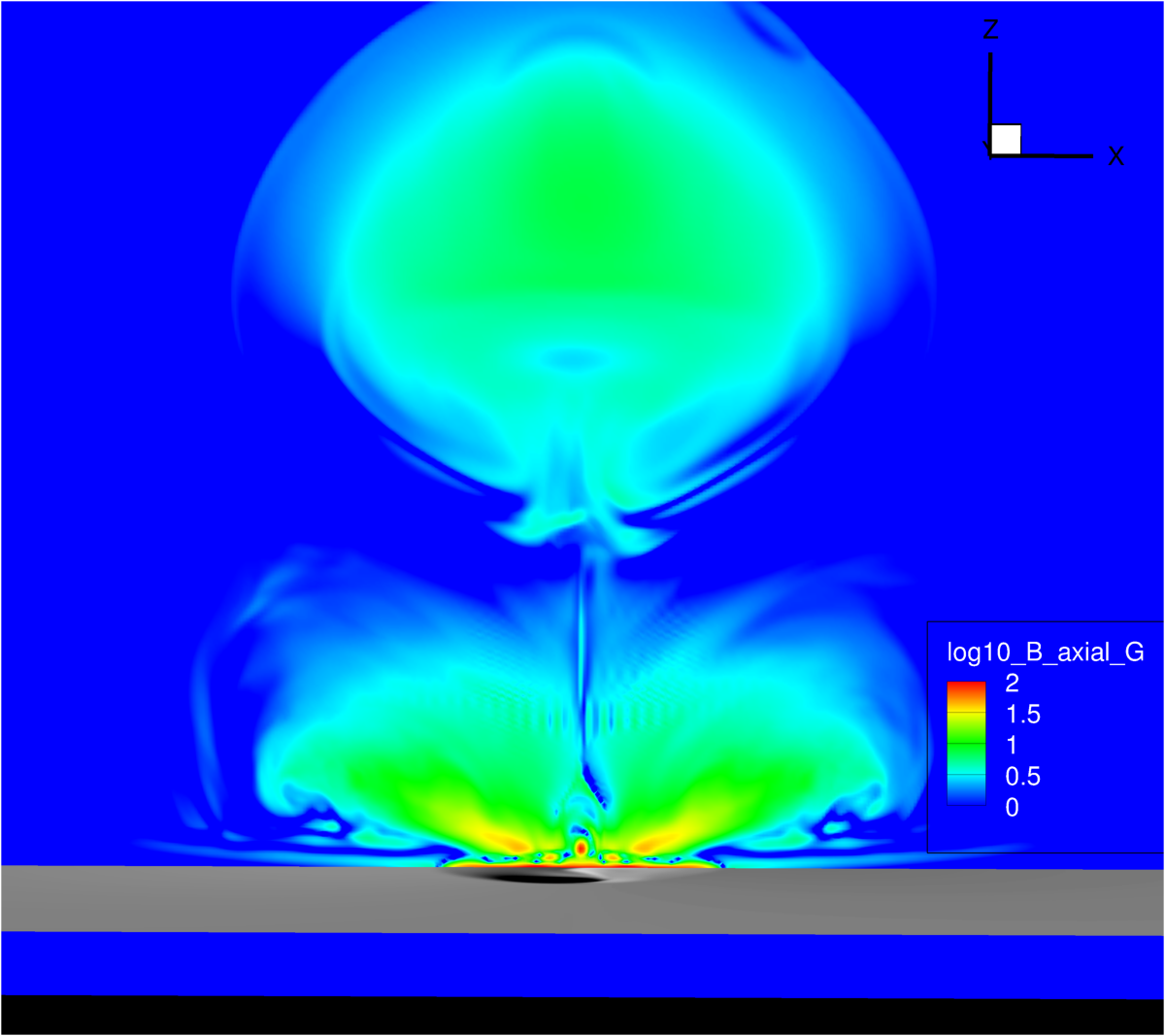} \\
\includegraphics[width=0.31\textwidth]{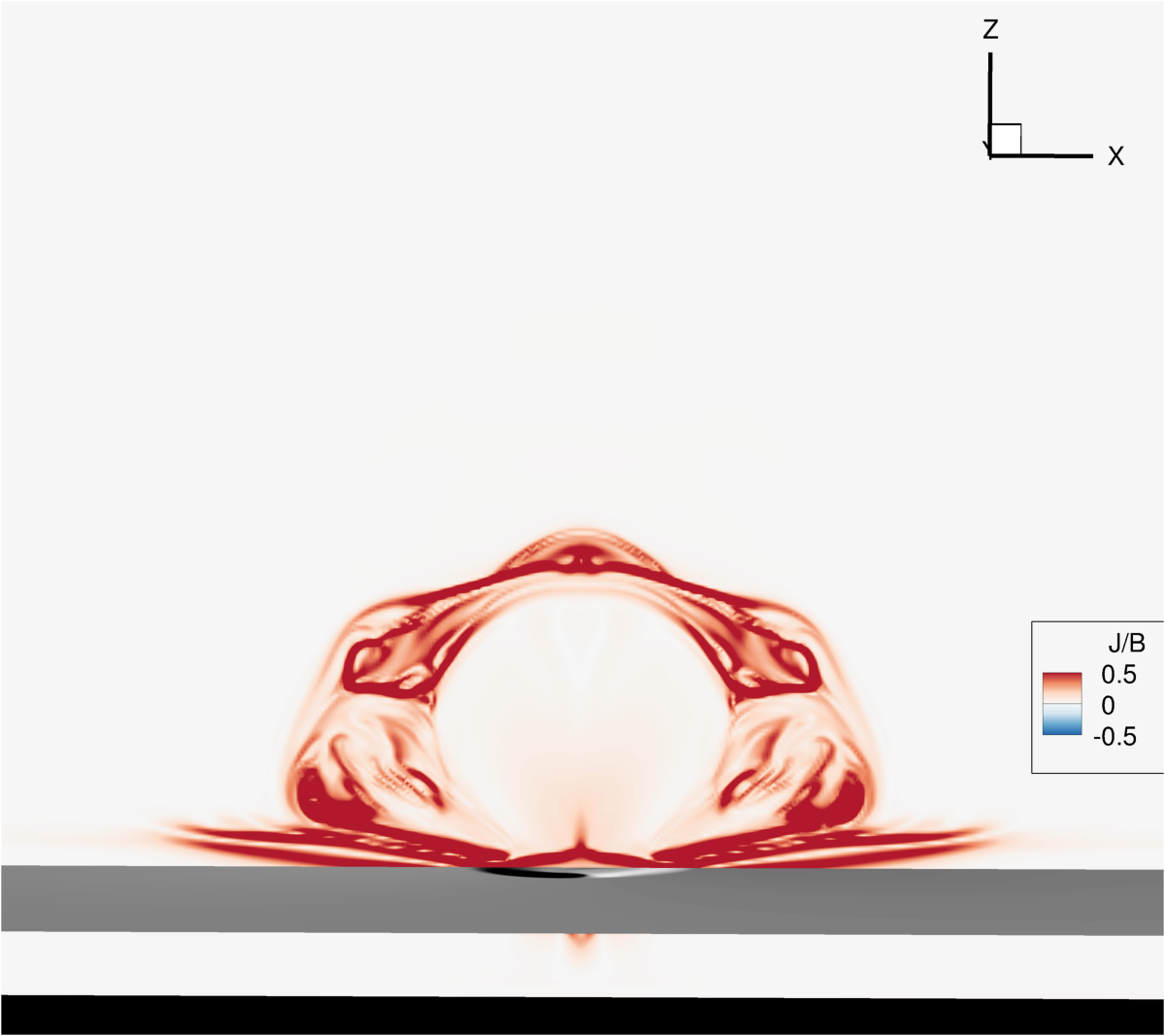}
\includegraphics[width=0.31\textwidth]{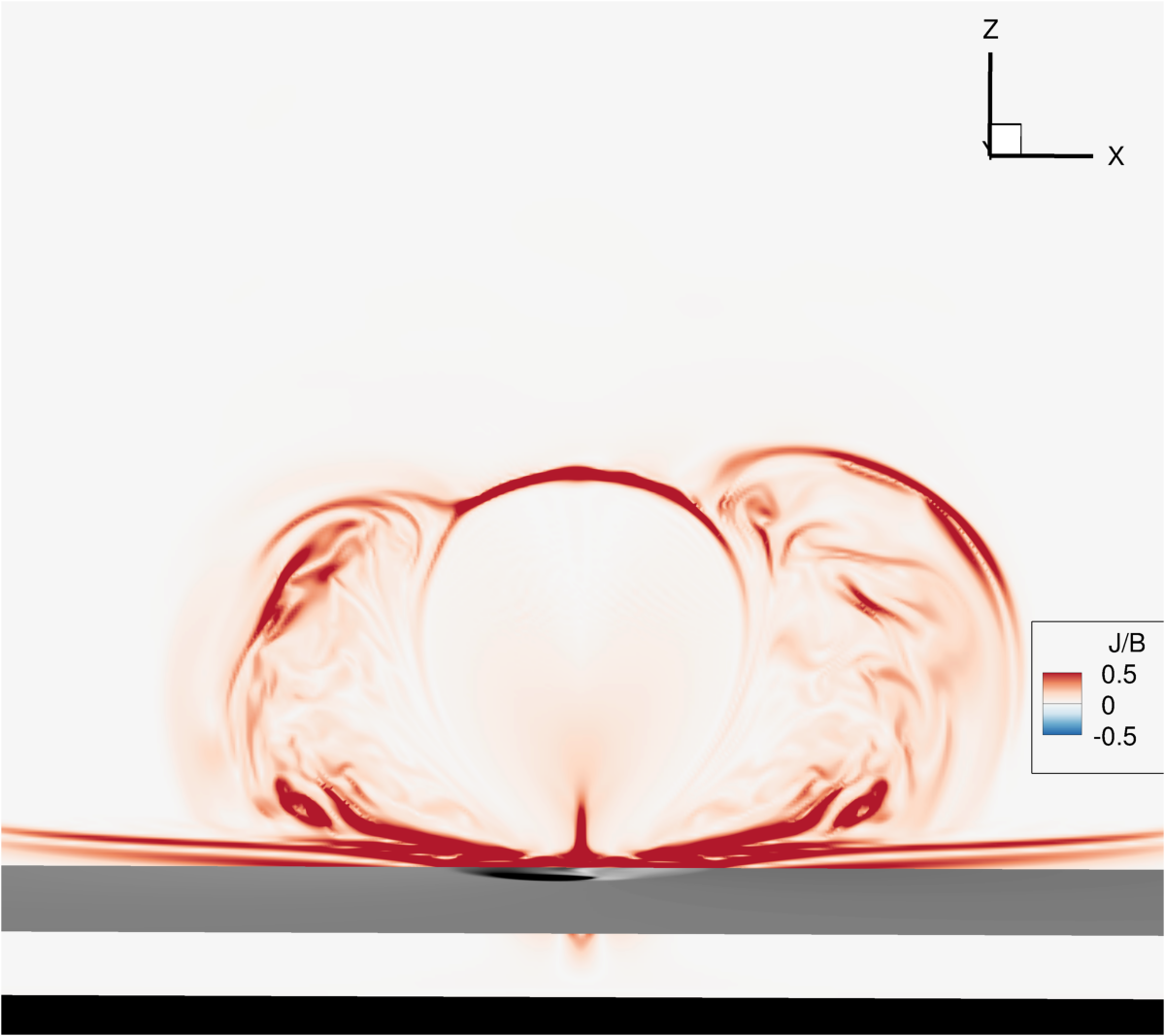}
\includegraphics[width=0.31\textwidth]{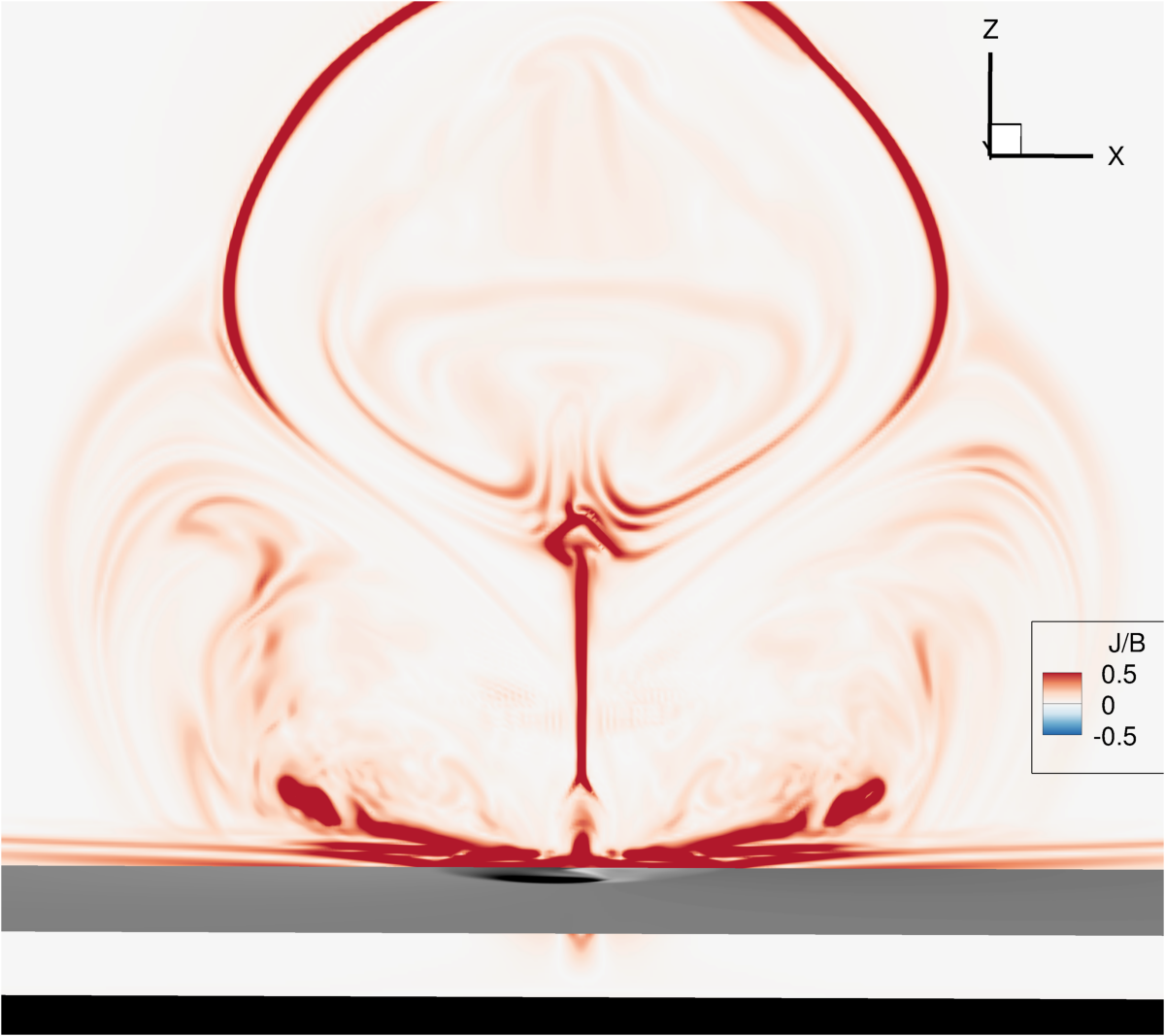} \\
\caption{ $|\mb{J}/J_0|/|\mb{B}/B_0|$ (bottom panels) and $\log_{10}B_{axial} (G)$ (top panels) in the mid-plane for Simulation 4, at the same times as Figure \ref{fig:Sim24_3Dview1}. Also  shown is the vertical magnetic field at the model photosphere, at an acute angle. 
\label{fig:Sim24_3Dview2}}
\end{center}
\end{figure}


Figures \ref{fig:Sim24_3Dview1}-\ref{fig:Sim24_3Dview2} show the overall evolution of the magnetic field in Simulation 4 ($\theta=4\pi/4$). This parameter choice is similar to the eruptive simulation in \citet{Leake_2014}, but the domain is larger in these simulations and the coronal field is a 3D dipole rather than a 2D non-potential arcade. Despite these differences, the qualitative evolution is the same, highlighting the robustness of these results to certain model parameters. 

Figure \ref{fig:Sim24_3Dview1} shows select fieldlines during the evolution of Simulation 4, seeded, at the same locations as in Figure \ref{fig:IC}, in the cross-sectional plane normal to the original flux rope axis. The fieldlines are colored based on their termination points, as a simple way to show their connectivity. The yellow lines originate and terminate at the side boundaries in the convection zone where the original flux rope exits/enters the domain, and so represent the emerging flux rope structure. The purple fieldlines connect at both ends to the lower boundary of the domain and so represent the overlying dipole field. The cyan lines connect at one end to the flux rope and one end to the dipole field, and so early on have  typically reconnected across the Breakout current sheet formed between the emerging twist component of the flux rope, and the overlying dipole. These reconnected fieldlines  form side lobes (see left panels) and contribute to a quadrupolar structure in the corona which is similar to the pre-supposed magnetic configuration of the Magnetic Breakout Model. Figure \ref{fig:Sim24_3Dview2} shows the value of  $|\mb{J}/J_0|/|\mb{B}/B_0|$ in the $y=0$ plane for Simulation 4 which highlights the Breakout current sheet, as well as a vertically aligned current sheet that appears later in the emergence at the center of the AR, which is discussed below.

The buoyant section of the convection zone flux rope rises to the surface and partially emerges, forming predominantly concave down fieldlines above the model surface. At this point, there are no concave up \JEL{flux rope} field lines \JEL{in the corona} and so the flux rope has not fully emerged. This partial emergence results in the 
  tadpole shaped surface magnetic field structure often observed in bipolar active regions \citep{Luoni_11,Poisson_15,Poisson_16}, and in compact active regions called delta spots \citep{Shi_1994}. This tadpole structure can be explained by the sequential emergence of initially twisted fieldlines, followed by fieldlines closer to the flux rope axis. By time  $t=142 ~ t_0$, a sheared arcade has formed and displaced the coronal field. The configuration at this point is  similar to that in the  Breakout Model after significant surface shearing. A current sheet has formed between the twist field of the emerging structure and the dipole field, primarily due to the antiparallel horizontal fields in the two systems ($\theta=4\pi/4$). 

As the magnetic rope continues to supply magnetic flux through the surface, the sheared arcade continues to expand, until the current sheet starts to significantly reconnect (yellow fieldlines reconnect with purple fieldlines to generate side lobes, cyan fieldlines). The Breakout reconnection removes the overlying fields of the dipole field, but also the twist field that emerged.  Both of these fields provide tension which restrict the rise of the arcade, and so the removal of those fields allows further expansion.

As can be seen in Figures \ref{fig:Sim24_3Dview1}-\ref{fig:Sim24_3Dview2}, the vertical rise of the sheared arcade results in the formation of a vertically-aligned current sheet below the arcade. This current sheet is formed by horizontal gradients in the vertical field, and we label this current sheet the \textit{flare current sheet}, as its geometry resembles the magnetic field in the classical CSHKP flare model \citep{Carmichael,Sturrock,Hirayama,Kopp}. Reconnection here involves yellow fieldlines reconnecting with each other across the current sheet, as well as cyan fieldlines which have foot-points at either end of the original flux rope, see Figure 9 of \citet{Leake_2014}. As a result of this reconnection, concave up fieldlines are seen \JEL{above the surface} and an isolated flux rope is formed, which is distinct from the original flux rope that only partially emerges into the corona. Continued magnetic Breakout reconnection occurs, removing more overlying field. The reconnection at the flare current sheet not only adds flux to the coronal flux rope but also removes the tethering of the yellow fieldlines of the arcade. This ``tether cutting" reconnection \citep{Sturrock_1989} has been indirectly \JEL{detected} in observed active regions prior to large flares and CMEs \citep{Yurchyshyn_2006}.
 \citet{Karpen_2012} showed that such a current sheet is necessary for a fast eruption of the sheared arcade/coronal flux rope. In the simulations presented here, it will be shown that reconnection in this current sheet results in a fast rise of the flux rope, which is labelled here as an eruption. Furthermore, the development of the flare current sheet is controlled by the amount of expansion of the emerging structure, and thus by the Breakout reconnection and the relative orientation of the emerging and coronal fields. The eventual evolution of the eruption is discussed as part of the parameter study below.

\begin{figure}
    \centering
    
    \includegraphics[width=0.3\textwidth]{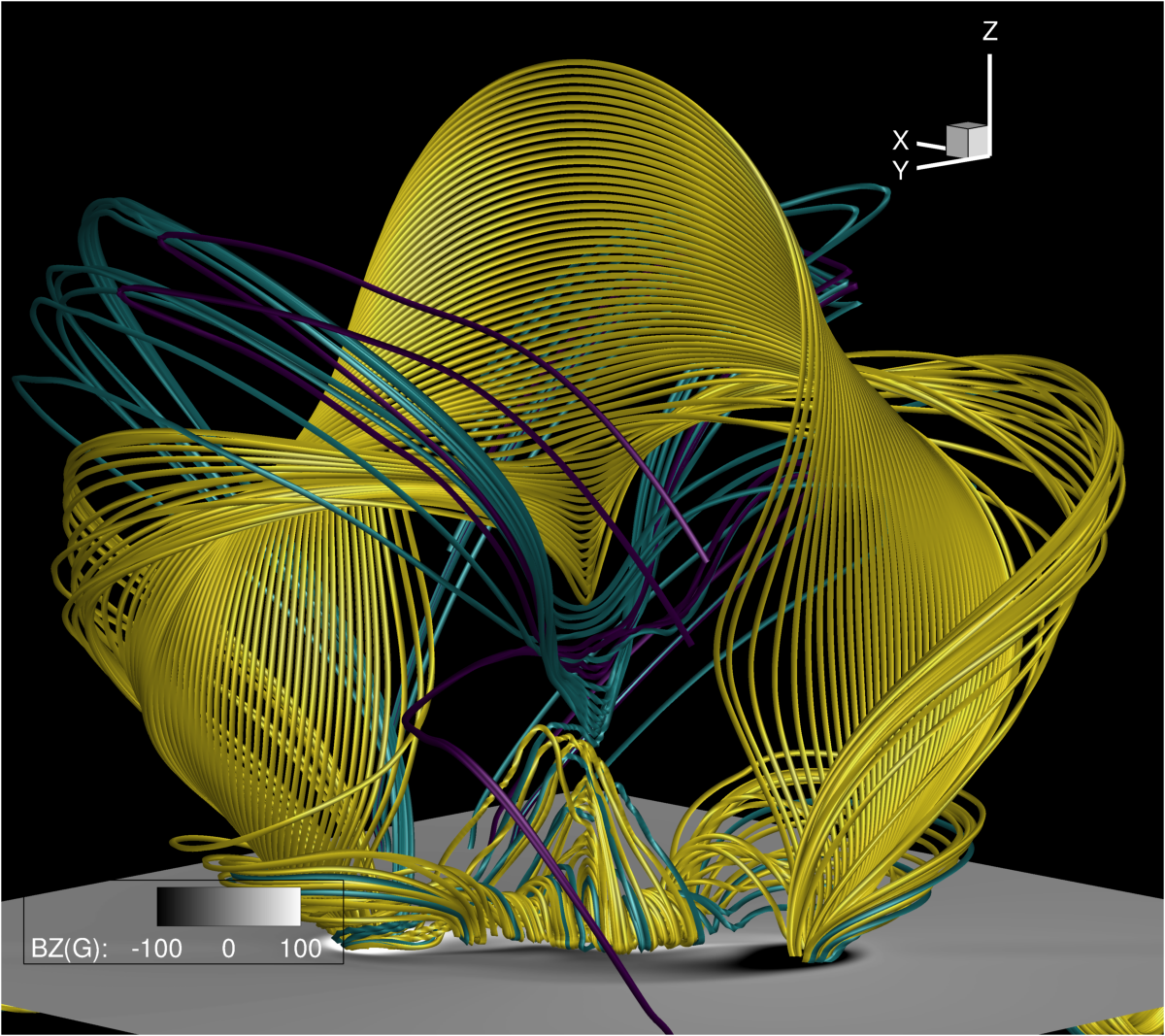}
    \includegraphics[width=0.3\textwidth]{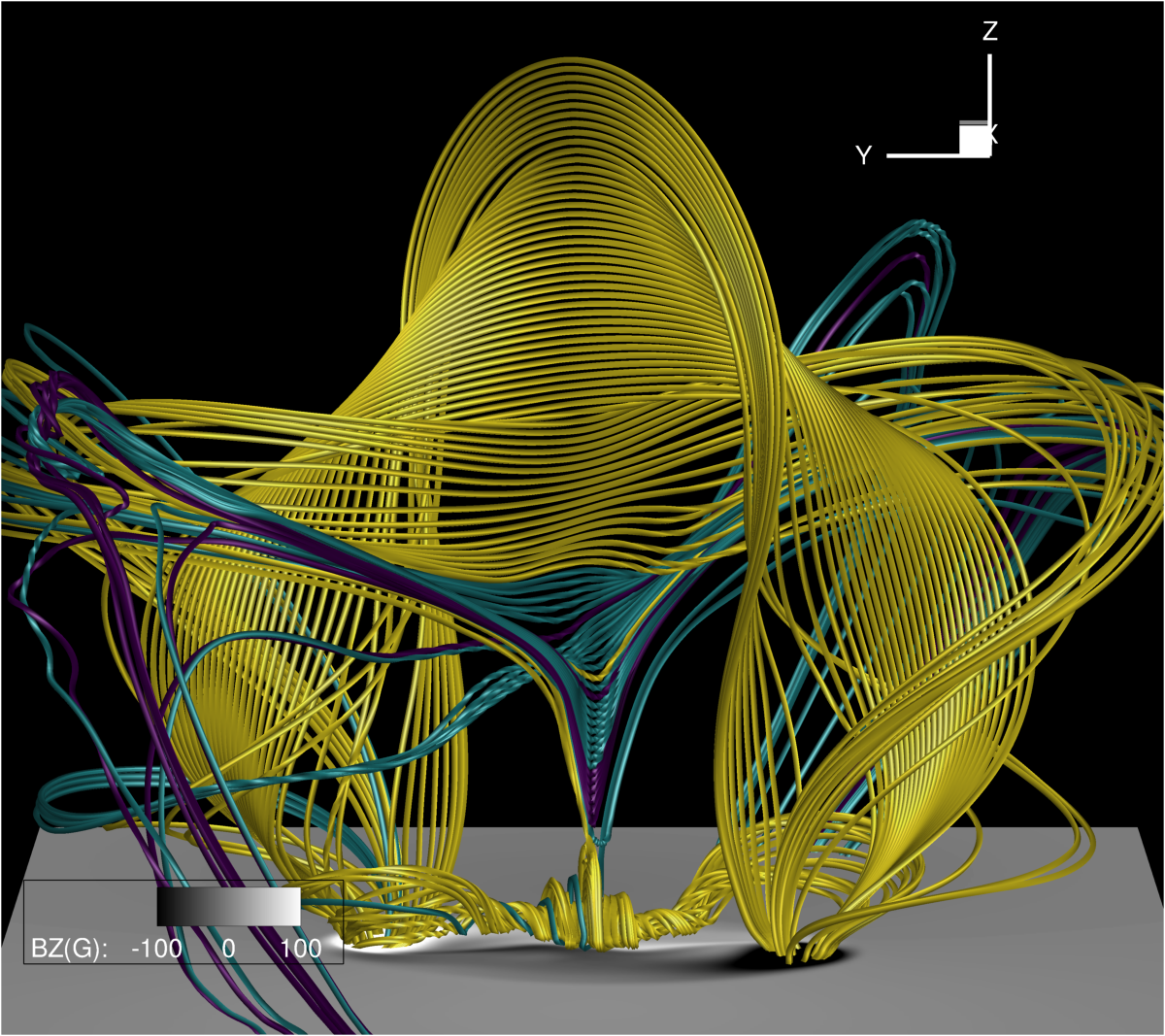}
    \includegraphics[width=0.3\textwidth]{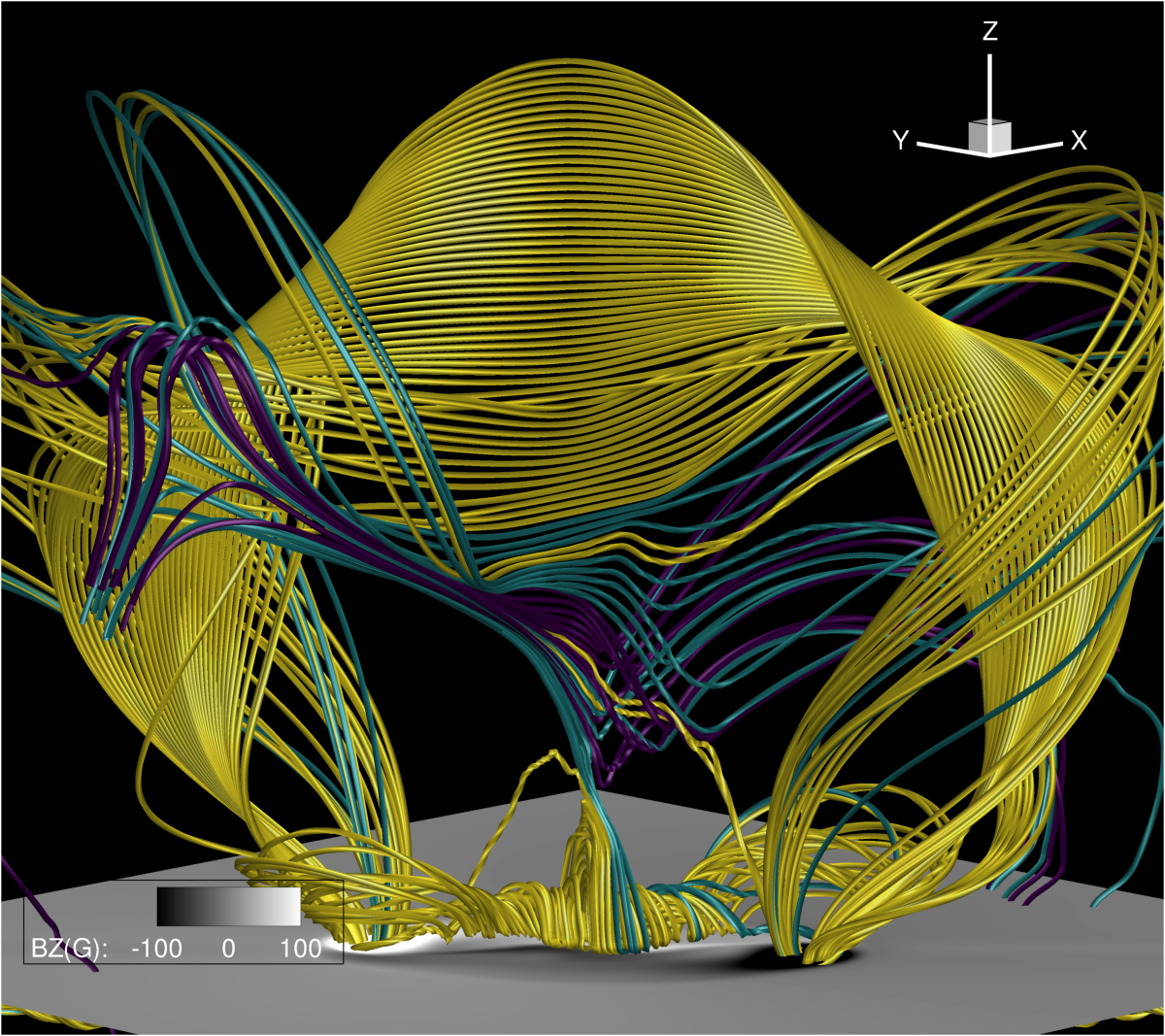} \\
    \includegraphics[width=0.3\textwidth]{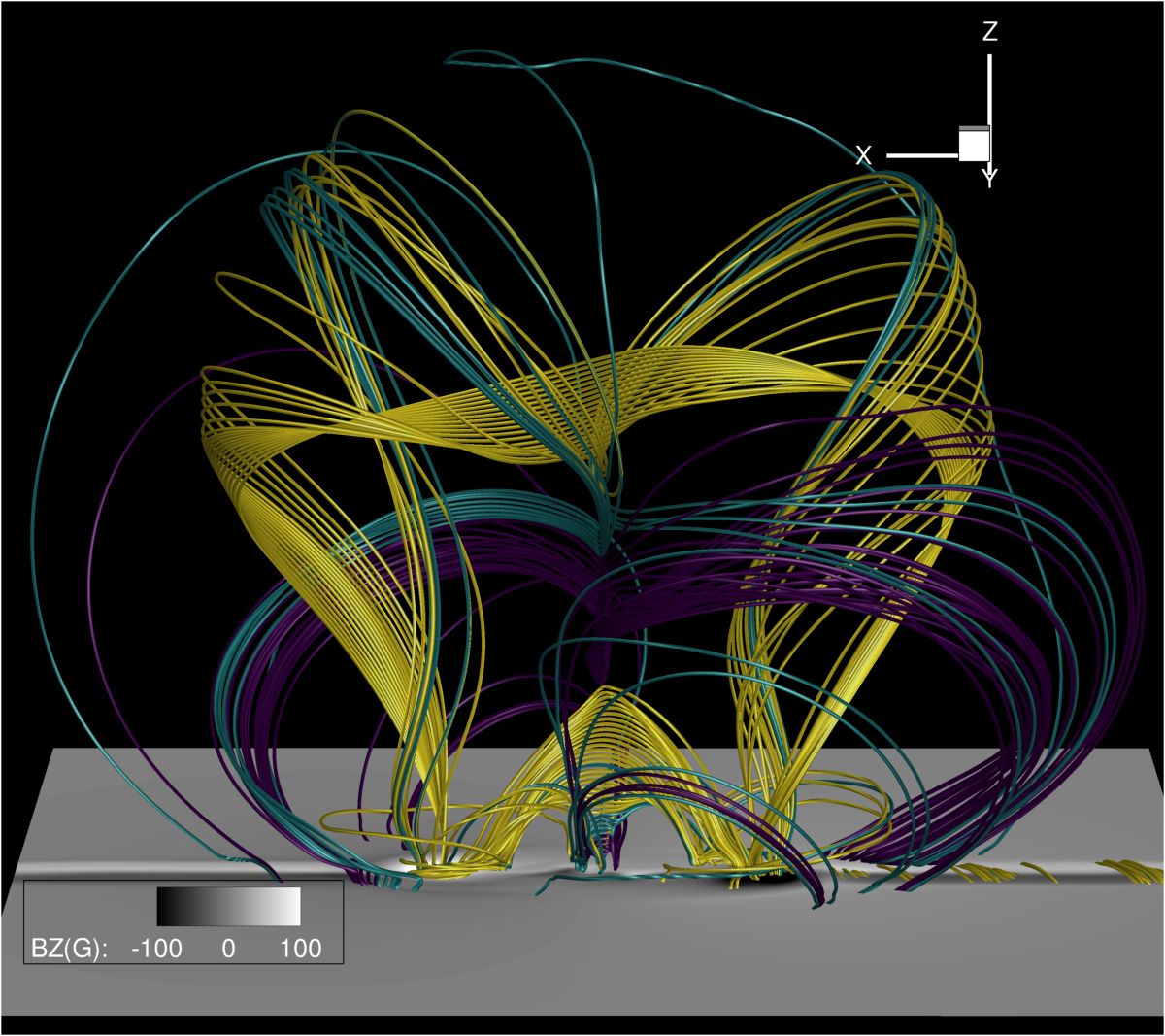}
    \includegraphics[width=0.3\textwidth]{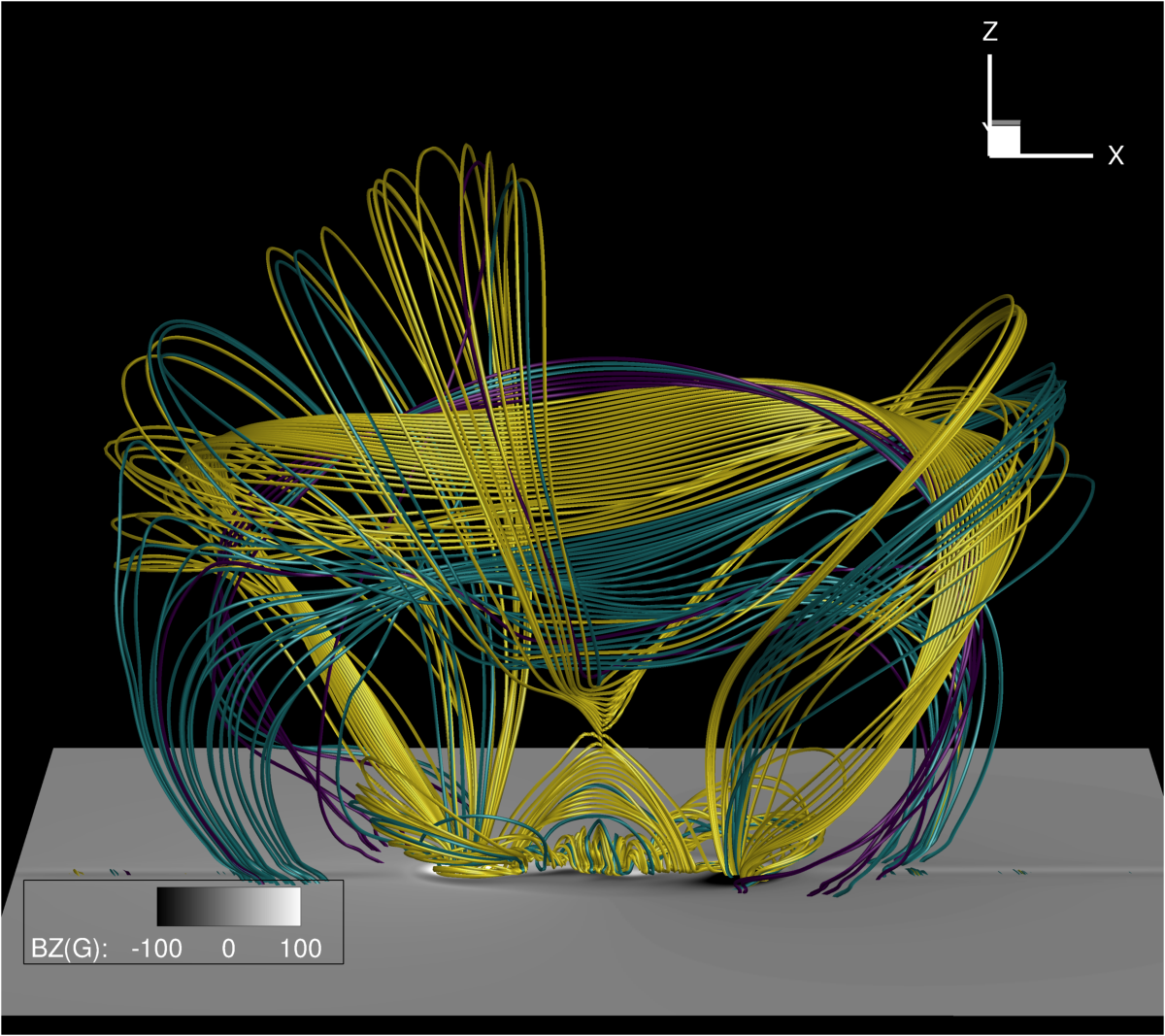}
    \caption{Selected fieldlines during the eruption in Simulations 3,4,5 (top row, left to right) and Simulations 2 and 6 (bottom row, left to right). The fieldlines are colored the same as in Figure \ref{fig:Sim24_3Dview1}. \label{fig:newropes}}
    
\end{figure}

Much discussion has occurred \citep[e.g.,][]{Patsourakos_2020} about the pre-eruptive magnetic configuration of solar eruptions, with different models for eruption relying on flux ropes (such as the torus instability) and a sheared arcade with a null point (magnetic Breakout). In these simulations, a transition from a sheared arcade to a coronal flux rope is observed as part of the eruption process. \RR{This can be seen in Figure \ref{fig:newropes} which shows selected fieldlines during the eruptions that occur in Simulations 2,3,4,5, and 6. The details of these simulations are discussed in the next subsection. The result of the strong flare reconnection which precedes the fast rise into the upper domain in these simulations is to create a new flux rope out of the sheared arcade, which itself was formed by the partial emergence of the original convection zone flux rope. The convection zone flux rope remains partially submerged near the surface.} The magnetic configuration and its relationship with different proposed models of solar eruptions will be considered below.

\subsection{Effect of relative orientation}

\begin{figure}
  \centering
  \setlength{\unitlength}{0.1\textwidth}
  \begin{picture}(10,10)
   \put(0,7.5){\includegraphics[width=0.31\textwidth]{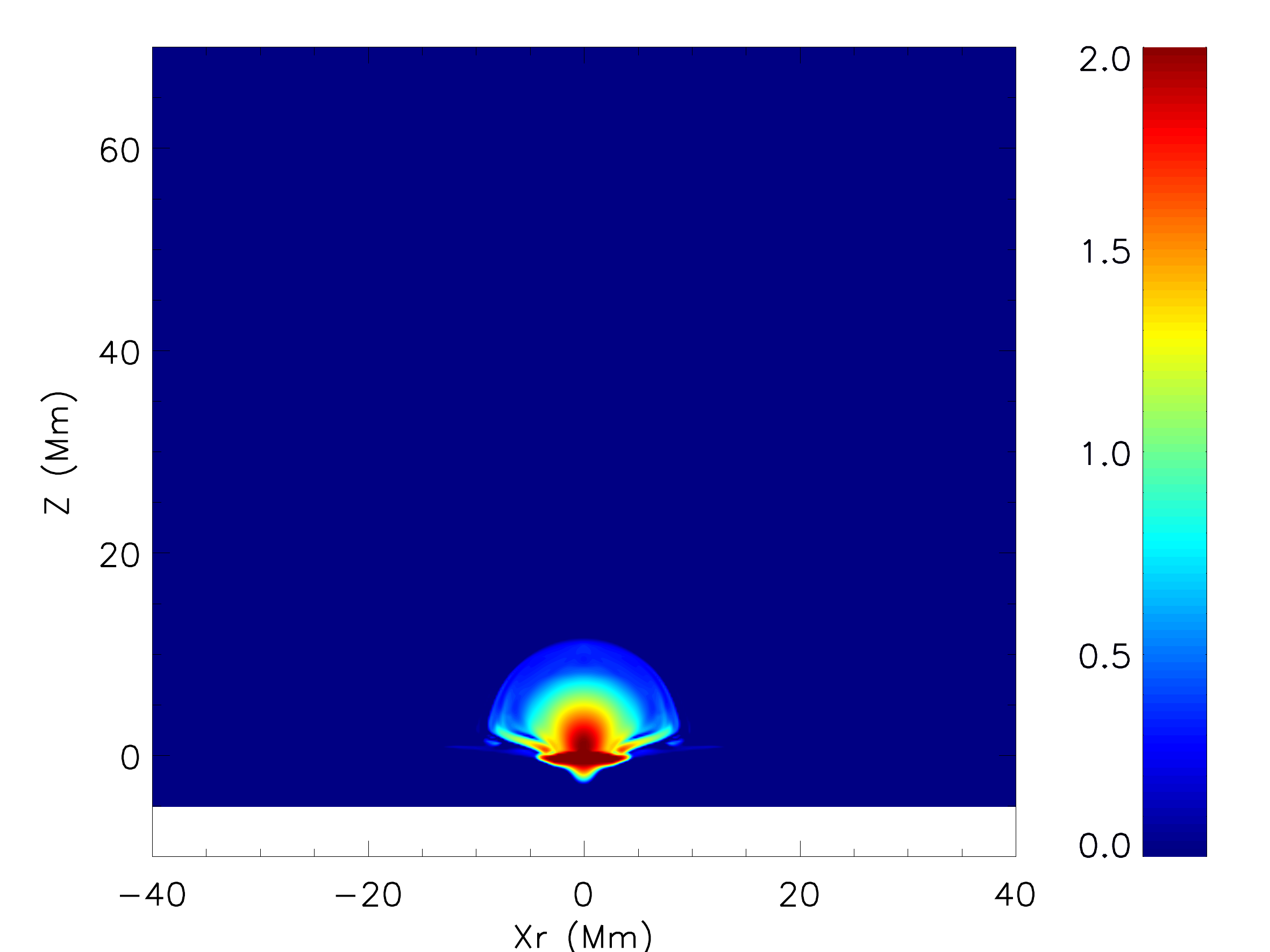}}
    \put(3,7.5){\includegraphics[width=0.31\textwidth]{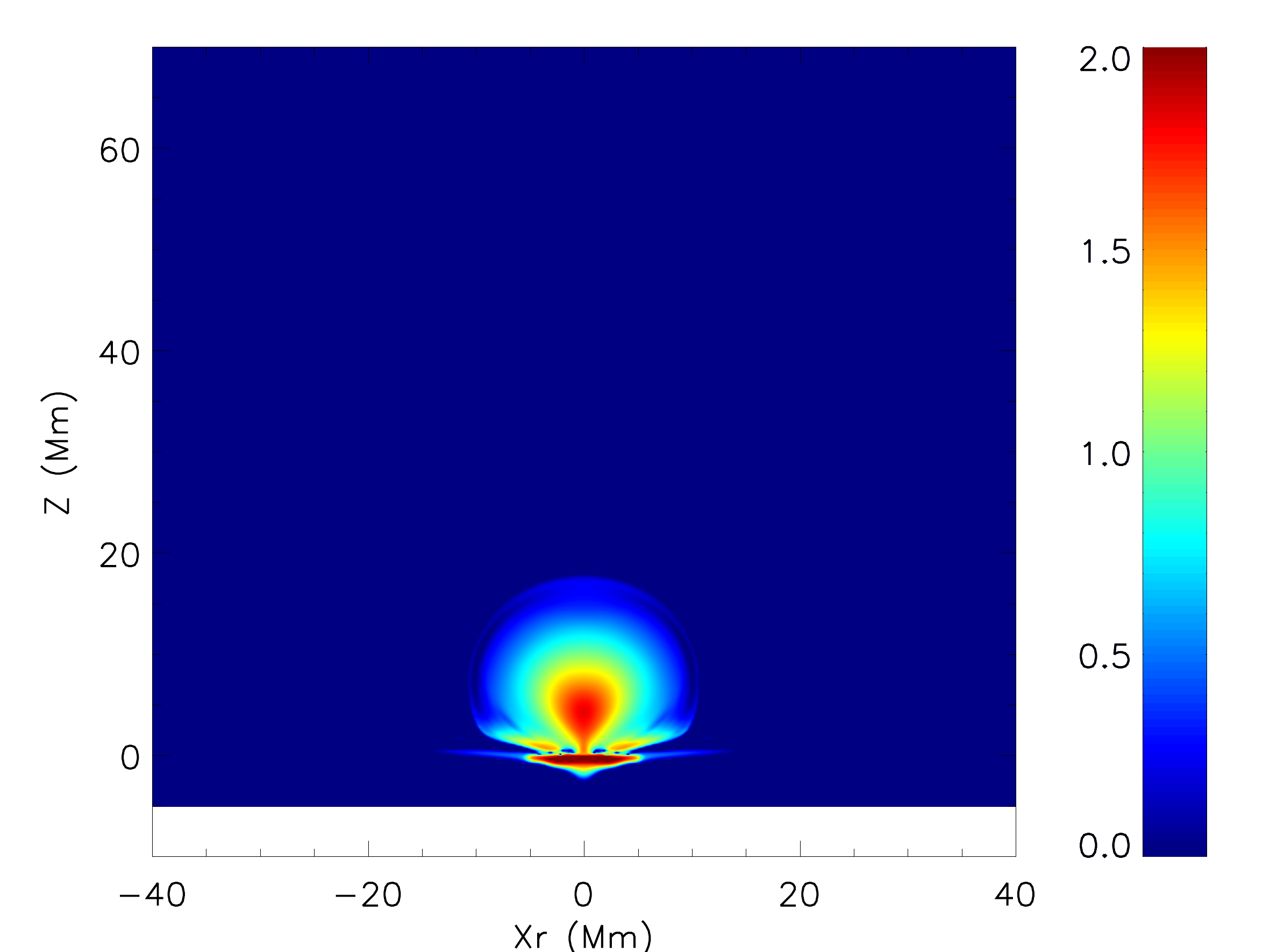}}
    \put(6,7.5){\includegraphics[width=0.31\textwidth]{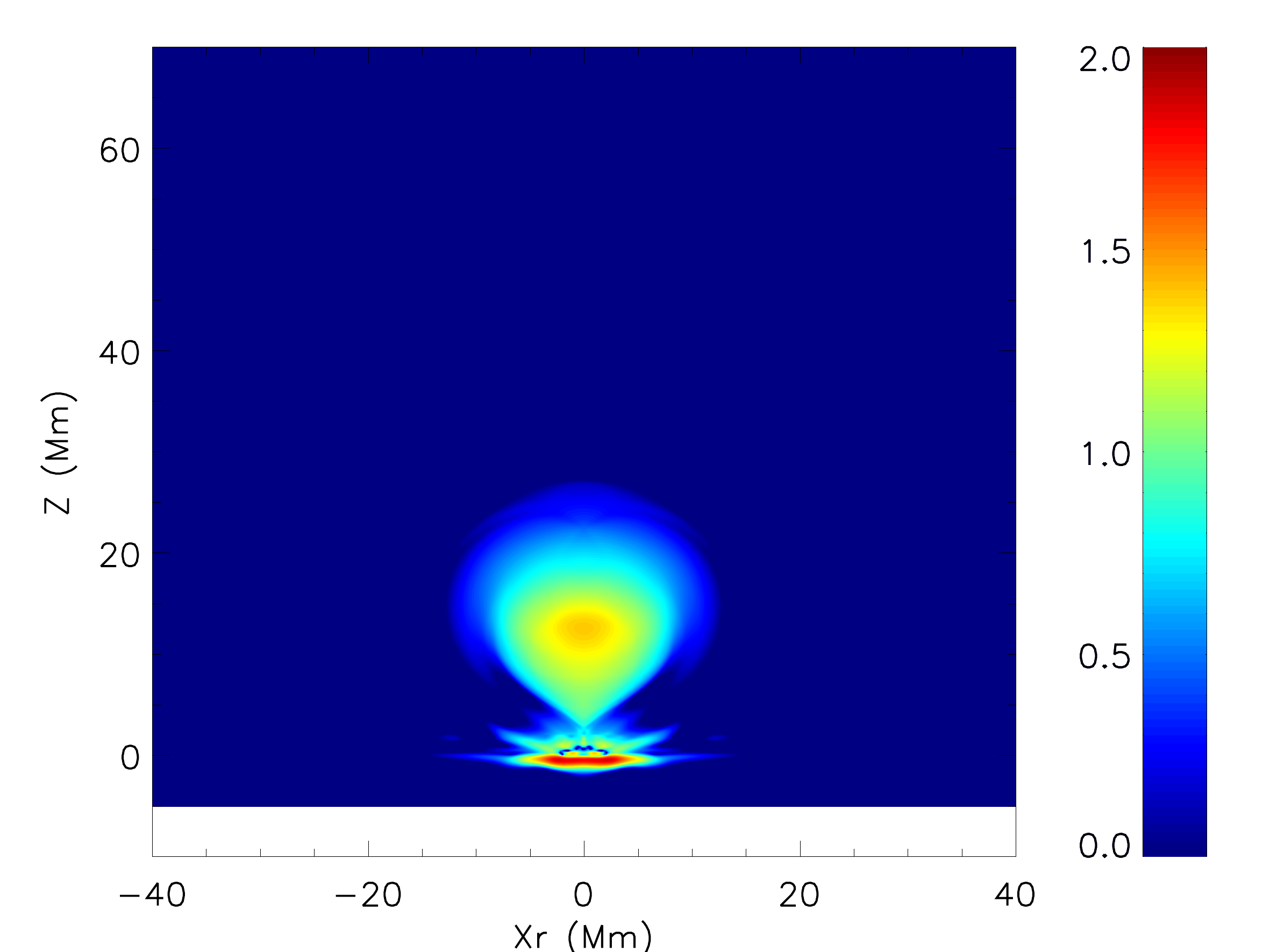}}
    \put(0.5,9.5){\colorbox{white}{{\color{olive} Sim 0: $t=100t_{0}$}}}
    \put(3.5,9.5){\colorbox{white}{{\color{olive} Sim 0: $t=170t_{0}$}}}
    \put(6.5,9.5){\colorbox{white}{{\color{olive} Sim 0: $t=400t_{0}$}}}
    \put(0,5){\includegraphics[width=0.31\textwidth]{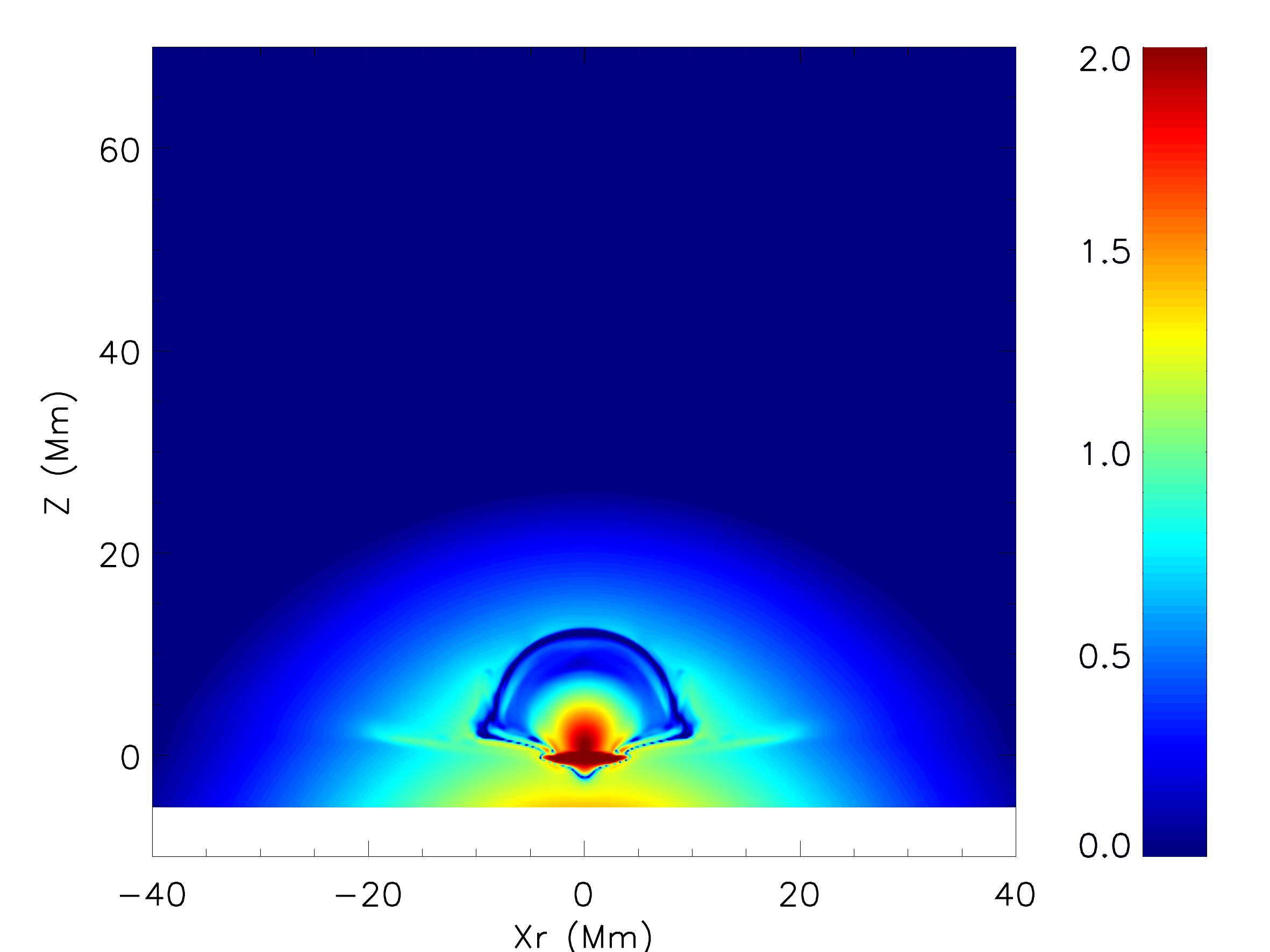}}
    \put(3,5){\includegraphics[width=0.31\textwidth]{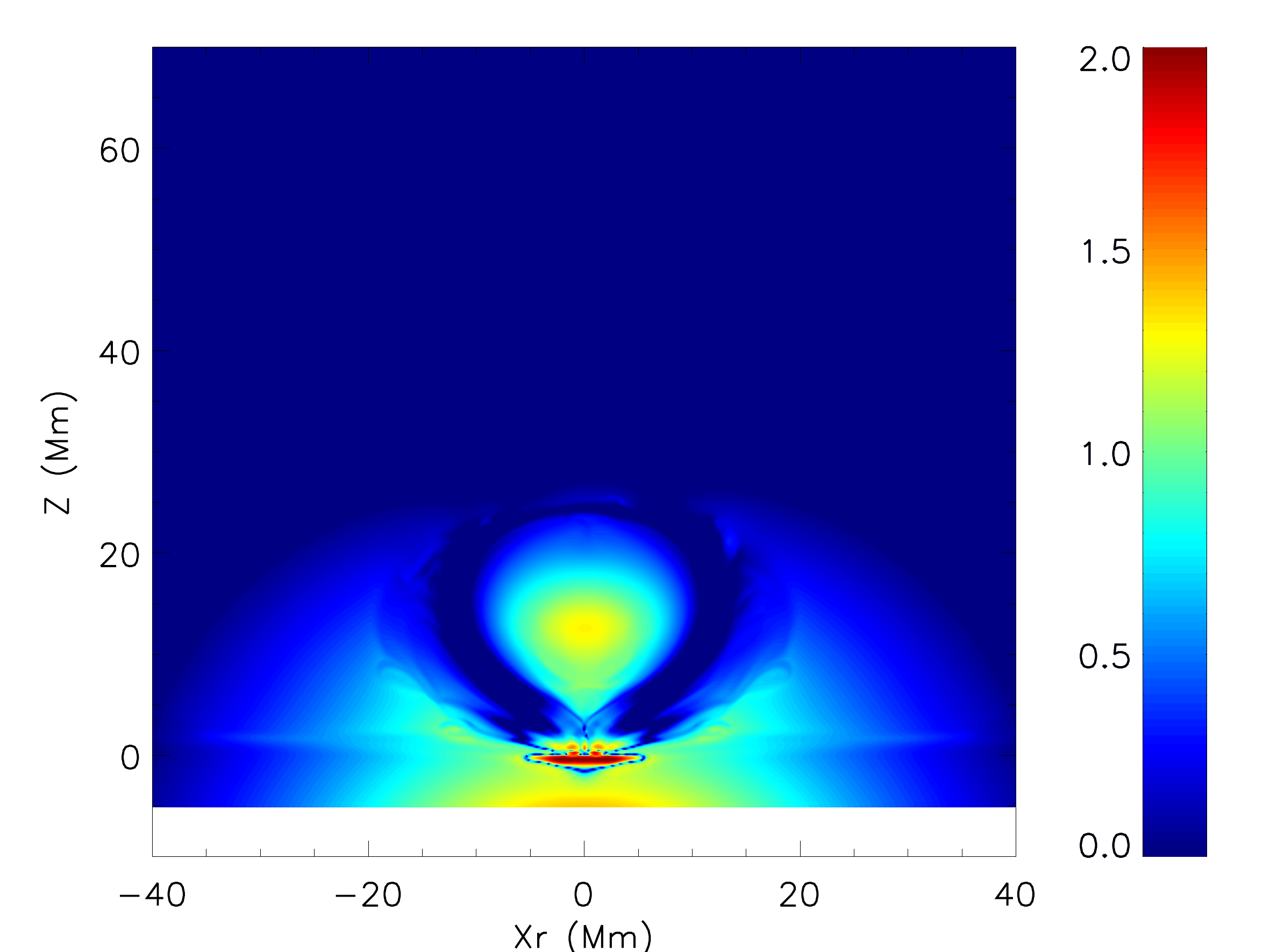}}
    \put(6,5){\includegraphics[width=0.31\textwidth]{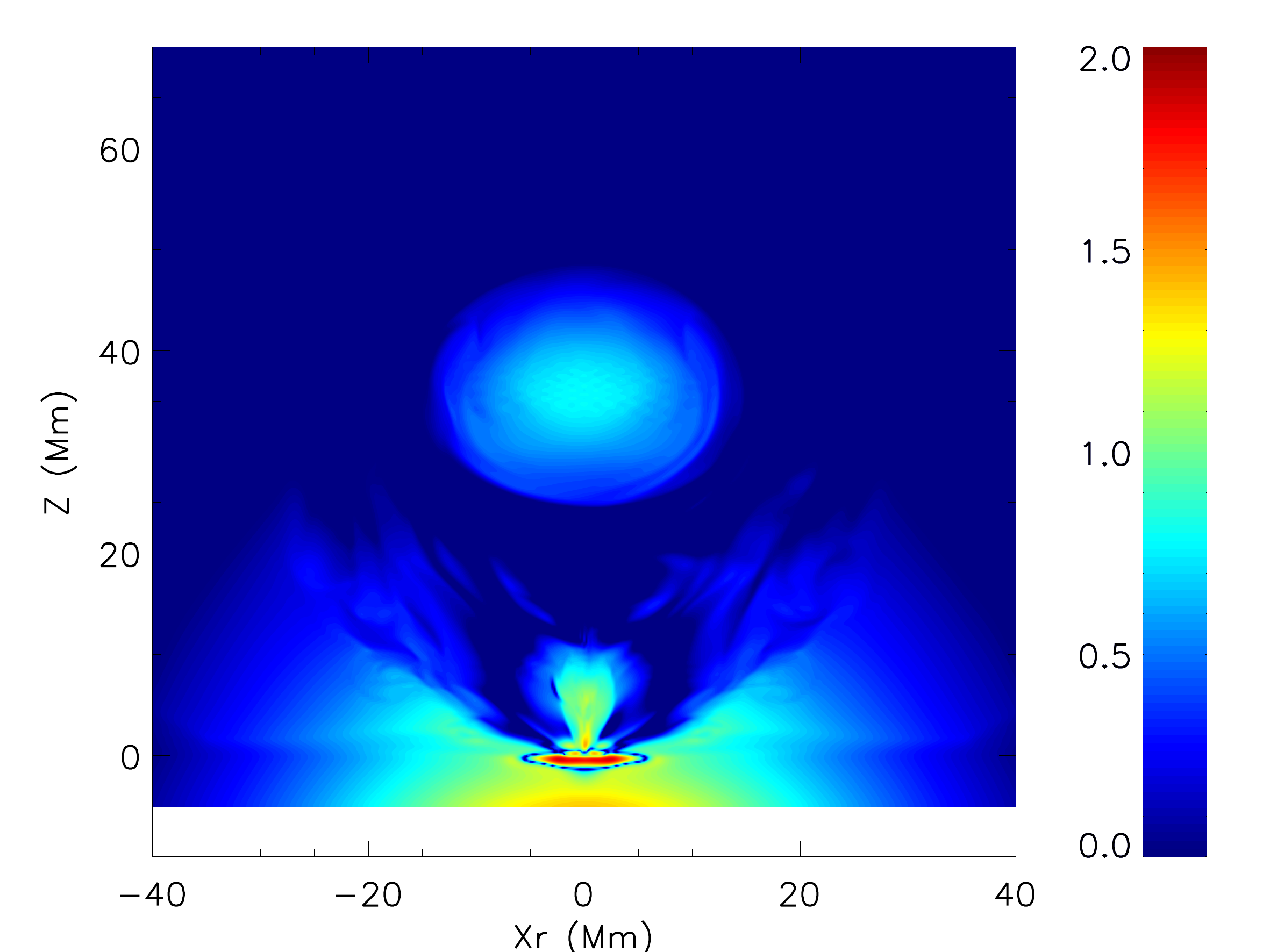}}
    \put(0.5,7){\colorbox{white}{{\color{blue} Sim 2: $t=100t_{0}$}}}
    \put(3.5,7){\colorbox{white}{{\color{blue} Sim 2: $t=240t_{0}$}}}
    \put(6.5,7){\colorbox{white}{{\color{blue} Sim 2: $t=380t_{0}$}}} 
    \put(0,2.5){\includegraphics[width=0.31\textwidth]{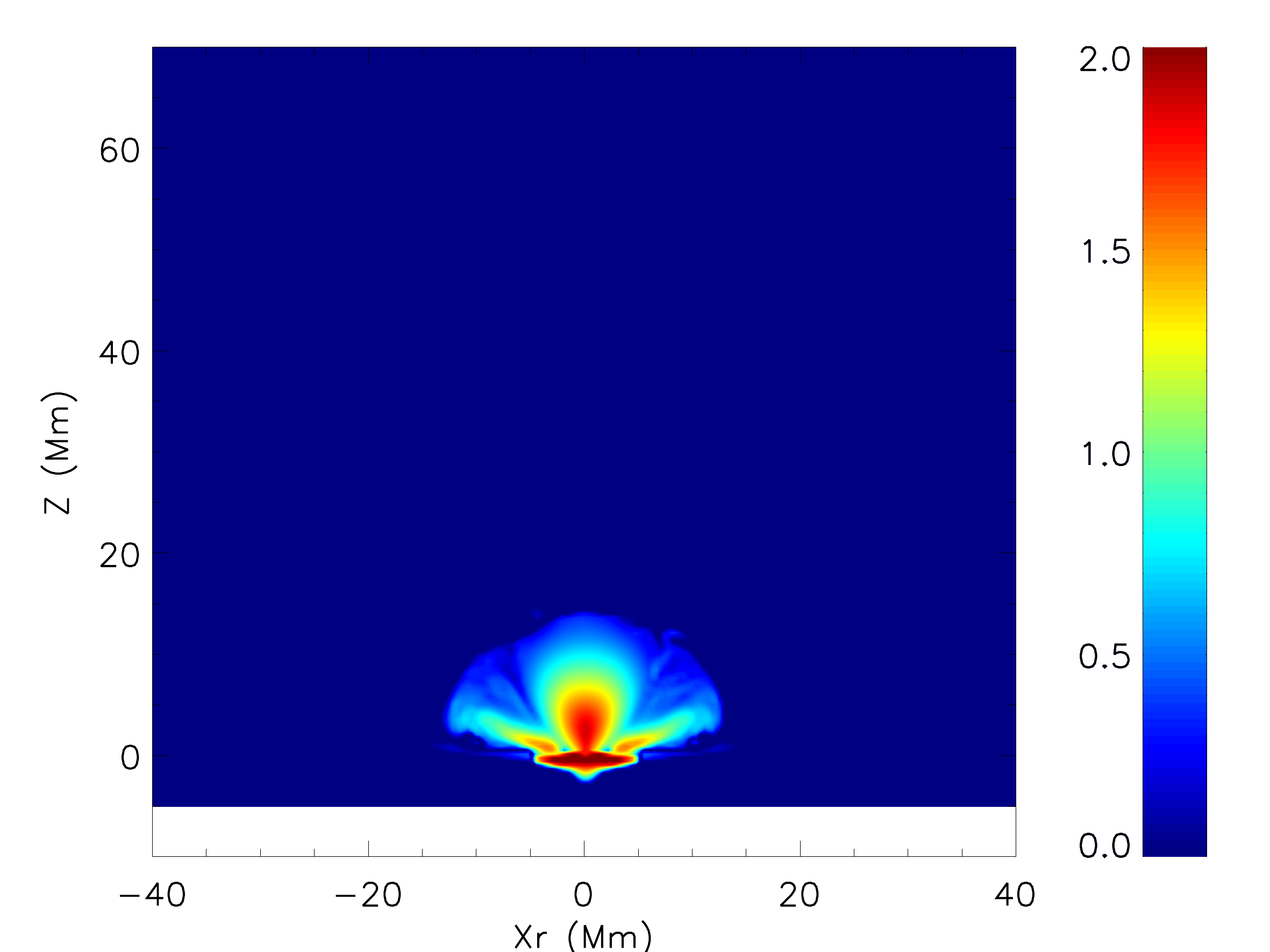}}
    \put(3,2.5){\includegraphics[width=0.31\textwidth]{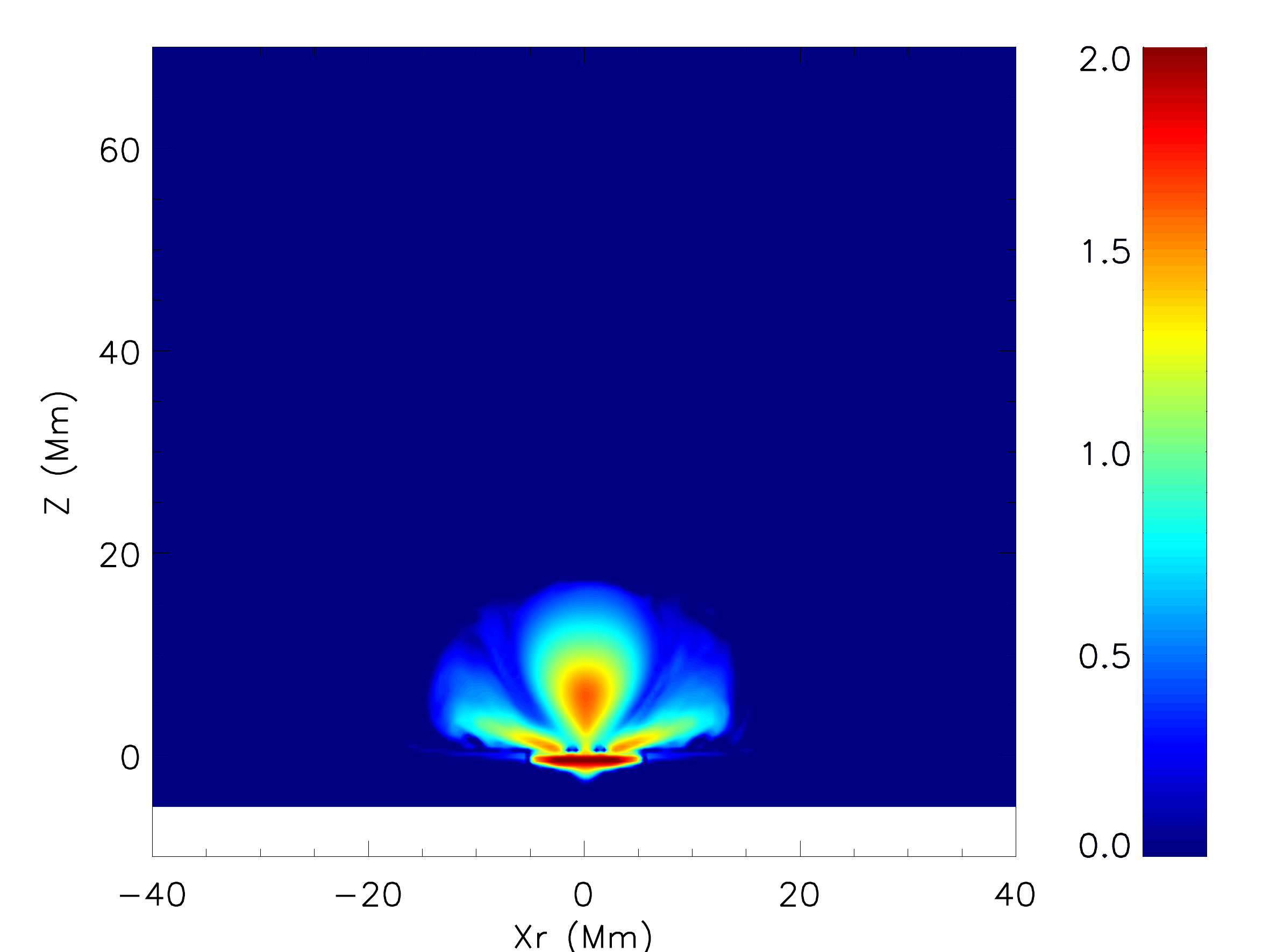}}
    \put(6,2.5){\includegraphics[width=0.31\textwidth]{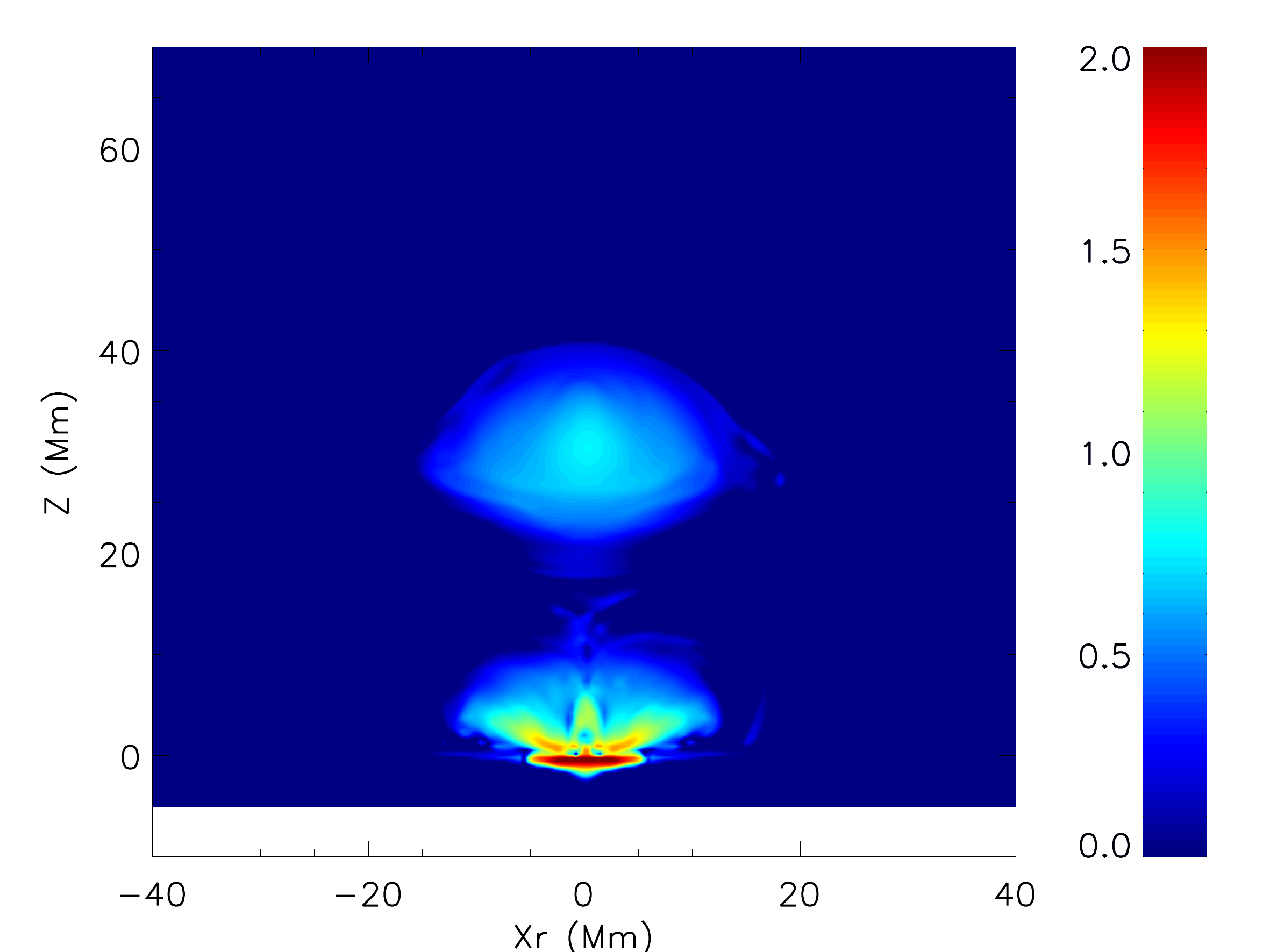}}
    \put(0.5,4.5){\colorbox{white}{{\color{orange} Sim 4: $t=130t_{0}$}}}
    \put(3.5,4.5){\colorbox{white}{{\color{orange} Sim 4: $t=160t_{0}$}}}
    \put(6.5,4.5){\colorbox{white}{{\color{orange} Sim 4: $t=190t_{0}$}}}
     \put(0,0){\includegraphics[width=0.31\textwidth]{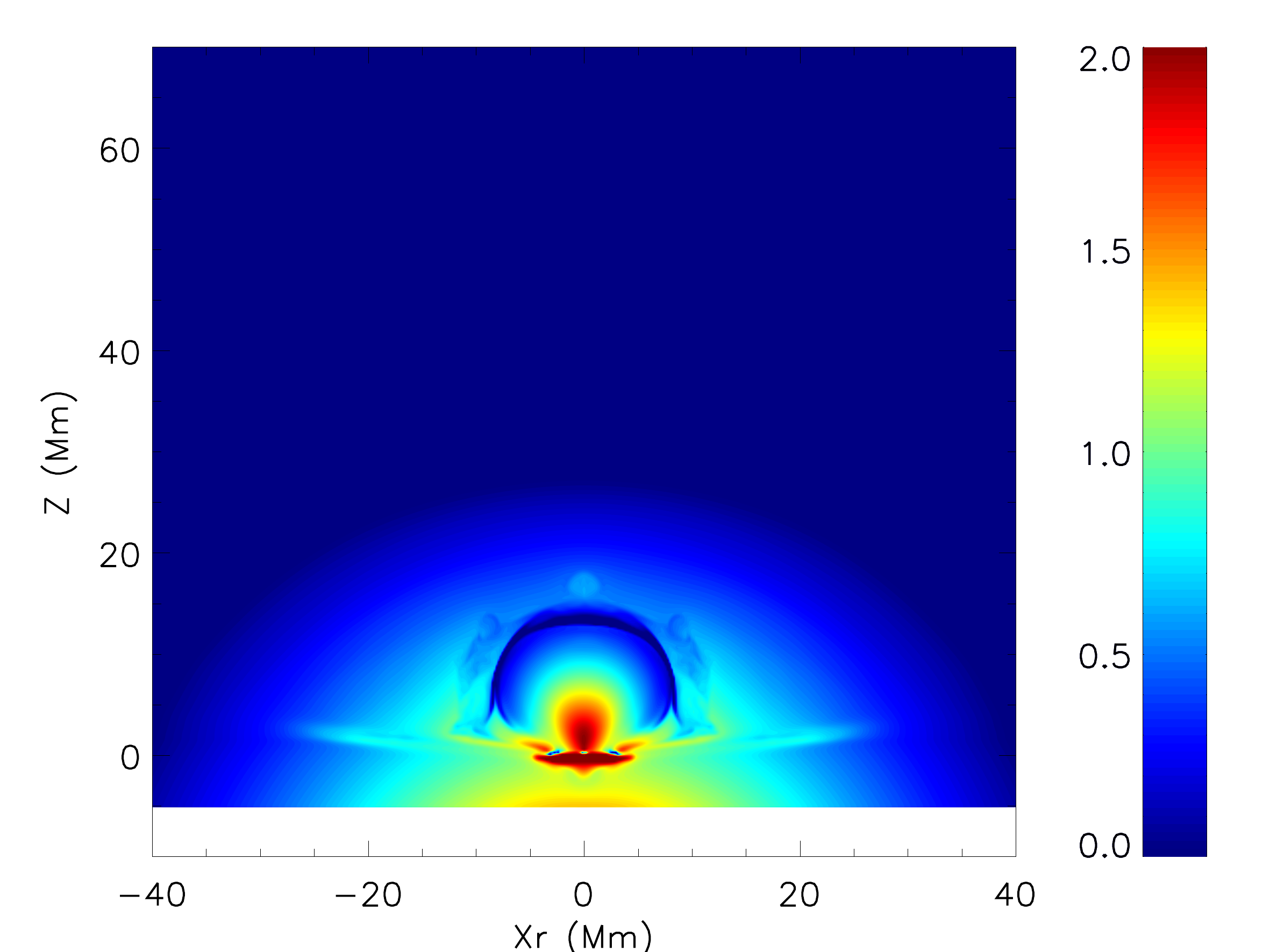}}
    \put(3,0){\includegraphics[width=0.31\textwidth]{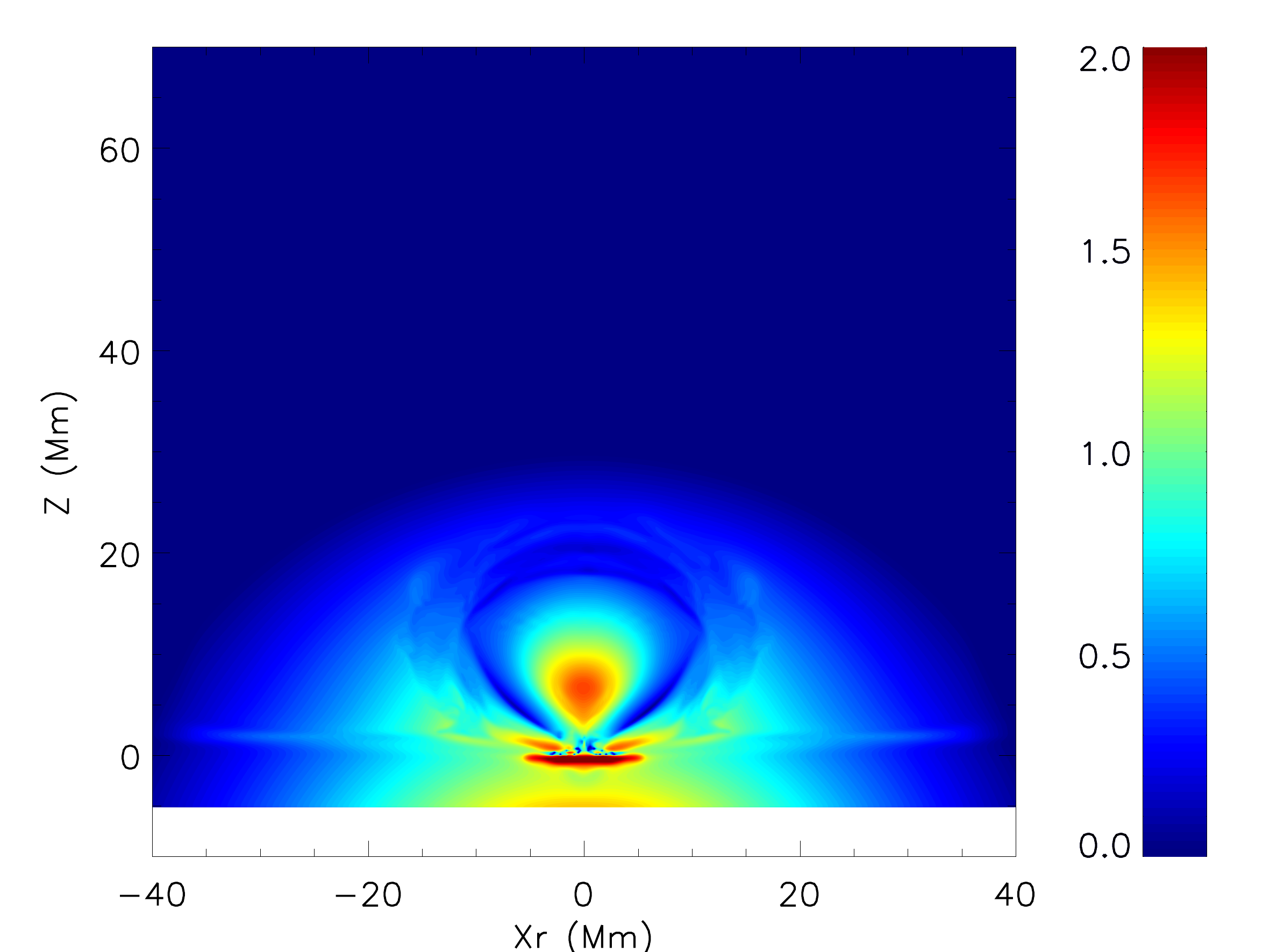}}
    \put(6,0){\includegraphics[width=0.31\textwidth]{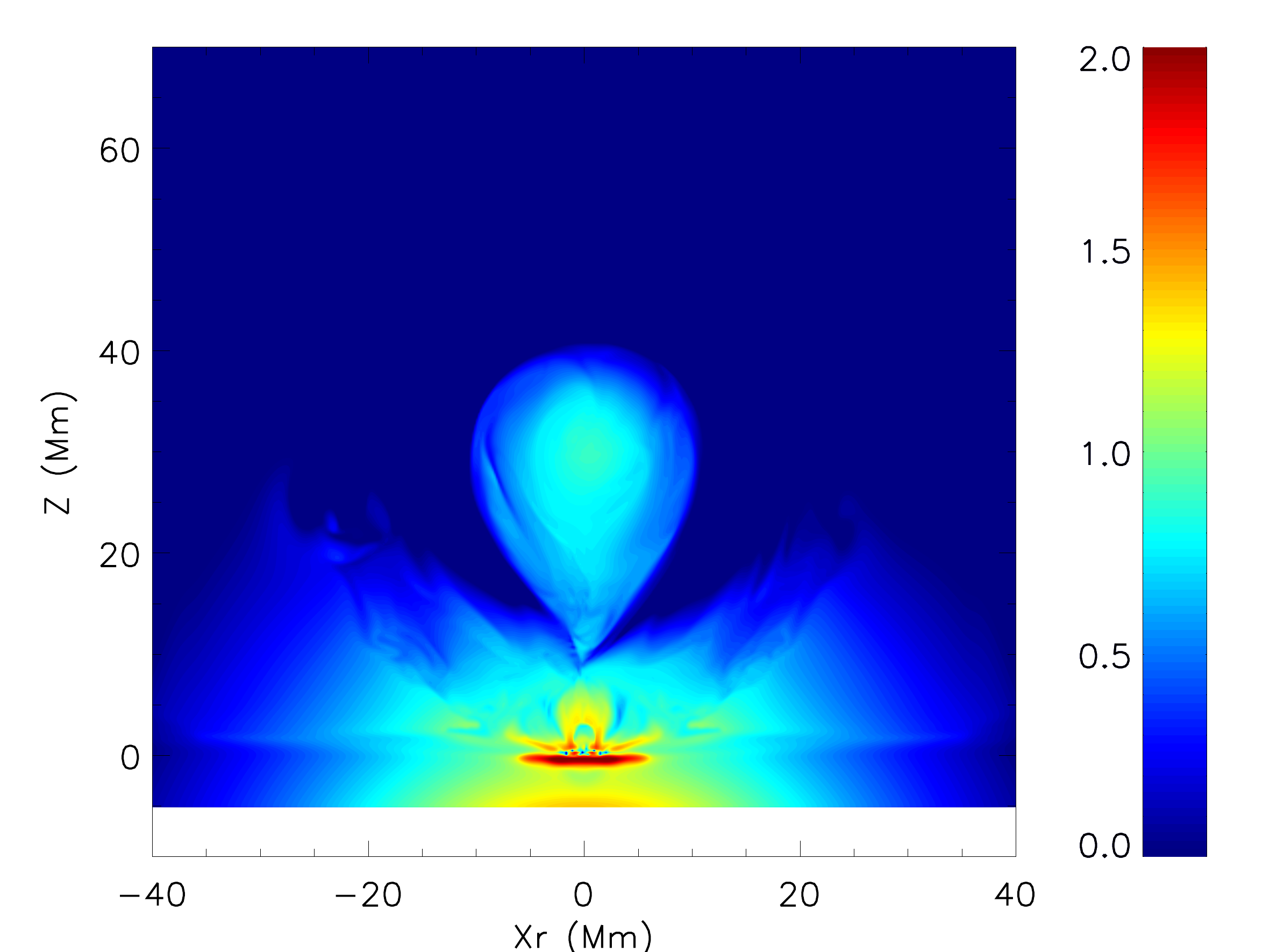}}
    \put(0.5,2){\colorbox{white}{{\color{NavyBlue} Sim 6: $t=120t_{0}$}}}
    \put(3.5,2){\colorbox{white}{{\color{NavyBlue} Sim 6: $t=190t_{0}$}}}
    \put(6.5,2){\colorbox{white}{{\color{NavyBlue} Sim 6: $t=260t_{0}$}}}
   \end{picture}
  \caption{Logarithm of the absolute value of the axial field (in G) in the cross-sectional plane halfway along the flux rope for Simulations (0,2,4,6) at three different times.\label{fig:B_axial}}
  \end{figure}
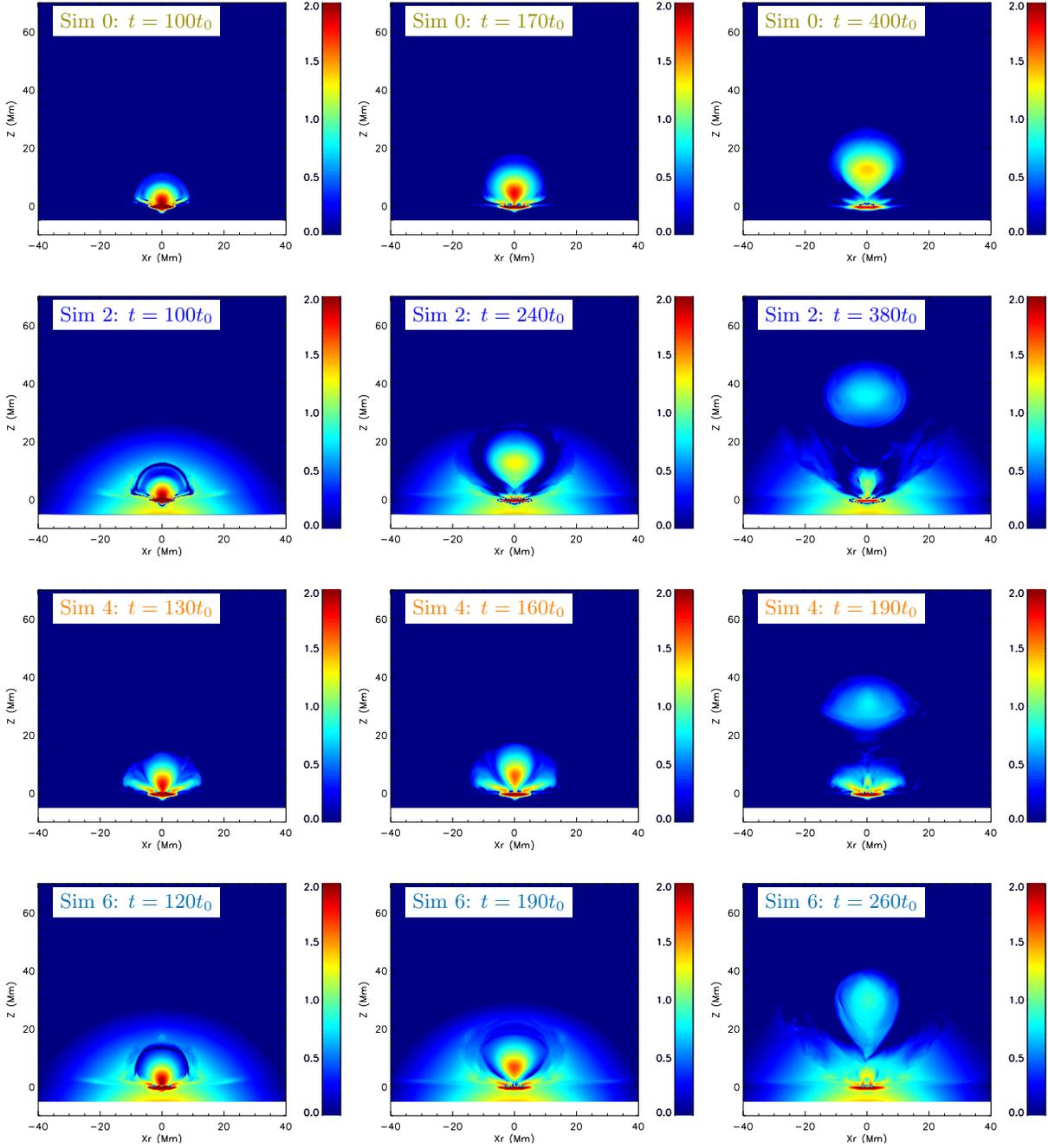

\begin{figure}
  \centering
  \setlength{\unitlength}{0.1\textwidth}
  \begin{picture}(10,10)
   \put(0,7.5){\includegraphics[width=0.31\textwidth]{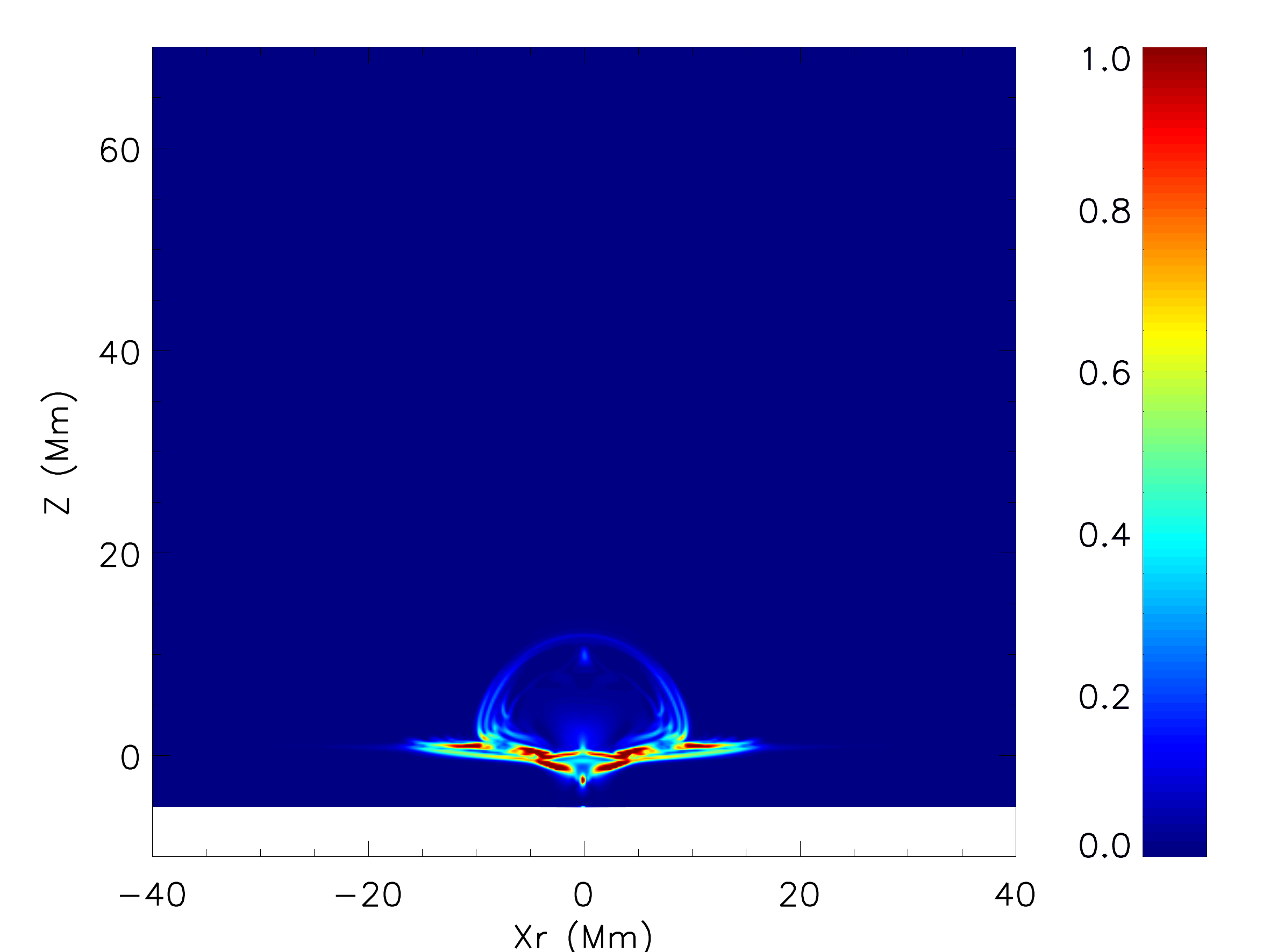}}
    \put(3,7.5){\includegraphics[width=0.31\textwidth]{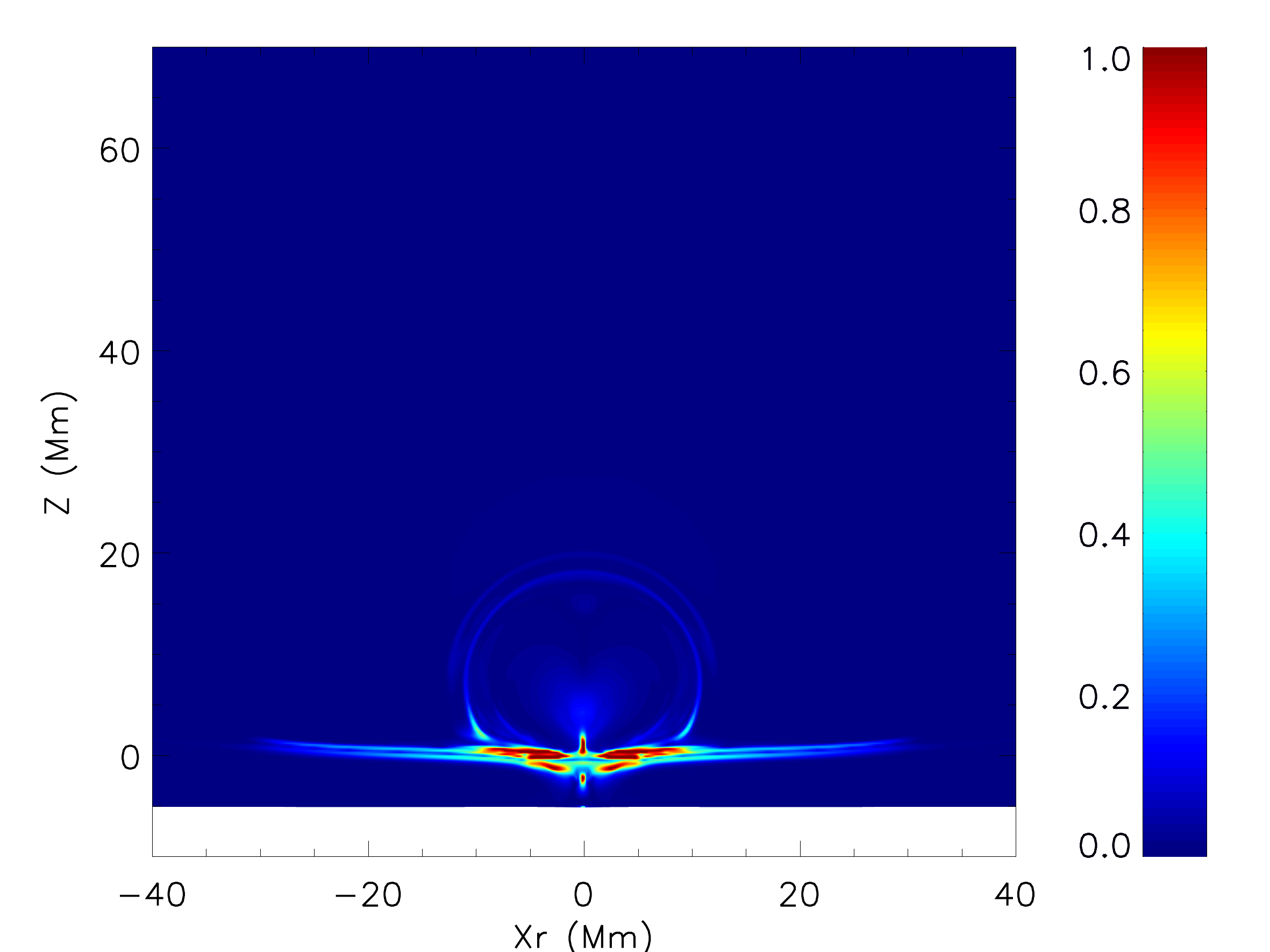}}
    \put(6,7.5){\includegraphics[width=0.31\textwidth]{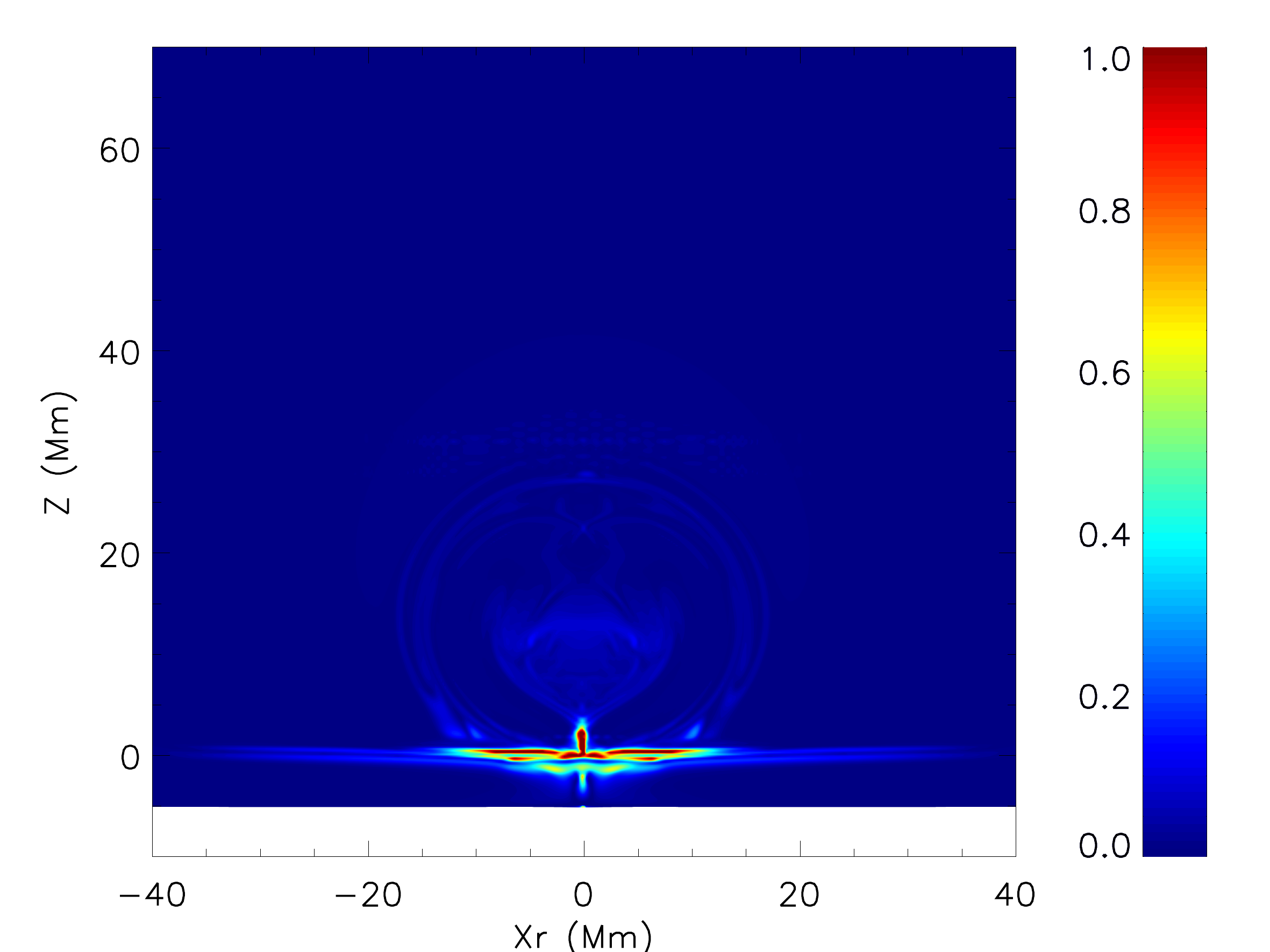}}
    \put(0.5,9.5){\colorbox{white}{{\color{olive} Sim 0: $t=100t_{0}$}}}
    \put(3.5,9.5){\colorbox{white}{{\color{olive} Sim 0: $t=170t_{0}$}}}
    \put(6.5,9.5){\colorbox{white}{{\color{olive} Sim 0: $t=400t_{0}$}}}
    \put(0,5){\includegraphics[width=0.31\textwidth]{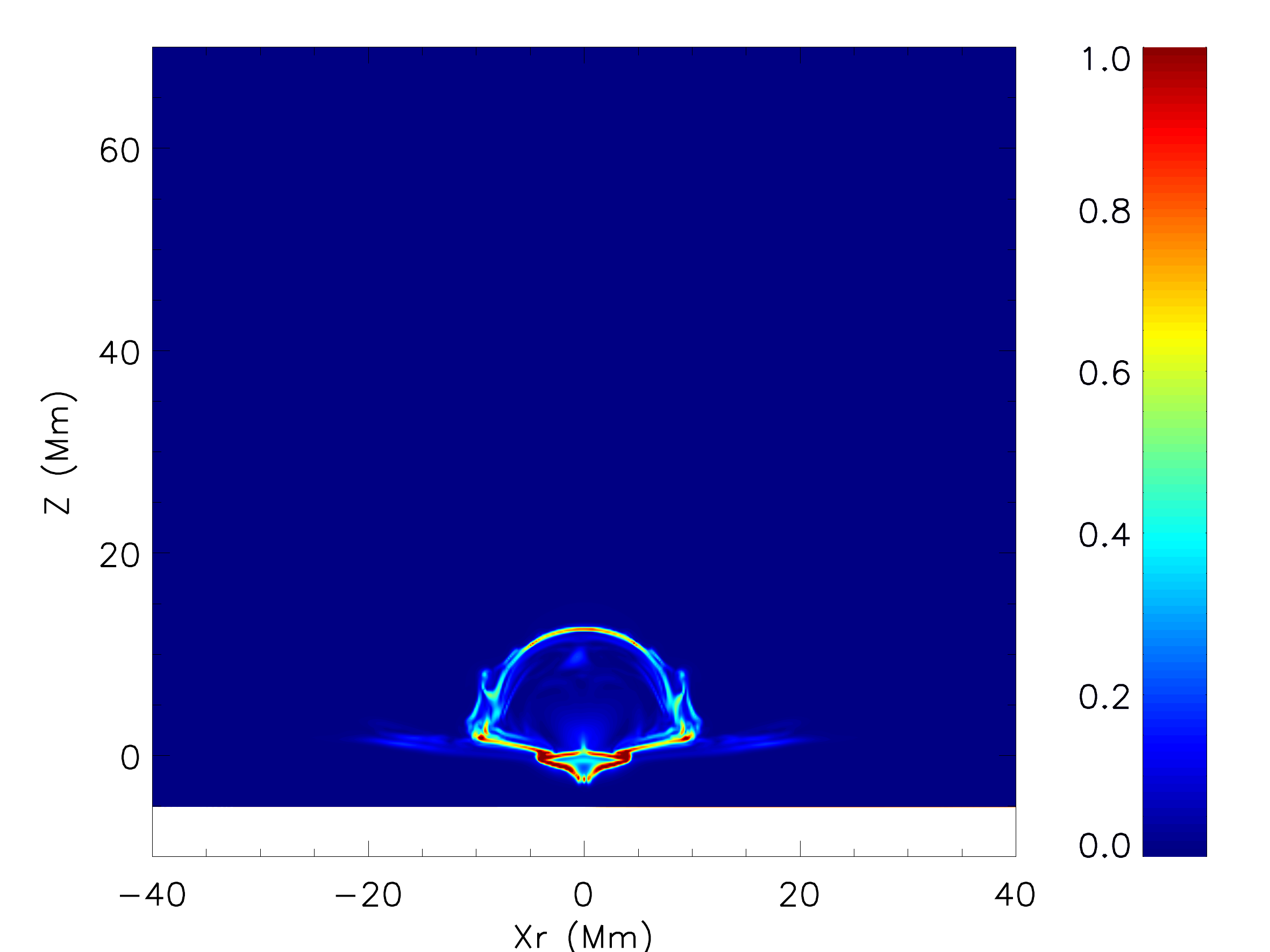}}
    \put(3,5){\includegraphics[width=0.31\textwidth]{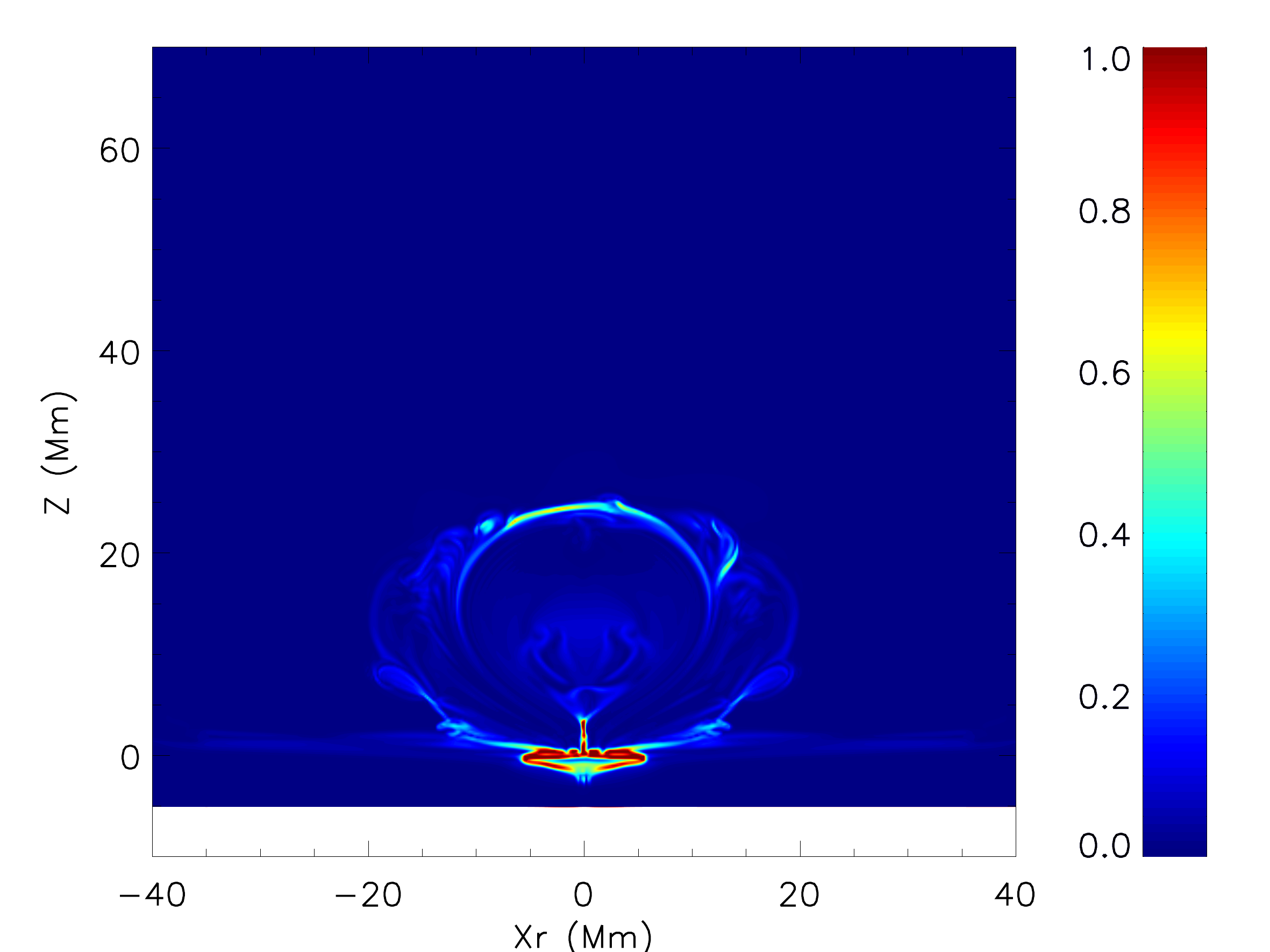}}
    \put(6,5){\includegraphics[width=0.31\textwidth]{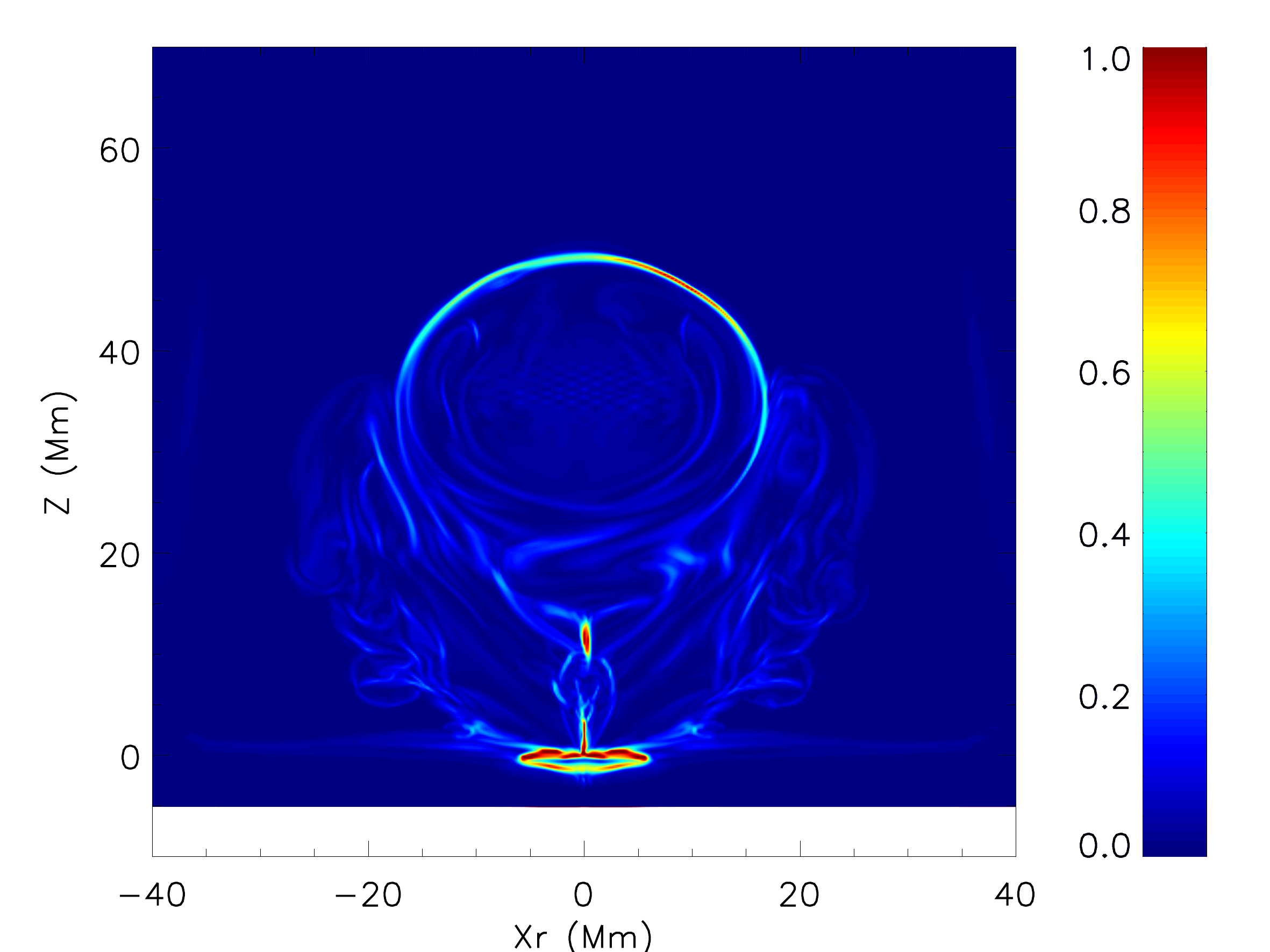}}
     \put(0.5,7){\colorbox{white}{{\color{blue} Sim 2: $t=100t_{0}$}}}
    \put(3.5,7){\colorbox{white}{{\color{blue} Sim 2: $t=240t_{0}$}}}
    \put(6.5,7){\colorbox{white}{{\color{blue} Sim 2: $t=380t_{0}$}}} 
    \put(0,2.5){\includegraphics[width=0.31\textwidth]{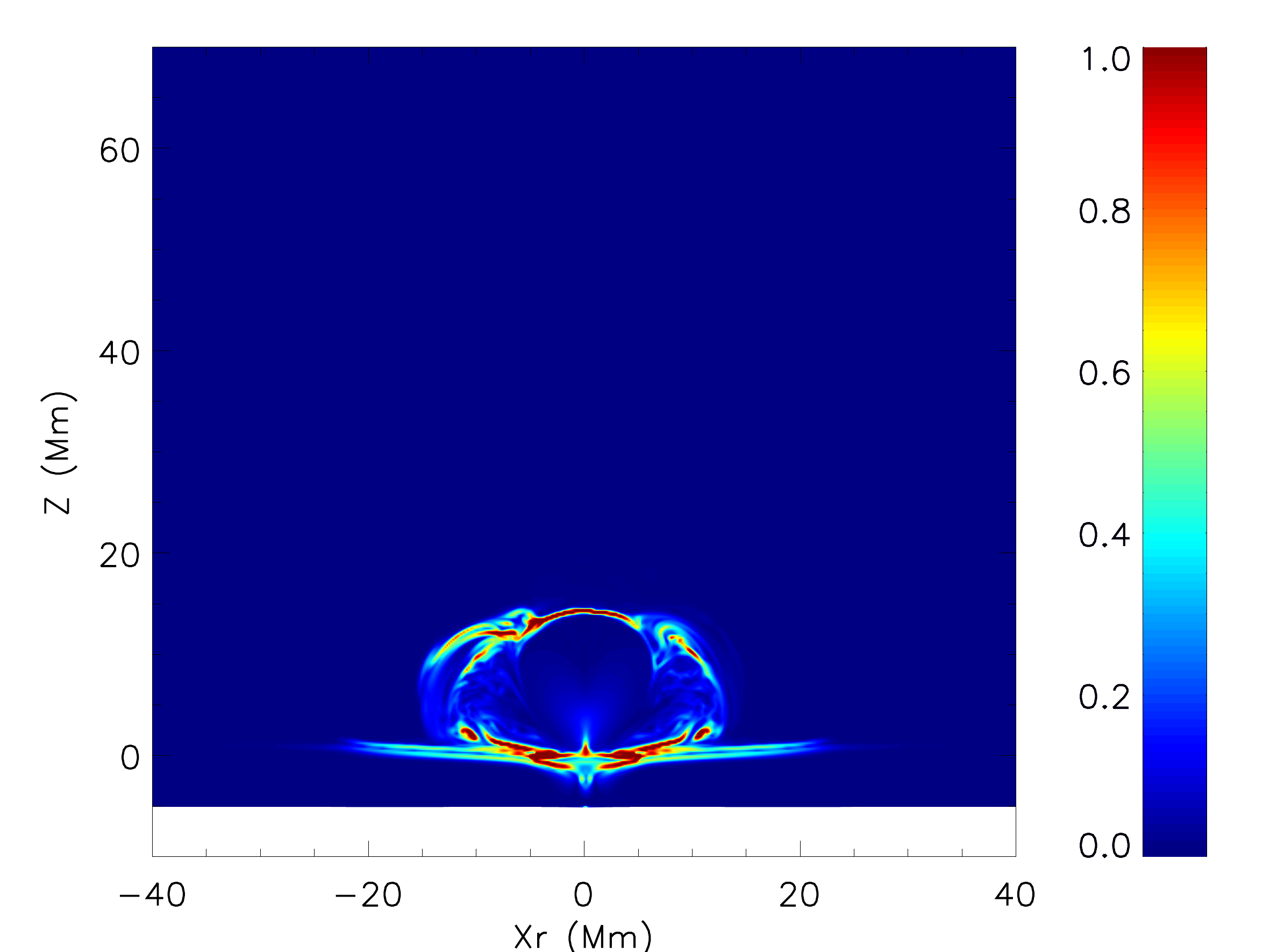}}
    \put(3,2.5){\includegraphics[width=0.31\textwidth]{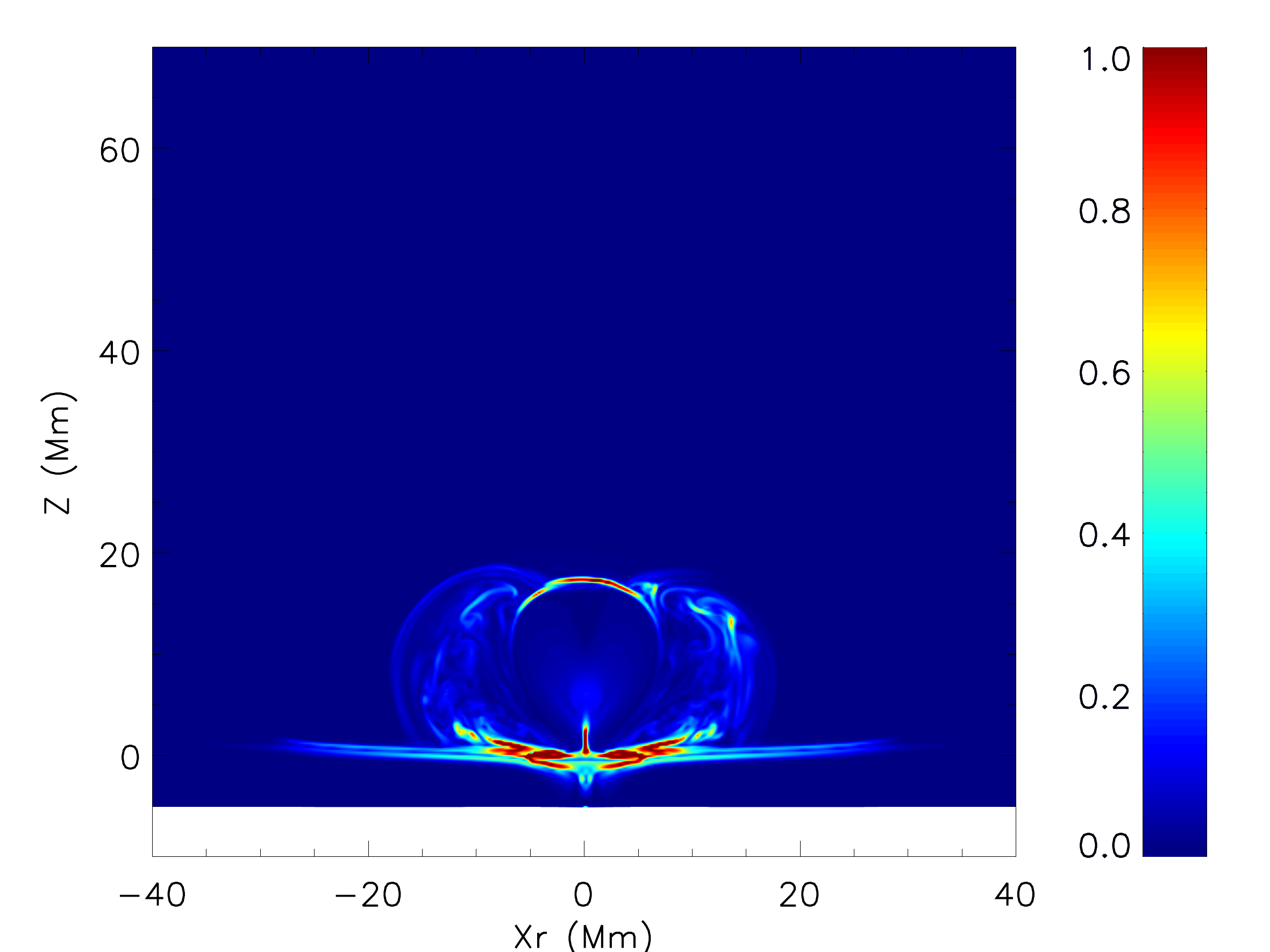}}
    \put(6,2.5){\includegraphics[width=0.31\textwidth]{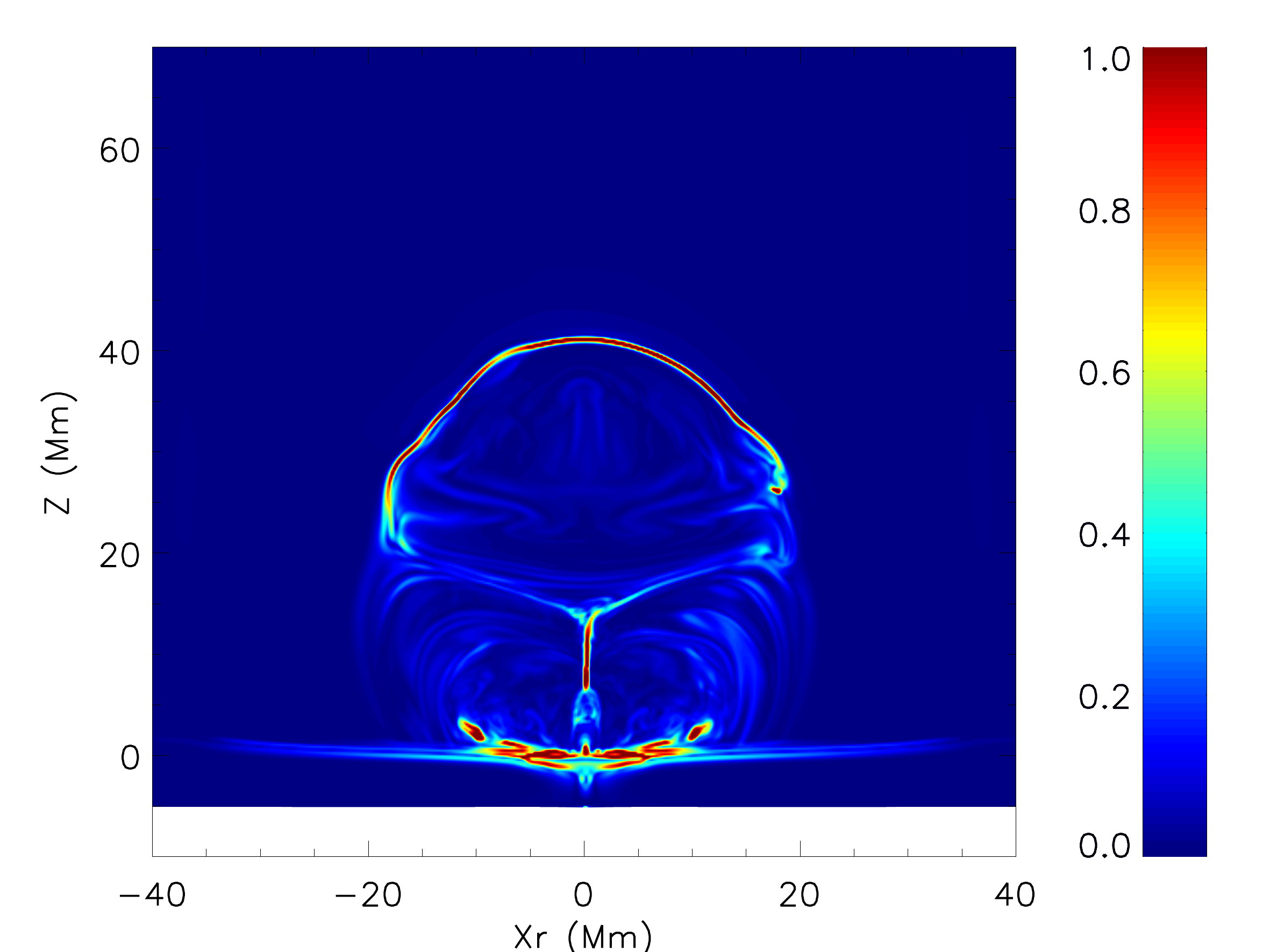}}
     \put(0.5,4.5){\colorbox{white}{{\color{orange} Sim 4: $t=130t_{0}$}}}
    \put(3.5,4.5){\colorbox{white}{{\color{orange} Sim 4: $t=160t_{0}$}}}
    \put(6.5,4.5){\colorbox{white}{{\color{orange} Sim 4: $t=190t_{0}$}}}
     \put(0,0){\includegraphics[width=0.31\textwidth]{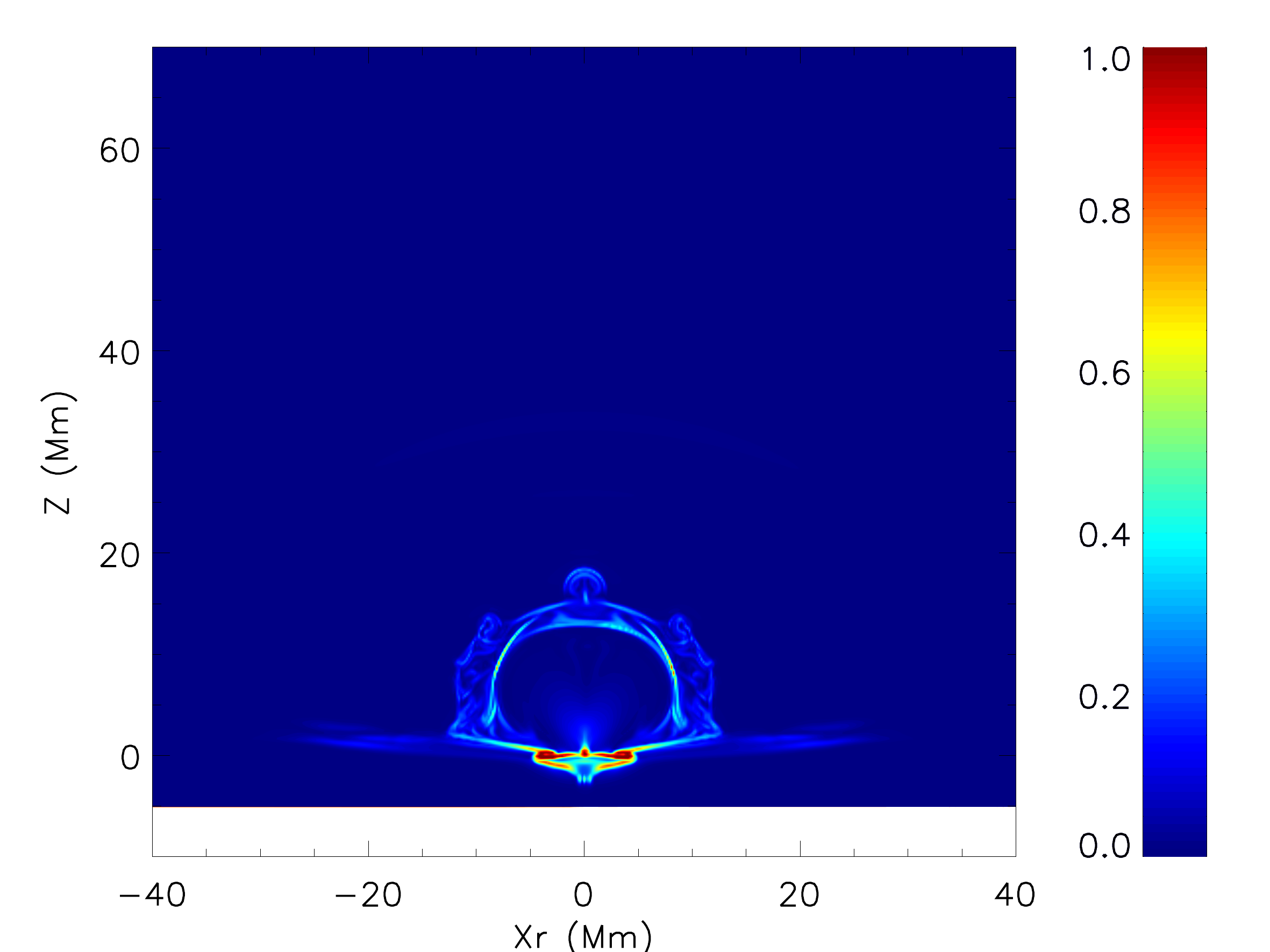}}
    \put(3,0){\includegraphics[width=0.31\textwidth]{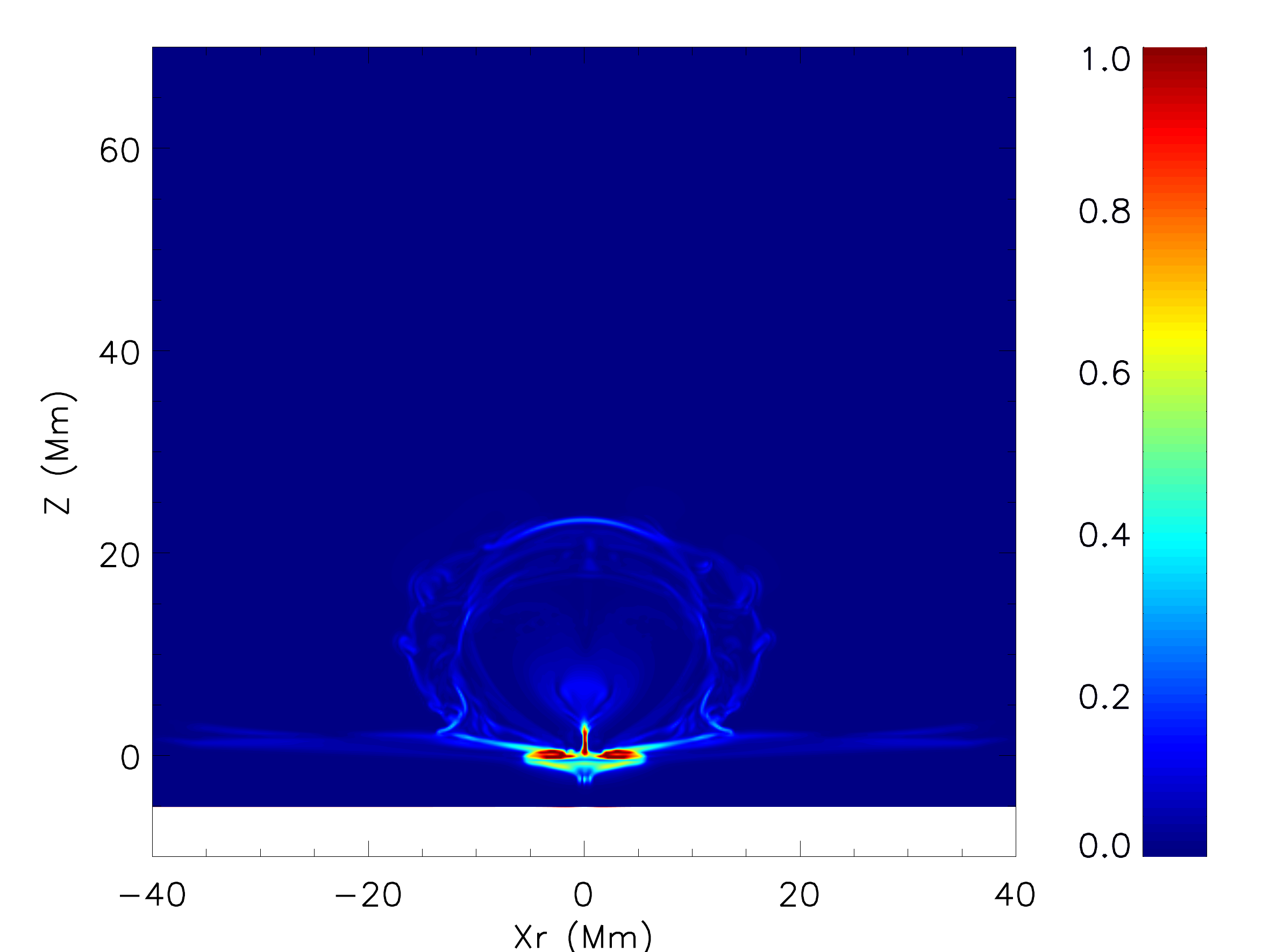}}
    \put(6,0){\includegraphics[width=0.31\textwidth]{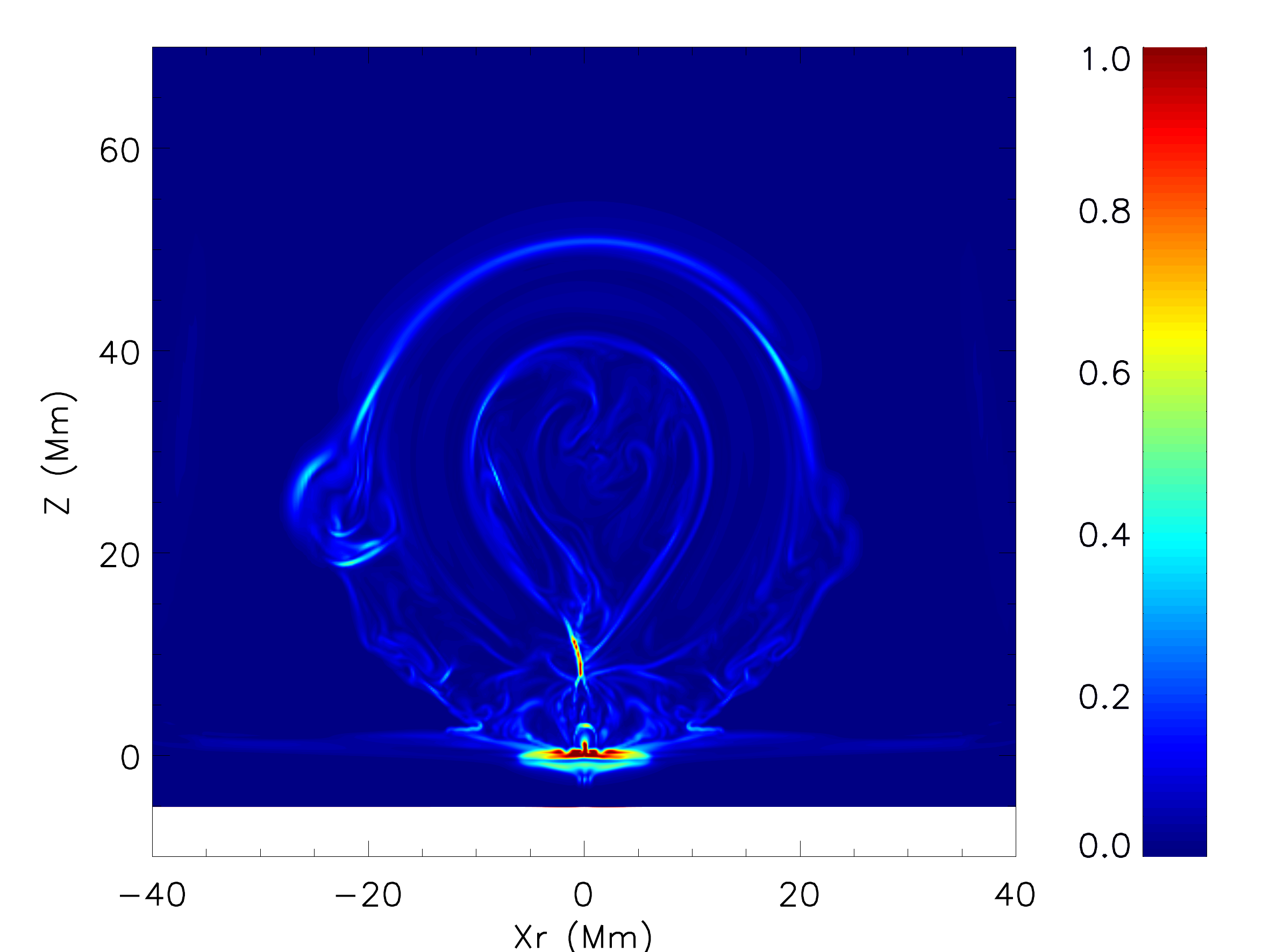}} 
    \put(0.5,2){\colorbox{white}{{\color{NavyBlue} Sim 6: $t=120t_{0}$}}}
    \put(3.5,2){\colorbox{white}{{\color{NavyBlue} Sim 6: $t=190t_{0}$}}}
    \put(6.5,2){\colorbox{white}{{\color{NavyBlue} Sim 6: $t=260t_{0}$}}}
   \end{picture}
  \caption{$|\mb{J}/J_0|/|\mb{B}/B_0|$ for Simulations (0,2,4,6) at three different times. \label{fig:JB}}
  \end{figure}

In this section, the effect of the parameter $\theta$, the relative angle between the emerging field and the coronal field, is investigated. Based on the simple description of the relative orientation of the emerging twist field and the dipole field in Figure \ref{fig:orientation}, it seems reasonable to predict that Simulations [7,0,1] ($\theta=[7,0,1,]\pi/4$) will result in a stable configuration, as there would not be a strong Breakout current sheet between the emerging twist field of the flux rope and the coronal dipole field. In a similar way, it seems reasonable to predict that Simulations [3,4,5] ($\theta=[3,4,5]\pi/4$) will behave in the same way as Simulation 4, as the relative orientations presented in Figure \ref{fig:orientation} is similar. Simulations 2 and 6 represent intermediate cases, where the initially emerging twist component is perpendicular to the dipole field, and it is more difficult to predict their behavior.

Despite the different relative orientations between the emerging field and the dipole field in the simulations, all eight simulations show similar signatures at and above the surface in the early phase (up to $t=120t_{0}$), with the same tadpole structure, concave down fieldlines, and a sheared arcade formed as a result of the partial emergence. It is important to note that ultimately all eight simulations do exhibit an internal vertical current sheet in the emerged region as was shown for Simulation 4. This is shown for Simulations [0,2,4,6] in Figures \ref{fig:B_axial}-\ref{fig:JB} which show the axial field and the quantity $|\mb{J}/J_0|/|\mb{B}/B_0|$, respectively (the other Simulations are not shown for brevity, as it will be shown below that the 8 simulations can be divided into groups). Reconnection across this current sheet results in concave up fieldlines above the surface, and the sheared arcade becomes a flux rope. In Simulation 4, discussed above, this transition to a coronal flux rope is close to co-temporal with an eruption, but this is not the case in all the simulations, as the relative orientation dictates how the emerging structure can continue to rise, and therefore dictates the evolution of the vertical current sheet formed low in the atmosphere during the emergence. 

\begin{figure}
\begin{center}
\includegraphics[width=0.4\textwidth]{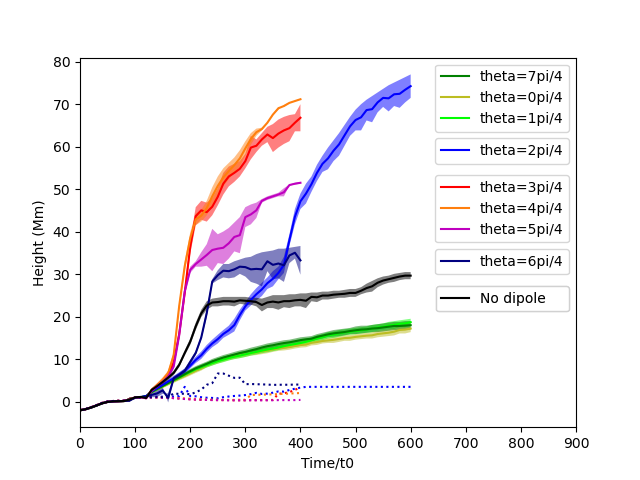}
\caption{Heights of flux ropes in the 8 simulations. Also shown is the simulation with no coronal magnetic field (black lines). {
For each simulation, the semi-transparent colored area corresponds to the range between the two methods used to estimate the flux rope location. The first uses local maxima in the axial magnetic field and the second uses changes in sign in the component of the horizontal field orthogonal to this axial field.} Also shown by the solid colored line for each simulation is the average of these two methods. \RR{The low-lying dashed lines are the height of the original flux rope after the flare reconnection has created a newly erupting flux rope.} \label{fig:heights}}
\end{center}
\end{figure}

\begin{figure}
\begin{center}
\includegraphics[width=0.48\textwidth]{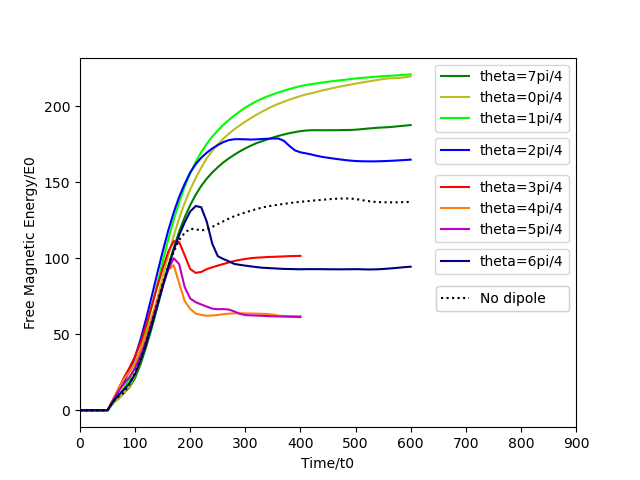}
\includegraphics[width=0.48\textwidth]{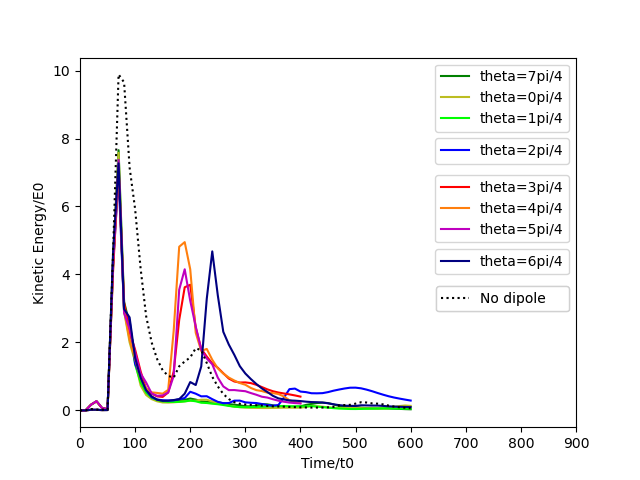}
\caption{ Free magnetic energy (left panel) and kinetic energy (right panel) in the corona (above $10 L_0= 1.7 Mm$) in all 8 simulations. Also shown is the simulation with no coronal magnetic field (dashed black line). \label{fig:energies}}
\end{center}
\end{figure}

Figure \ref{fig:heights} presents the height of the coronal flux rope for all 8 simulations. For each simulation, the height of the flux rope is estimated by two methods.  The first finds the local maximum in the axial magnetic field along the vertical line $x=y=0$, and the second finds the location on the same line where the component of the horizontal field orthogonal to this axial field changes sign. For a symmetric cylindrical flux rope these two methods would result in the same location, but not in general.

\RR{Figure \ref{fig:heights} shows estimates of the flux rope locations as a function of time for all the simulations, using these two methods, along with the average of the two methods. During the early stages of the simulation, as the original flux rope rises to the surface, the two measurements above are very close, as the rope is not significantly distorted. During the emergence of this rope into the atmosphere, its axis remains stuck at the photosphere and only the concave-down fieldlines emerge, and looking only above the surface, one sees a sheared arcade.
 For the stable simulations, where the overlying field inhibits reconnection (7,0,1) only one flux rope axis location is found for the entirety of the simulation, but that does not mean that the original rope emerges bodily into the corona in these simulations. As discussed in \citet{Manchester_2004,Fan_2009,Leake_2013}, and \citet{Leake_2014}, the partial emergence of the original rope is followed by rotational motions and shearing which rotate the fieldlines of the original axis around a new axis.  For the eruptive simulations (2,3,4,5, and 6) however,  the flare reconnection that ultimately occurs converts the observed sheared arcade into a new erupting flux rope, leaving behind the original flux rope  near the surface, as shown in Figure \ref{fig:newropes}. This results in two locations where the component of the horizontal field orthogonal to the axial field changes sign, and the lower of the two locations is shown by the dashed line in Figure \ref{fig:heights}.}


 Figure \ref{fig:energies} shows the free magnetic energy and kinetic energy in the corona in the same 8 simulations. The coronal volume is defined by $z>10L_{0}$. The magnetic free energy is defined by the difference between the total magnetic energy in the coronal volume and the magnetic energy  of 
a potential field given by $\mb{B}=\nabla\phi$ with the boundary condition $\hat{\mb{n}}\cdot\mb{B} = d\phi/dn$ at the 4 side boundaries, the top boundary, and the plane $z=10 ~ L_0$. The process of emergence for $t<120t_{0}$ manifests in an increase in the free magnetic energy in the corona, and a large peak in kinetic energy, due to the upward flows and the lifting of convection zone material by the emerging magnetic fields. During this period, the 8 simulations behave in a similar fashion. The emerging field, which is the same in all 8 simulations, dominates the evolution, while the overlying dipole field has little effect on the evolution until later in the simulations. 

\begin{figure}[h!]
  \centering
  \setlength{\unitlength}{0.1\textwidth}
  \begin{picture}(10,10)
 \put(0,5){\includegraphics[width=0.49\textwidth]{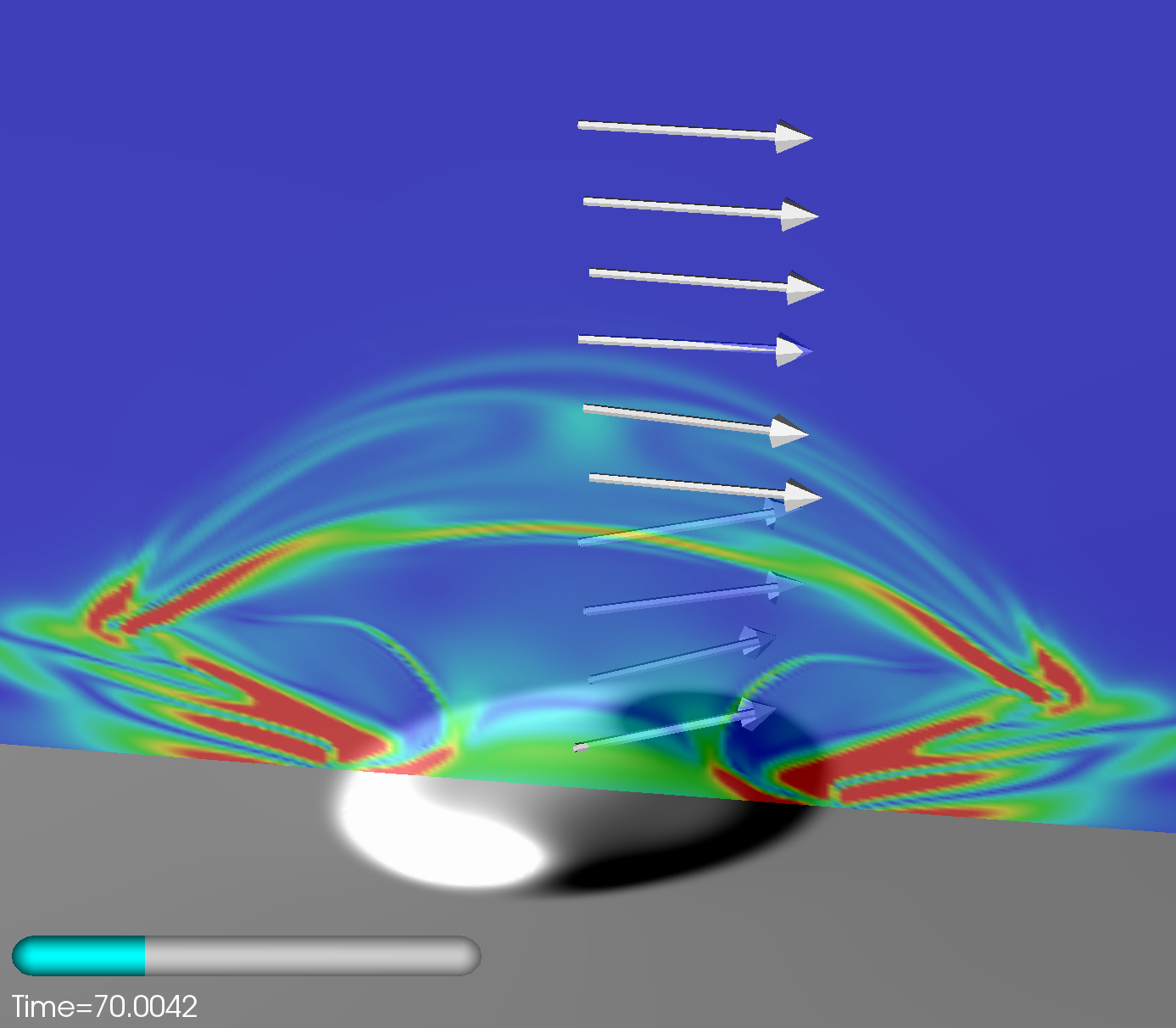}}
 \put(5,5){\includegraphics[width=0.49\textwidth]{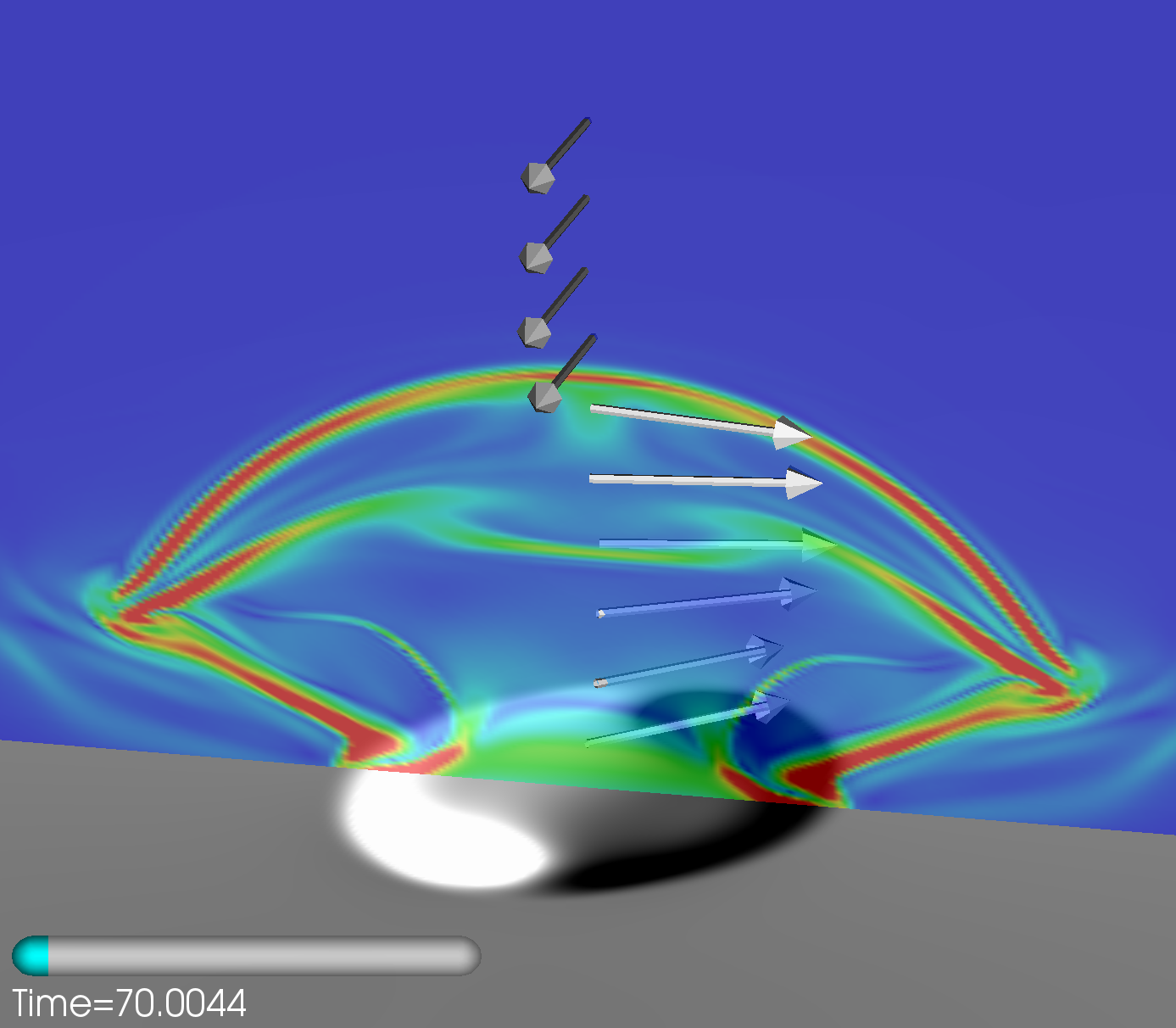}}
 \put(0,0.5){\includegraphics[width=0.49\textwidth]{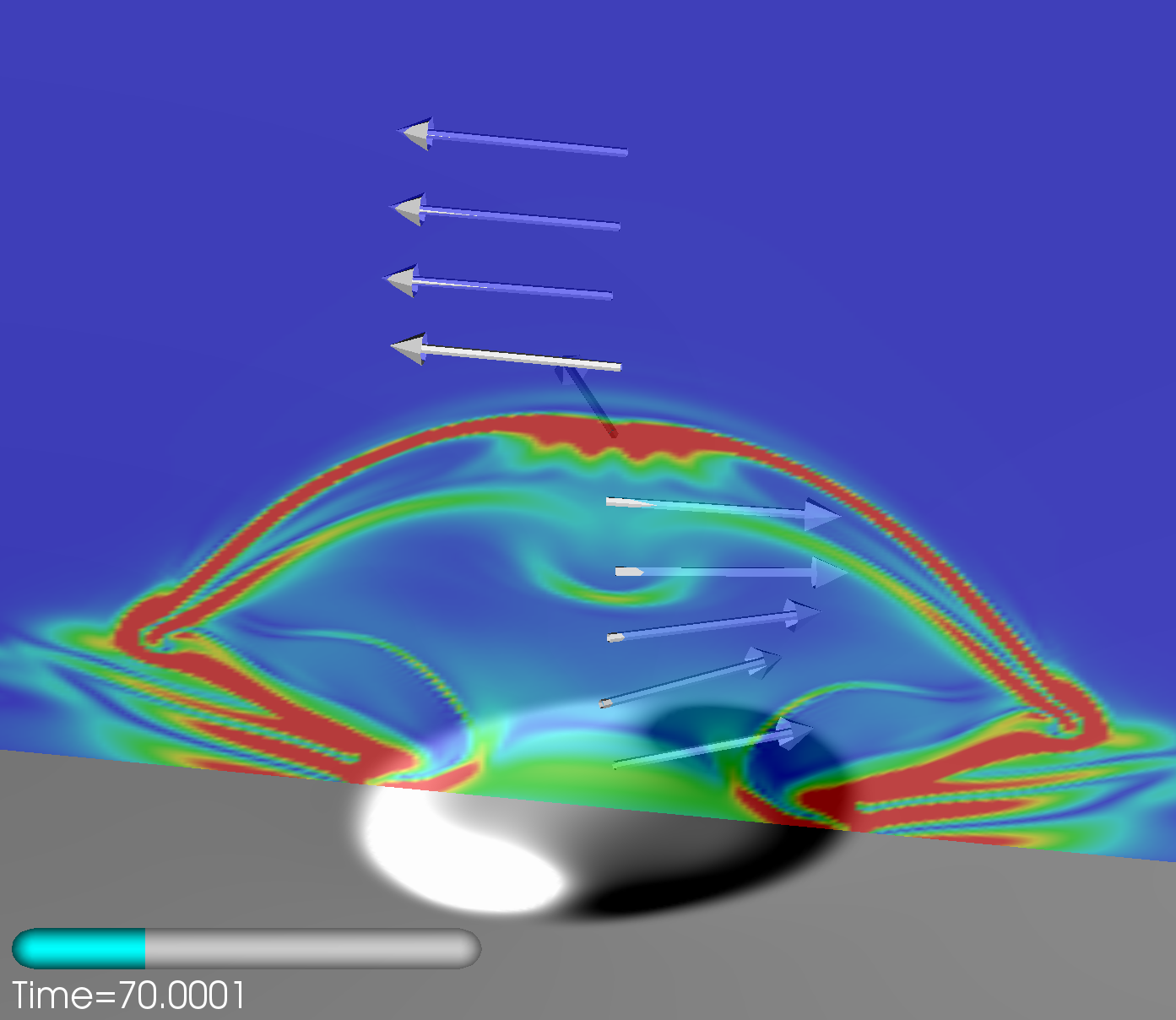}}
 \put(5,0.5){\includegraphics[width=0.49\textwidth]{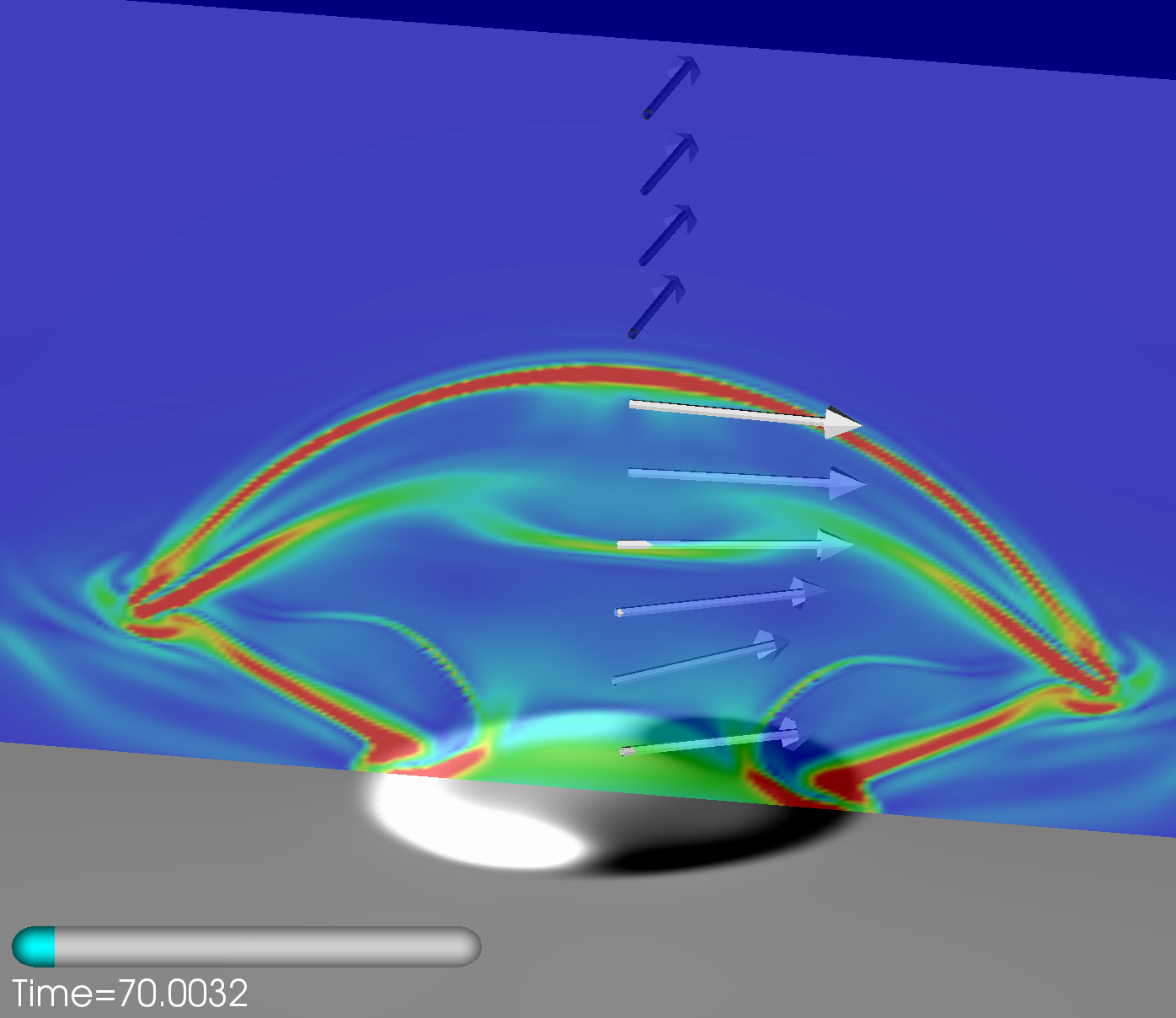}}
 \put(0.2,9){\colorbox{white}{{\color{olive} Sim 0: $t=70t_{0}$}}}
 \put(5.2,9){\colorbox{white}{{\color{blue} Sim 2: $t=70t_{0}$}}}
 \put(0.2,4.5){\colorbox{white}{{\color{orange} Sim 4: $t=70t_{0}$}}}
 \put(5.2,4.5){\colorbox{white}{{\color{NavyBlue} Sim 6: $t=70t_{0}$}}}
 \end{picture}
 \caption{ Direction of field in Simulations 0 (top left), 2 (top right), 4 (bottom left), 6 (bottom right), along with $B_z$ in the photosphere,  and $|\mathbf{J}/J_0|/|\mathbf{B}/B_0|$ color contours. The color scales for $B_z$ and $|\mathbf{J}/J_0|/|\mathbf{B}/B_0|$ are the same across the four plots. 
 \label{fig:angle_arrows}}
 \end{figure}
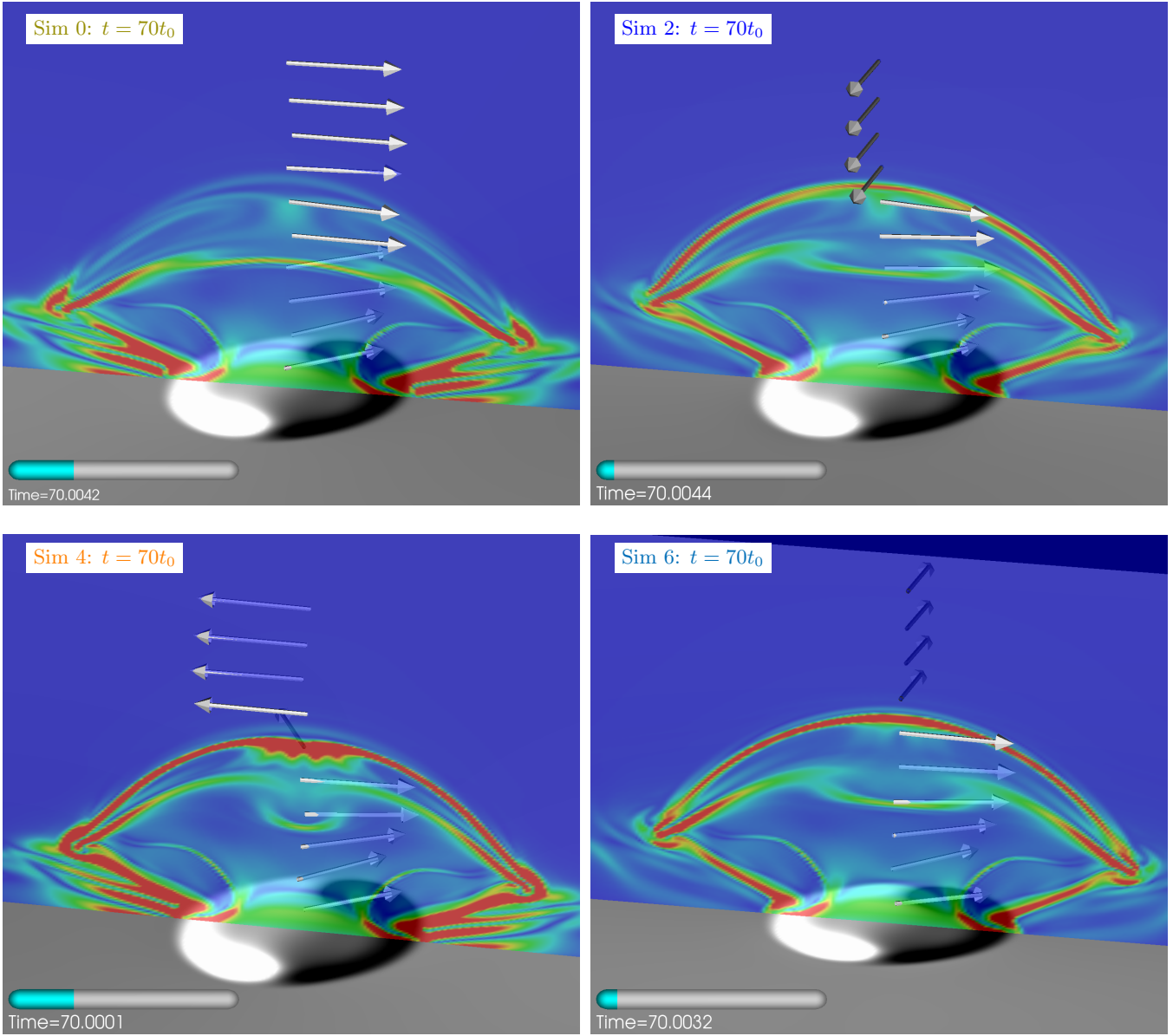

After $t=120t_{0}$ the 8 simulations diverge in their behavior. The three simulations [7,0,1] ($\theta=[7,0,1]\pi/4$) show similar behavior to each other.  These simulations are colored various shades of green in Figures \ref{fig:heights}-\ref{fig:energies}. As can be seen in Figure \ref{fig:orientation}, the emerging twist component of the flux rope is mainly aligned with the dipole field, and so there is very little \JEL{Breakout reconnection}. 
This is shown in the top left panel of Figure \ref{fig:angle_arrows}  for Simulation 0 ($\theta=0$). This figure shows the value of $|\mb{J}/J_0|/|\mb{B}/B_0|$ in the mid-plane and the direction of the magnetic field. The current between the emerging field and dipole field is relatively weak. 

 As a result of the lack of Breakout reconnection, the coronal structure formed by the emergence remains stable, with both the dipole field and the twist component of the magnetic field that emerges first constraining the axial component that emerges later. The coronal kinetic energies in these 3 simulations (7,0,1), shown by green lines in Figure \ref{fig:energies}, right panel, decreases with time after $t=150t_{0}$, and the coronal magnetic free energy increases relatively slowly near $t=600t_{0}$. This behavior is the same as in the stable simulation in \citet{Leake_2013}, and is as predicted by the consideration of the parameter $\theta$ in Figure \ref{fig:orientation}.

In contrast to the stable simulations ($\theta=[7,0,1]\pi/4$), the three Simulations [3,4,5] ($\theta=[3,4,5]\pi/4$) 
 all show a \JEL{second} large increase in kinetic energy \JEL{at around 150-300 $t_0$}, an associated drop in free magnetic energy in the corona, and a fast rise of the coronal flux rope, which here is identified as an eruption. As can be seen in Figure \ref{fig:orientation} the twist component of the flux rope that emerges early has a significant component antiparallel to the dipole field for the cases $\theta=[3,4,5]\pi/4$. Figure \ref{fig:angle_arrows}, bottom left panel, shows the formation of the Breakout current in Simulation 4 due to this sharp change in the magnetic direction with height. As a result of this Breakout current sheet, the emerging field can rise higher into the corona until the vertically aligned current sheet below gives rise to  reconnection and a fast rise of the structure into the upper domain, as discussed for Simulation 4 above.  The fast rise does not last until the end of the simulation, and hence these eruptions are identified as ``confined", which is discussed later.


Figures \ref{fig:heights} and \ref{fig:energies} also show the results for a simulation with no coronal dipole field, but the same rope orientation as Simulation 0, see the black dotted lines in those Figures. The stabilization of the flux rope and the lack of kinetic energy increase at later times highlights the fact that without any coronal field to interact with, the emerging structure will not erupt, as discussed in \S \ref{sec:intro}.

\RR{It is worthwhile here to relate these results to those of similar studies which also report on the behaviour of coronal flux ropes formed by flux emergence.} \RR{\citet{2008ApJ...674L.113A,2012A&A...537A..62A} and \citet{Leake_2013}
reported on the height evolution of coronal flux ropes formed by flux emergence into a field-free corona. All these previous studies had smaller simulation domains than this study (top boundaries at 19-25 Mm compared to 90 Mm in this study). For comparable initial conditions these earlier studies showed that the newly formed coronal flux rope levelled off  between 12 and 17 Mm which is less than the 30 Mm height at which the coronal flux rope levels off in this study. While this shows that the boundary conditions affect where the coronal flux rope levels off in the earlier simulations, the fact that the 30 Mm height is much less than the 90 Mm top boundary in this study suggests that these coronal flux ropes can indeed be considered stable, and that flux emergence into a field-free corona is unlikely to drive any eruptive behavior.}

\RR{\citet{2008ApJ...674L.113A,2010A&A...514A..56A, 2012A&A...537A..62A}, and \citet{Leake_2014} also all reported on the height evolution of coronal flux ropes formed by flux emergence into a coronal field of opposite sign to the emerging field.  \citet{2008ApJ...674L.113A,2010A&A...514A..56A, 2012A&A...537A..62A} used a horizontal constant coronal field, while \citet{Leake_2014} used a spatially varying field representing a decaying arcade field in the corona. For comparable initial conditions for the sub-surface flux rope but varying coronal field configurations and strengths, these earlier studies all show similar behavior with 
 the coronal flux ropes rising rapidly at 120-200 $t_0$ to heights near the top boundaries at 19-25 Mm (depending on the study). In this study, the evolution is in agreement in the early phase, and the eruptive flux ropes continue rising in the larger domain but are eventually confined at heights of 50-70 Mm. The question of the cause of the confinement (boundary conditions, stabilization of the solution, etc) is discussed in \S\ref{sec:eruption_dynamics}. }

\RR{It is also worth noting that \citet{2012A&A...537A..62A} varied both the direction and strength of the ambient horizontal coronal field in a slightly larger domain than \citet{2008ApJ...674L.113A,2010A&A...514A..56A} with a top boundary at 25 Mm. For the three cases where the ambient field was antiparallel or orthogonal to the top edge of the emerging field, the resultant coronal flux rope experienced a fast rise to the top of the box at 25 Mm, but weakening the strength of the ambient field for the orthogonal cases resulted in a confined coronal flux rope. While these earlier studies use a different coronal field and smaller domain than in this study, it is clear that the interplay of orientation and strength of coronal field needs to be considered to get the full picture. In this study the coronal magnetic field is arched, and so the coronal field has magnetic tension that is not present in the studies with a horizontal coronal field, but the free parameters of this model magnetic field will be varied to investigate this interplay in future studies.}

As can be seen in Figures \ref{fig:heights}-\ref{fig:energies}, Simulations 2 and 6 behave very differently from each other. Simulation 6 behaves in a similar fashion to simulations [3,4,5] ($\theta=[3,4,5]\pi/4$) with a fast rise and a confined eruption, and an associated decrease in free magnetic energy and rise in kinetic energy in the corona.
Simulation 2 initially behaves like Simulations [7,0,1] ($\theta=[7,0,1]\pi/4$), showing little to no increase in kinetic energy during the time (150-300 $t_0$) when  simulations 3-6 are erupting into the upper corona. However, the coronal flux rope formed during the emergence appears to slowly rise into the corona until about $t=350$ and then undergoes a short period of faster rise, with a drop in  free magnetic energy and a small rise in kinetic energy in the corona,  before slowing down again. Thus it shows a later confined eruption. The overall evolution of the two Simulations [2,6] can be seen in Figures \ref{fig:B_axial} and \ref{fig:JB}. 

Based on the initial consideration of the parameter $\theta$, this divergent behavior of Simulations 2 and 6 is not obvious. Why does Simulation 6 show an early confined eruption as in Simulations [3,4,5] while Simulation 2 initially undergoes a slow rise and then a relatively weak confined eruption? Figure \ref{fig:angle_arrows} shows $|\mb{J}/J_0|/|\mb{B}/B_0|$ and the direction of the magnetic field. The top right panel shows Simulation 2, while the bottom right panel shows Simulation 6. In both cases, the magnetic field rotates approximately $\pi/2$ across the region between emerging flux rope and dipole field. However, the current sheet appears stronger in Simulation 6, and this is consistent with the early eruption seen in Simulation 6 compared to the slow rise in Simulation 2. Further investigation of the magnetic field around the Breakout current sheet is clearly required. 

\begin{figure}
\begin{center}
\includegraphics[width=0.75\textwidth]{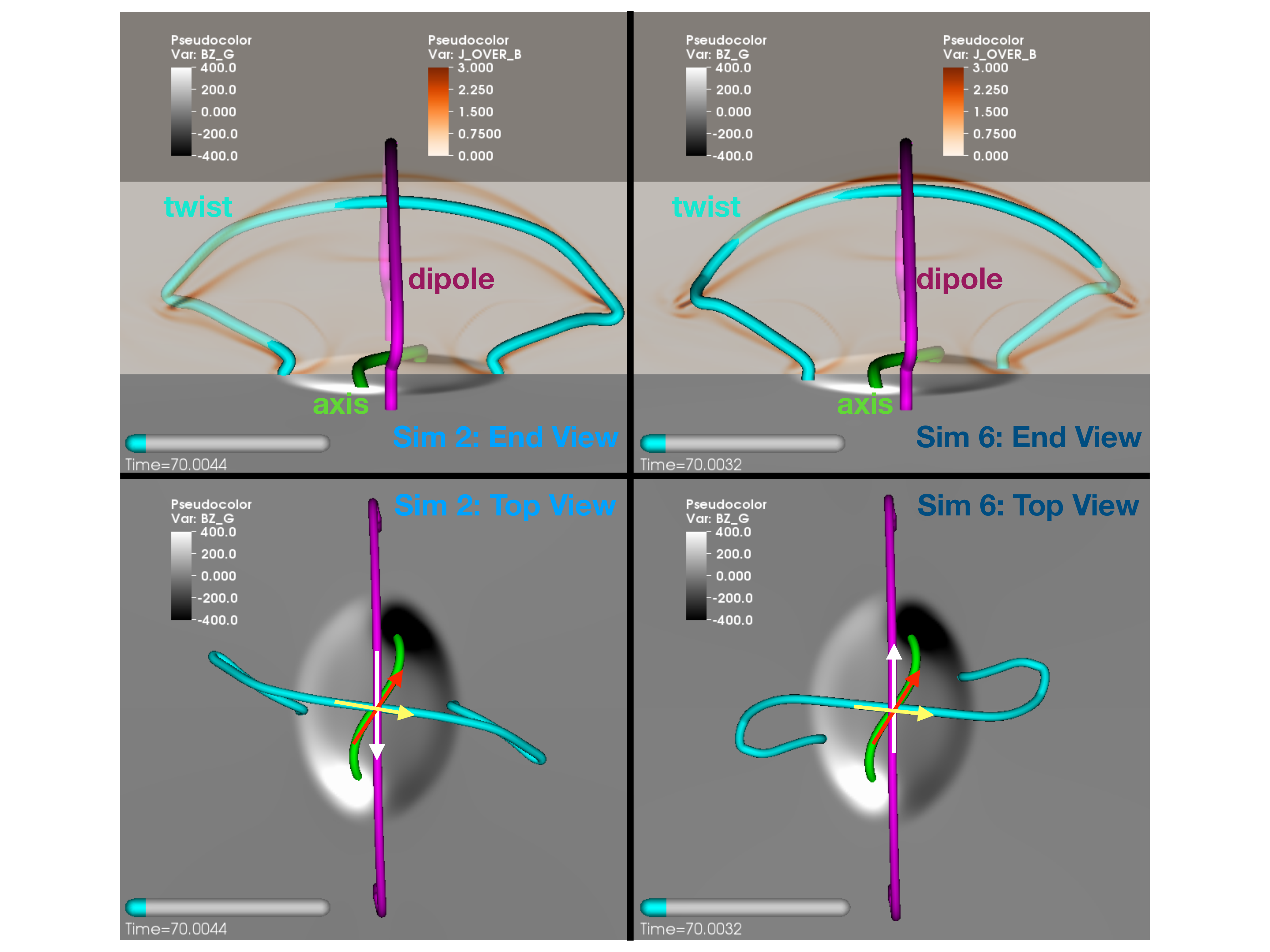}
\caption{Magnetic fieldlines, $B_{z}$ (in G) at the model photosphere, and $|\mathbf{J}/J_0|/|\mathbf{B}/B_0|$ in the cross-sectional plane in Simulation 2 and 6. The green field-line is the axis of the convection zone rope and the red line is the axial field's direction. The cyan fieldline is a sample twisted fieldline away from the rope's axis, and the yellow arrow is the direction of that field-line where it crosses over the axis. The magenta fieldline is from the dipole field above the rope and the white arrow is the direction of the dipole field.  \label{fig:Sim22_Sim26_writhe}}
\end{center}	
\end{figure}

Figure \ref{fig:Sim22_Sim26_writhe} shows select fieldlines in Simulations 2 and 6, at two different angles. The viewpoints are rotated so that the direction of the flux rope axis is the same in each figure, which means the dipole field (represented by the magenta field-line) has opposite orientation in the left and right panels. The green line is the field-line associated with the original axis of the convection zone flux rope, and as can be seen experiences a right handed writhe during the emergence.  As described in \citet{Knizhnik_2018}, the rope is marginally kink unstable, but becomes slightly kink unstable as it rises and expands (increasing the strength of the twist field relative to that of the axial field, to conserve flux). This causes it to writhe in a right handed sense, for this right handed twist. The yellow field lines represent the twist component of the emerging field, and because of the writhe that occurs during emergence, the local direction of the field-line (yellow arrow) experiences a rotation around the $z$ axis.  As a result, on the left panel, it is seen that the twist field in Simulation 2 now has a small component parallel with the dipole field, while the twist field in Simulation 6, right panel, has a  small component antiparallel to the dipole field. This rotation associated with the writhe of the emerging field explains the different behavior in Simulations 2 and 6. Simulation 2 now has a twist field with a component parallel to the dipole field, similar to Simulations [7,0,1] and so initially (but not later) is a stable situation. Simulation 6 has a small twist field component antiparallel to the magnetic field, and so results in an early eruption, as in Simulations [3,4,5].



\subsection{The Eruption Dynamics}
\label{sec:eruption_dynamics}
\begin{figure}
\begin{center}
\includegraphics[width=0.25\textwidth]{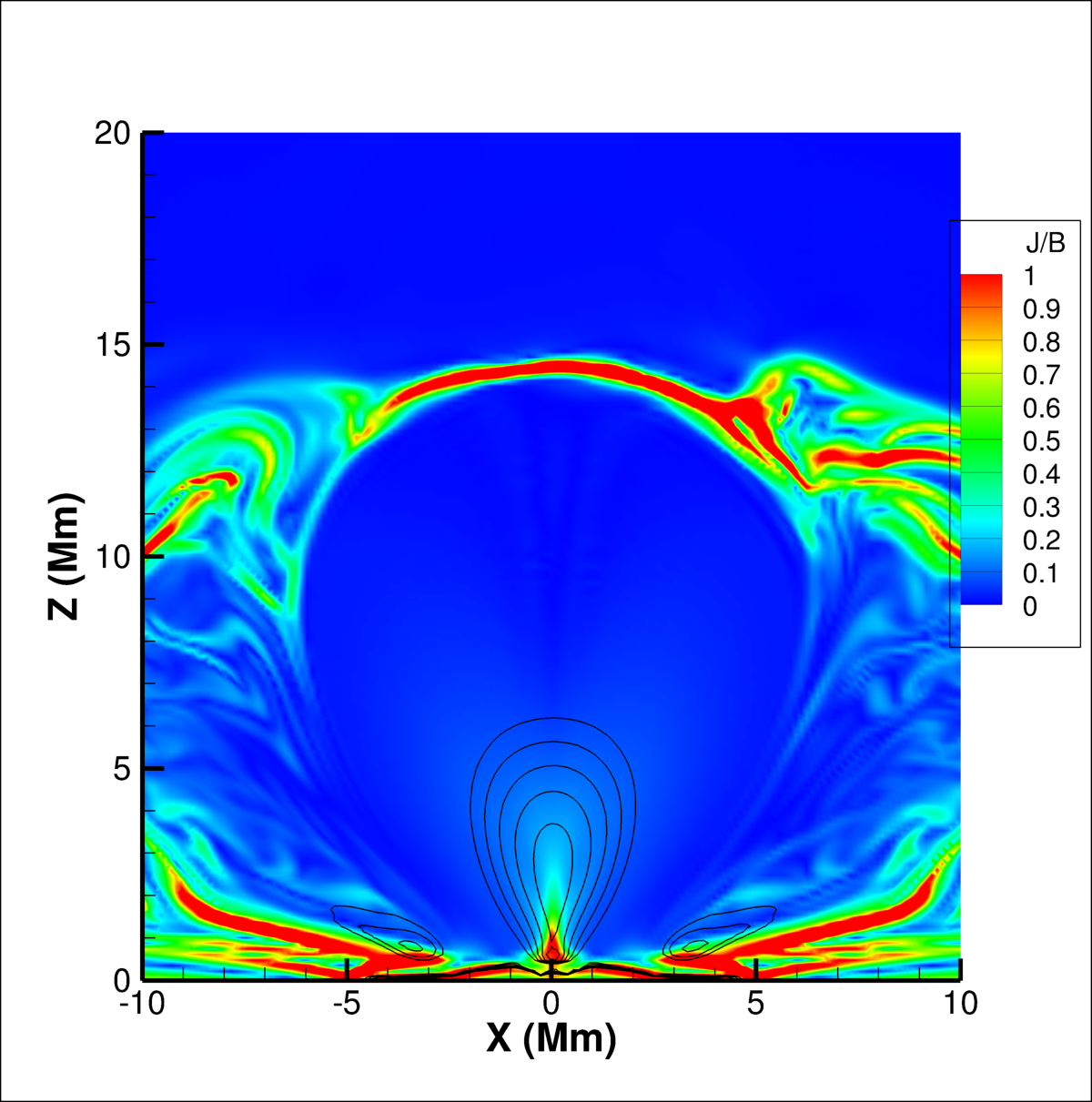}
\includegraphics[width=0.25\textwidth]{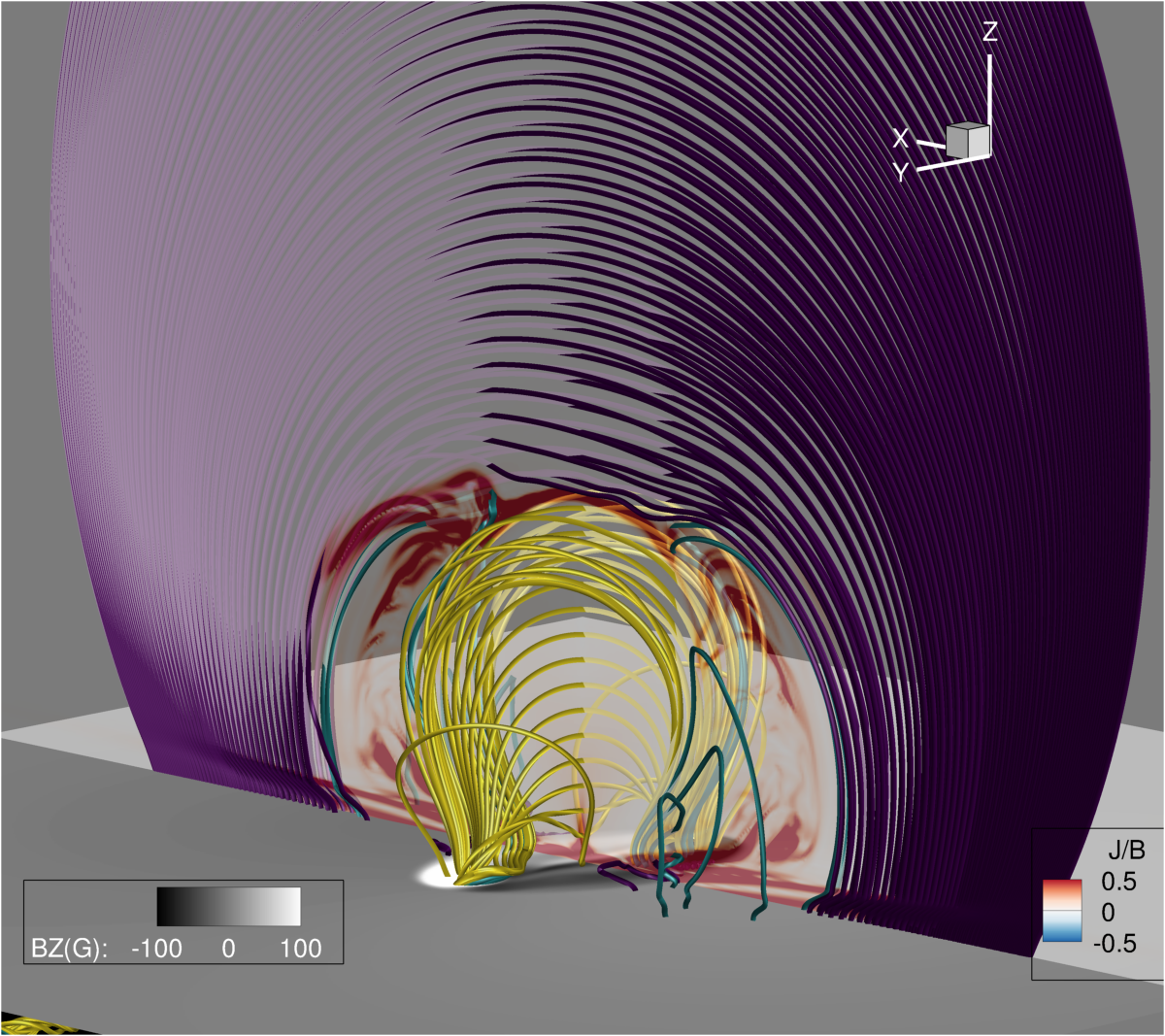}
\includegraphics[width=0.3\textwidth]{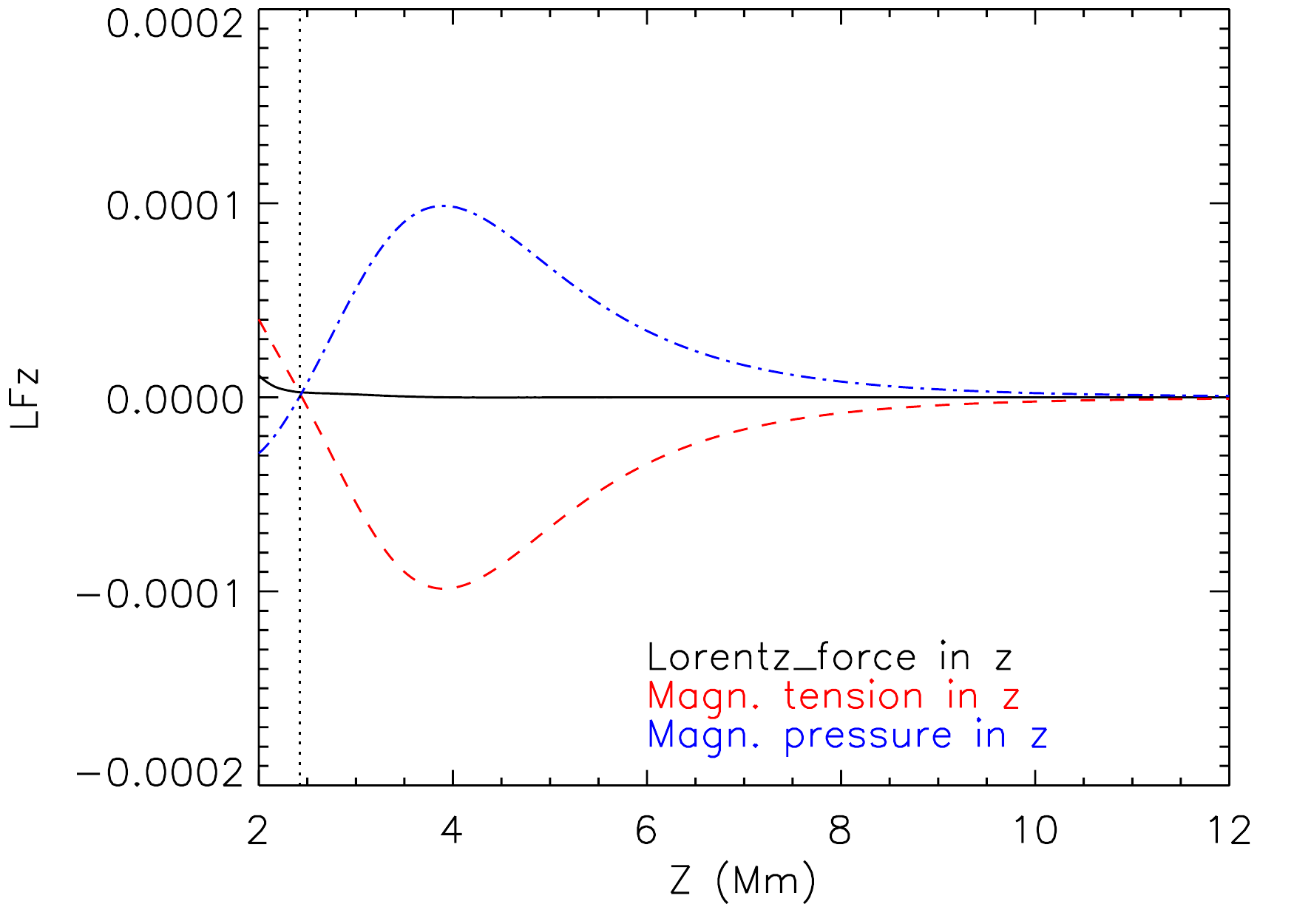} \\
\includegraphics[width=0.25\textwidth]{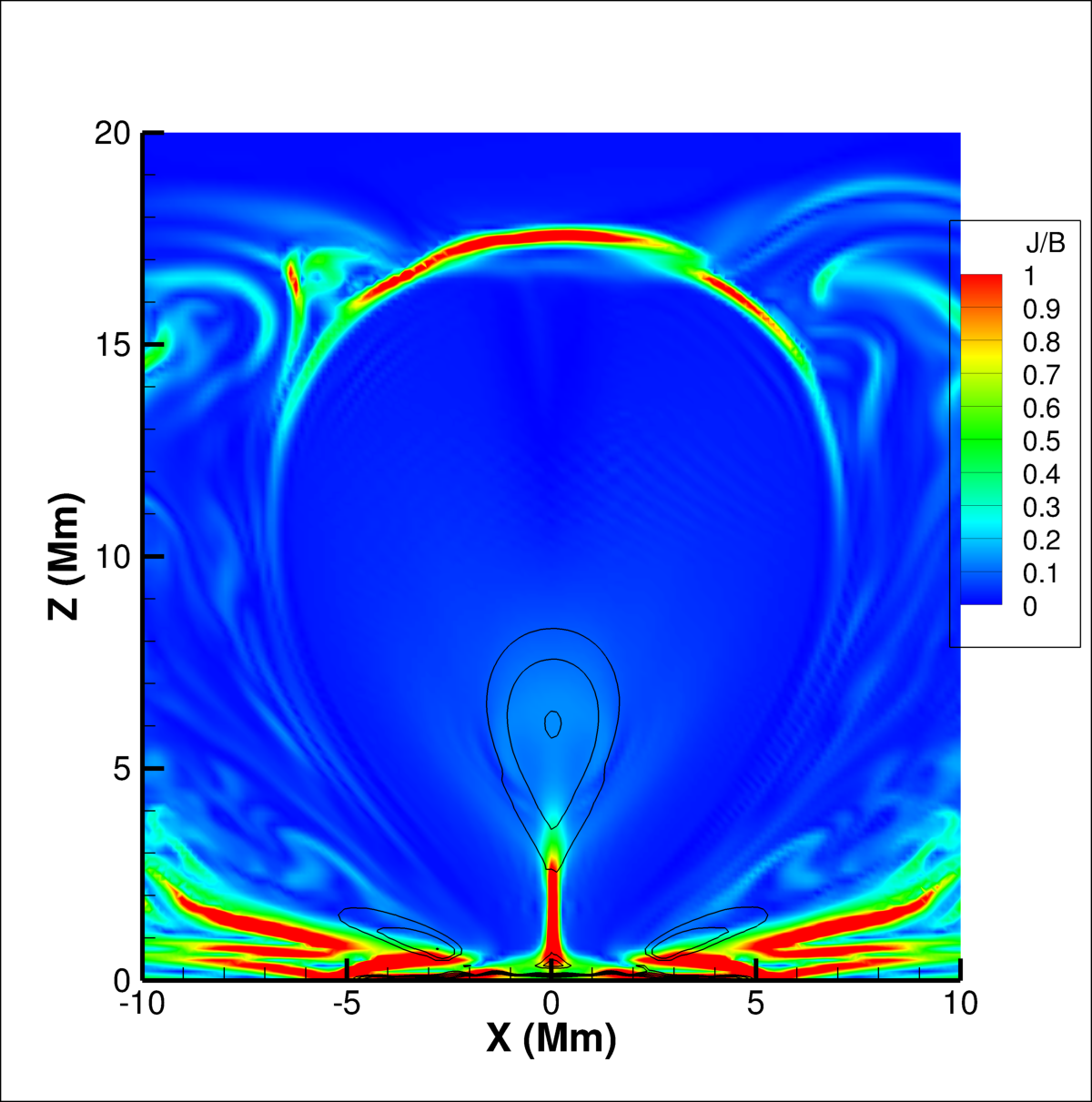}
\includegraphics[width=0.25\textwidth]{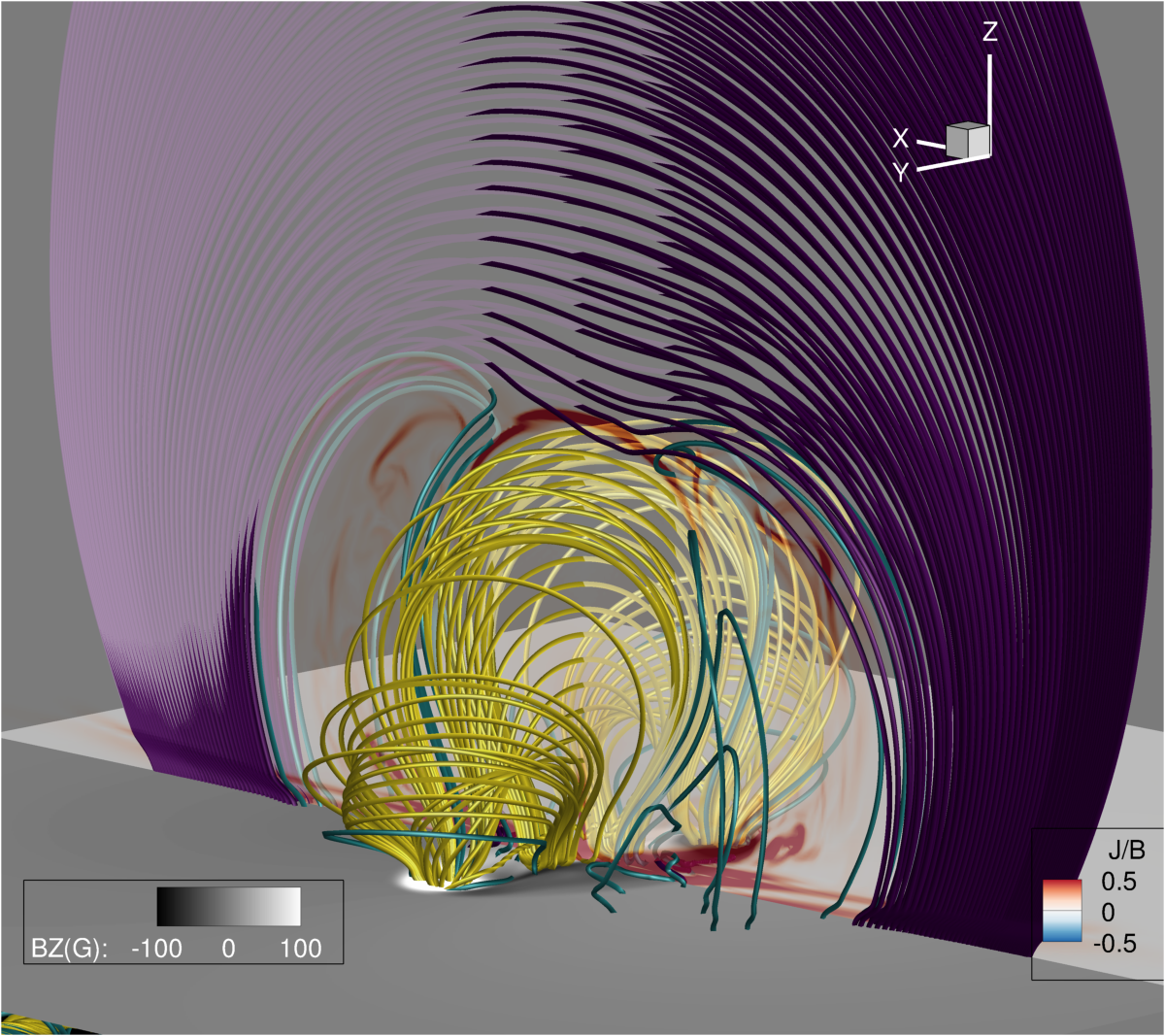}
\includegraphics[width=0.3\textwidth]{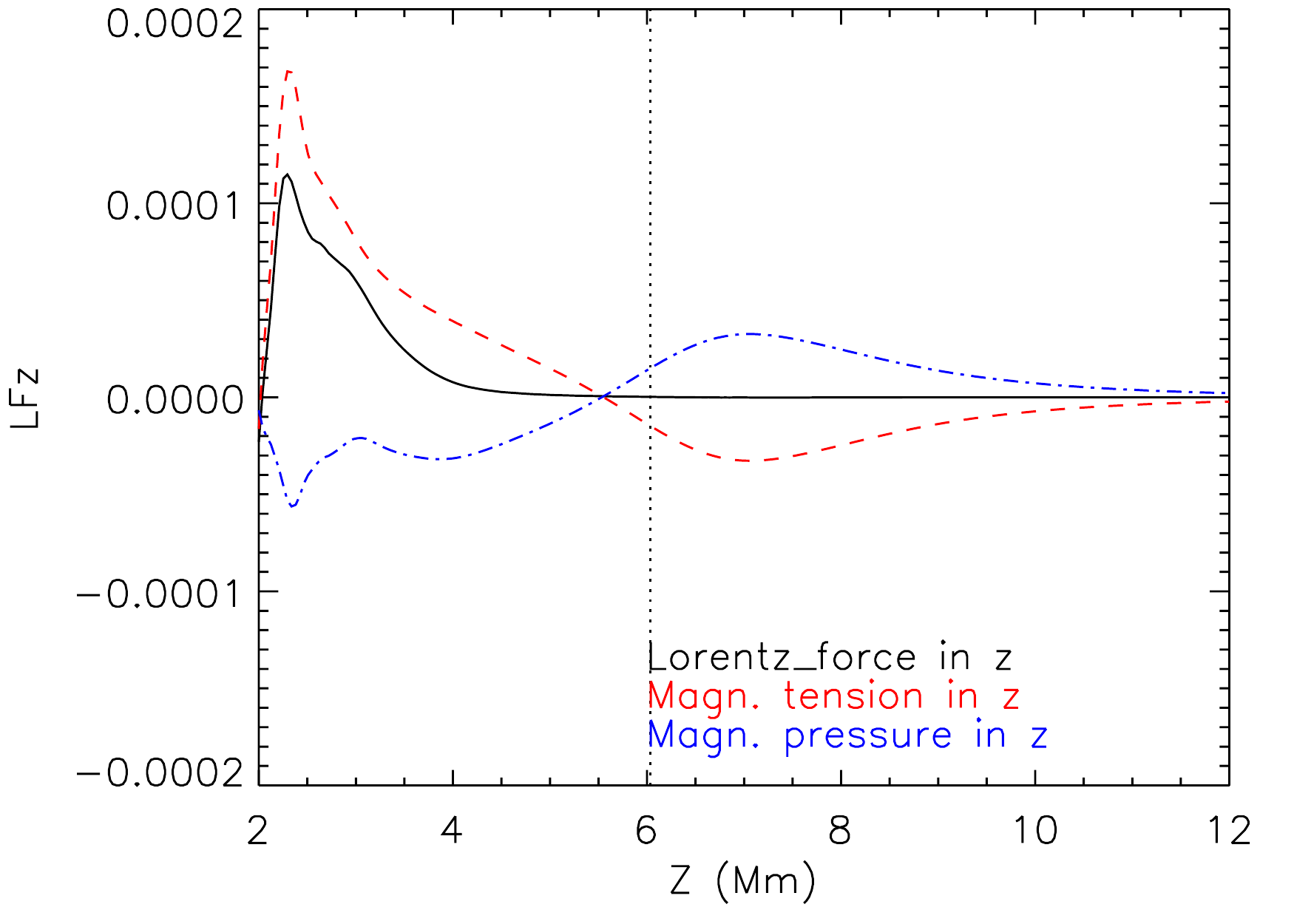}
\caption{Top panels: $t=130 ~ t_{0}$ in Simulation 4.  Left: $|\mathbf{J}/J_0|/|\mathbf{B}/B_0|$ in the mid-plane, along with specific contours of $Log_{10}(|B_{axial}(G)|)$ in the same plane, to aid with location of the flux-rope. Middle:  multiple fieldlines in the same format as in Figure 3. Right: \JEL{$z$}-components of the Lorentz force. Vertical dotted lines show the height at which the axial magnetic field indicated by contours in the left panels is locally maximal along a vertical line (x=y=0). Bottom panels: The same but for $t=160 ~ t_{0}$ in Simulation 4. 
\label{fig:forces}}
\end{center}
\end{figure}

Simulations 2,3,4,5, and 6 appear to be confined eruptions, in that the height of the flux rope reaches a close-to-constant value before the end of the simulations. 
This result is expected for our system in which the side boundaries are impenetrable, so that horizontal cross-sectional area of the domain is fixed with height. Given a constant area, then any overlying flux that is stretched upward by the eruption will have approximately constant field strength; consequently, the energy required to stretch the field upwards increases linearly and indefinitely with height \citep{Sturrock_1991}. Eventually the eruption runs out of energy and stops at some new equilibrium height. Such behavior has been seen in previous studies of CME eruption in Cartesian domains, for example, the loss-of-equilibrium models \citep{Forbes1991,Lin02}.  In order to test the conclusion that confinement is simply due to the geometry of the system, we performed additional simulations for $\theta=4\pi/4$ with a higher top boundary at $z=114.0 ~ \textrm{Mm} $ compared to $z=90.2 ~ \textrm{Mm}$, and 
velocity damping region starting at $z_d=99.7 ~ \textrm{Mm}$ compared to $z_d=78.9 ~ \textrm{Mm}$. A difference of only a few percent was seen in the resultant final height of the erupting flux rope. While this does not completely rule out other possibilities, it does strongly suggest that the eruption confinement is an intrinsic property of the solution. 

 In order to understand better the physics of the eruption dynamics, the forces in the momentum equation were examined. Looking above 2 Mm, it is reasonable here to just focus on the Lorentz force, as the gas pressure and gravity forces rapidly fall off with height in the stratified atmosphere. Figure \ref{fig:forces} shows the magnetic pressure and magnetic tension components of the vertical Lorentz force just before and during the eruption in Simulation 4. 

For context, we note the role of magnetic forces in the more well-known coronal-only eruption models. In \citet{Kliem_2006}, the evolution of a coronal flux rope is due to a balance, or a loss of balance, between the hoop force and the strapping/tension force, with the hoop force due to a gradient in twist field along the torus' minor axis.  In simulations that operate in the Breakout paradigm it is proposed that downward magnetic tension balances upward magnetic pressure until the tension is released by Breakout reconnection, so the upward magnetic pressure force starts to win and drives a runaway process, which results in the rapid flare reconnection and an eruption.

In the simulations presented here, there appears to be a close balance between magnetic pressure and magnetic tension in the sheared arcade and during its transition to a flux rope, as shown at $t=130 ~ t_{0}$ (top panels of Figure \ref{fig:forces}). This is to be expected, because if these forces were substantially out-of-balance, Alfvenic motions would quickly result, and these are rarely seen except  during the actual eruption.  The reversal of the two components of the Lorentz force at approximately 2.25 Mm suggest that the transition from sheared arcade to coronal flux rope has already occurred. There is a small residual Lorentz force that drives the slow expansion, but it is difficult to determine if the Breakout reconnection is removing magnetic tension as proposed above. 

  By $t=160 ~ t_{0}$, the flare reconnection has already developed in the low-lying vertically aligned current sheet. As discussed above and in \citet{Leake_2014} this produces bi-directional flows and retracting highly curved fieldlines in either direction, both upward and downward. These highly curved fieldlines that are exhausted by the flare reconnection produce a large upward tension force at a height of 2.25 Mm. This is the only force that appears able to drive the fast rise of the newly formed flux rope into the corona. \RR{This result is consistent with the force analysis of simulations of the evolution of coronal flux ropes formed by flux emergence by \citet{2012A&A...537A..62A}.}  At higher points there is still a balance in the erupting flux rope between magnetic pressure and magnetic tension.  At later times, this upward tension force is reduced, and there is no obvious force that can drive any further expansion. 

  After a period of rapid rise, the effect of the flare-reconnection on the rise of the flux rope appears to diminish, likely for a few reasons. First and foremost, the shear component of the reconnecting flux decreases so that there is less upward pressure to drive further reconnection. The flare-reconnected fieldlines are moving to a relaxed state, and so have reduced upwards force \citep[See, e.g., cartoon in ][]{Welsch_2018}. Furthermore, the flare-reconnection slows down or stops, and so stops generating upward forces. Finally, the reconnection gets farther from the eruption and so will likely have less of an effect on the eruption. It should also be noted that the erupting structure is continually losing flux due to the ongoing Breakout reconnection, and so it could be that the upward directed Lorentz force driving the initial eruption decreases due to this loss of flux. 
It is left to a future study to further analyze the magnetic forces and fluxes in these simulations along with coronal-only simulations of eruptions.

A key point is that in all simulations presented, the emerged magnetic field undergoes a transition from sheared arcade to flux rope in the corona, and thus it is important to discuss the results in the context of the Torus Instability coronal eruption model \citep{Kliem_2006}. The emergence dynamics  in the simulations presented here result in a coronal flux rope such that the cross product of the rope's axial current and its  internal strapping field creates a downward strapping force.  In the stable simulations of [7,0,1] the coronal field adds to this strapping field, and so in the context of the torus instability, the strapping force is balancing the hoop force.  These simulations do not produce a Breakout current sheet, and so the strapping field cannot be removed by reconnection. Other approaches must be used to remove strapping field, such as parasitic emergence of new flux, as studied in \citet{Dacie_2018}, consistent with the observations of in \citet{Feynman_1995}. 
Of course, these approaches also rely on reconnection rather than a purely ideal instability to induce eruption. In the unstable simulations [3,4,5] the strapping field of the pre-existing coronal field is in the opposite orientation, \JEL{but its tension still exerts a downward force on the underlying flux rope. Such a configuration, with a reversal in the strapping field above a flux rope, and a corresponding separatrix surface and magnetic null, is not a configuration that is addressed by torus instability models.} It is therefore not clear what the implications of this topology, if any, are for the ideal torus instability. As this separatrix will reconnect and generate a Breakout eruption if it can, a flux preserving code such as FLUX \citep{Rachmeler10} would be required to investigate this issue of whether an ideal Torus Instability eruption could also occur in such a configuration.


\section{Discussion}
\label{sec:discussion_new}

The results above have several important implications for understanding solar eruptions, but perhaps the most critical one is that they confirm the basic argument made in the Introduction that the photosphere-corona interaction must be included when modeling the energy buildup. The cases 0, 1, and 7 above did not lead to eruption even though the dynamics of the emergence were quite similar in all cases. A sheared filament channel formed along with some low-lying reconnection that imparted twist to the filament channel. The filament channels did not erupt in cases 0, 1, and 7 because they did not have sufficient free energy to push open the overlying flux. If, instead, we had simply imposed a continuous shear and/or cancellation at the photosphere, then it is likely that an eruption would have eventually occurred for all the cases above. In fact, recent simulations show exactly this result, namely that a sufficiently large shearing will inevitably produce eruption even in a Cartesian system with the field of a single bipole \citep{Jiang2021}. The key point of our simulations is that one is not free to choose arbitrarily large shear or twist; these must be derived from self-consistent modeling of the photosphere-corona interaction. 

Another critical implication of our simulations is that the interaction between the filament channel flux and the overlying strapping field is essential to determining eruption or non-eruption. We note from the results above that the filament channel structure formed by emergence, itself, is very similar in all cases. This is fully expected, because the photosphere has a much higher energy density than the corona, consequently the coronal magnetic field has a negligible influence on photospheric processes such as flux emergence or convection. The difference between the eruptive and non-eruptive cases is due solely to how the various flux systems interact \textbf{in the corona}.

Finally, our simulations imply that reconnection is the fundamental process for eruption onset. We find that breakout reconnection is responsible for the filament channel flux rising sufficiently to form a large coherent vertical current sheet, and that the onset of fast reconnection in this current sheet is responsible for the fast upward acceleration of the eruption. None of our simulations shows evidence of fast Alfv\'enic motions prior to the onset of fast flare reconnection. In fact, a vertical current sheet forms inside the filament flux very early during the emergence process and some reconnection occurs there in all cases, even the non-eruptive ones. The difference between the non-eruptive and eruptive cases is that in the former this reconnection remains slow or stops completely, whereas in the latter, this reconnection transitions to an explosive rate. The breakout reconnection enables this transition. We conclude that, at least for energy buildup via flux emergence, reconnection is the eruption driver.

In summary, this paper reports on a parameter study of the relative orientation between emerging magnetic flux from the solar convection zone and pre-existing coronal flux, and its effect on the eruptivity of the formed active region. Numerical simulations were performed, varying the relative angle between the twist field of emerging magnetic flux ropes and a coronal dipole field. When the relative orientation is conducive to breakout reconnection between the two flux systems, the resultant active region exhibits an eruption, while for the opposite cases, the resultant AR is stable. In all cases a coronal flux rope is formed. 

As indicated by earlier studies, emerging magnetic field is confined by its own envelope field, and eruption is unlikely unless Breakout magnetic reconnection between emerging fields and pre-existing fields can remove this envelope field. Furthermore, the relative orientation between the two flux systems determines the amount of this Breakout reconnection and is therefore a key parameter in determining the likelihood of eruption. When there is a significant angle between the two flux systems the results are in accordance with simple intuition. 
However, for the two situations where the coronal field is perpendicular to the envelope field of the emerging fields, further analysis is required, and the writhe of the emerging ropes during the early emergence determines the amount of Breakout reconnection and hence the later evolution. 

The eruptions here are characterized by a rapid rise into the upper atmosphere, a sharp peak in kinetic energy in the corona, and a drop in the free magnetic energy. The erupting coronal flux rope is ultimately confined in these simulations. While the upper boundary condition is closed and does not permit the free advection of the eruption, the fact that the ropes appear to stabilize before the upper boundary and that moving the boundary upwards does not significantly change the result implies that the reason for the confined nature of the eruptions is primarily the solution itself.

Preliminary analysis of the forces suggests an approximate balance of the Lorentz forces (magnetic pressure and tension) with the initial eruption driven by the concave up fieldlines escaping the flare reconnection region. Further analysis is required to see how the system evolves in the context of the Torus Instability and the Breakout Model, and what the role of the Breakout and flare reconnection is in adding/removing magnetic flux to the erupting structure.

\section{Acknowledgements}

James Leake was sponsored by the NASA LWS, HGI, and ISFM programs. Mark Linton was sponsored by The Office of Naval Research and by the NASA LWS and NASA HSR programs.
Spiro Antiochos was sponsored by the NASA LWS program. Numerical simulations were performed and archived on NASA High End Computing Capability resources, and the data produced by these simulations, along with the data analysis tools used, are available upon request. 

\bibliography{bibliography}

\end{document}